\documentclass[a4paper, 12pt]{article}
\usepackage[russian]{babel}
\usepackage[cp866]{inputenc}
\usepackage{graphicx}
\usepackage{amsfonts}
\usepackage{amsmath}
\usepackage{amssymb}

\begin{document}

\setcounter{figure}{0}
\setcounter{table}{0}
\setcounter{footnote}{0}
\setcounter{equation}{0}

\baselineskip 8.5mm
\thispagestyle{empty}
{\small \it ISSN 0038-0946, Solar System Research, 2013, Vol. 47, No. 5, 
pp. 386-402. c Pleiades Publishing, Inc., 2013.
Original Russian Text c E.V. Pitjeva, 2013, published in Astronomicheskii 
Vestnik, 2013, Vol. 47, No. 5, pp. 419-435}

\bigskip

UDK 521.172:523.2

\centerline{\large\bf Updated IAA RAS Planetary Ephemerides-EPM2011}
\centerline{\large\bf and Their Use in Scientific Research}

\smallskip
\centerline{\bf\copyright 2013 Ј. \ E. V. Pitjeva}
\centerline{\it Institute of Applied Astronomy, Russian Academy of Sciences,} 
\centerline{\it nab. Kutuzova 10, St. Petersburg, 191187 Russia}
\centerline{\small Received December 20, 2012}

\smallskip\noindent

{\bf Abstract} -The EPM ({\bf E}phemerides of {\bf P}lanets and the {\bf M}oon) 
numerical ephemerides were first created in the 1970s in support of Russian space 
flight missions and since then have been constantly improved at IAA RAS. In the 
following work, the latest version of the planetary part of the EPM2011 numerical 
ephemerides is presented. The EPM2011 ephemerides are computed using an updated 
dynamical model, new values of the parameters, and an extended observation 
database that contains about 680 000 positional measurements of various types 
obtained from 1913 to 2011. The dynamical model takes into account mutual 
perturbations of the major planets, the Sun, the Moon, 301 massive asteroids, 
and 21 of the largest trans-Neptunian objects (TNOs), as well as perturbations 
from the other main-belt asteroids and other TNOs. The EPM ephemerides are 
computed by numerical integration of the equations of motion of celestial bodies 
in the parameterized post-Newtonian n-body metric in the BCRS coordinate system 
for the TDB time scale over a 400-year interval. The ephemerides were oriented 
to the ICRF system using 213 VLBI observations (taken from 1989 to 2010) of 
spacecraft near planets with background quasars, the coordinates of which are 
given in the ICRF system. The accuracy of the constructed ephemerides was 
verified by comparison with observations and JPL independent ephemerides DE424.

The EPM ephemerides are used in astronavigation (they form the basis of the {\it 
Astronomical Yearbook} and are planned to be utilized in GLONASS and LUNA-RESURS 
programs) and various research, including the estimation of the solar oblateness, 
the parameters of the rotation of Mars, and the total mass of the asteroid main 
belt and TNOs, as well as the verification of general relativity, the secular 
variations of the Sun's mass and the gravitational constant, and the limits on 
the dark matter density in the Solar System.

The EPM ephemerides, together with the corresponding time differences TT - TDB 
and the coordinates of seven additional objects (Ceres, Pallas, Vesta, Eris, 
Haumea, Makemake, and Sedna), are available at ftp://quasar.ipa.nw.ru/incoming/EPM.

\smallskip
DOI: 10.1134/S0038094613040059

\setcounter{page}{2}

\bigskip

\centerline{HISTORICAL INTRODUCTION}
\smallskip

Until the coming of the space age in the 1960s, the classic analytical theories 
of planetary motion developed by Le Verrier, Hill, Newcomb, and Clemens, which 
were fully consistent with optical observations in terms of accuracy, were being 
constantly refined in accordance with the development of astronomical practice.

However, the launch of the first satellites exposed the demand for a more 
accurate calculation of the coordinates and the speeds of planets. Deep-space
experiments and the introduction of new observational techniques (lunar and 
planetary ranging, trajectory measurements, etc.) required the development of
planetary ephemerides that would be far more accurate than the classical ones. 
On the other hand, it was the new observational facilities that made it possible 
to develop ephemerides of the new generation.

The errors of the current best ranging observations do not exceed several meters, 
which makes it necessary to compute the ranging correctly up to the 12th 
significant digit. An appropriate model of the motion of celestial bodies is 
required to achieve such high precision. The construction of a proper model that 
would take into account all the significant factors is a serious problem, and the 
current most feasible way to solve it is to perform numerical integration of the
equations of motion of the planets and the Moon on a computer.

In the late 1960s several research groups in the United States and Russia 
developed numerical theories to support space flights. American groups worked at 
the California Institute of Technology and the Massachusetts Institute of 
Technology. Russian high-precision numerical ephemerides of planets (Akim et al.,
1986) were created as a result of the research carried out at the Institute of 
Applied Mathematics, the Institute of Radio Engineering and Electronics and the
Space Flight Control Center, and the Institute of Theoretical Astronomy, where 
N. I. Glebova, G. I. Eroshkin, and a group led by G. A. Krasinsky developed 
theories independently. This work was continued at the Institute of Applied 
Astronomy (IAA), where a series of EPM (Ephemerides of Planets and the Moon) 
ephemerides was produced. In order to provide technological support for such 
research, a large group of developers working at the IAA under the direction of
G. A. Krasinsky created a unique software system called ERA (Ephemeris Research 
in Astronomy) that uses a high-level language targeted at astronomical and 
geodynamical applications. This ensures the flexibility of the system, which is 
being constantly upgraded, and considerably simplifies the development of various 
applications. The two dynamical models of planetary motion that are being 
developed in the series of DE (Development Ephemeris, JPL) (Standish, 1998; 2004; 
Folkner, 2010; Konopliv et al., 2011) and EPM (Krasinsky et al., 1993; Pitjeva, 
2001; 2005a; 2012) ephemerides are currently the most complete, have the same 
precision, and are faithful to modern radio observations. For the reasons of 
technological independence, researchers at the Institut de Mecanique Celeste et 
de Calcul des Ephemerides (IMCCE) have started constructing their own numerical 
planetary ephemerides INPOP (Fienga et al., 2008; 2011) in 2006. The history of 
the creation of planetary ephemerides, the EPM2004 ephemeris and the differences 
between the DE and EPM ephemerides are discussed in greater detail in a paper by 
Pitjeva (2005a). In the present work the planetary part of the latest, updated 
version of the EPM ephemerides (EPM2011) and its use in various scientific 
investigations are discussed.

\bigskip
\centerline{EPM DYNAMICAL MODEL OF PLANETARY MOTION}
\smallskip

Construction of high-precision planetary ephemerides that are needed for space 
experiments, and would guarantee the meter-level accuracy of modern 
observations, requires creating a proper mathematical and dynamical model of the 
motion of planets, which takes into account all the significant perturbing 
factors on the basis of general relativity (GR).

The motion of the barycenter of the Earth-Moon system is appreciably perturbed 
by the Moon itself. The Moon's orbit is subject to perturbations from the
asphericity of the gravitational potentials of the Earth and the Moon, which 
makes it necessary to characterize the positions of the equators of the Earth and 
the Moon with respect to an inertial coordinate system (i.e., take into account 
the impact of precession, nutation, and physical libration) with sufficient 
accuracy. The resonant behavior of the coupling between orbital and rotational 
motions of the Moon makes it essential to reconcile various theories in a unified 
dynamical model. As a consequence, modern numerical theories are built by 
simultaneous numerical integration of the equations of motion of all planets and 
the Moon's physical libration, while also taking into account the perturbations 
on the figure of the Earth due to the Moon and the Sun and the perturbations on 
the figure of the Moon due to the Earth and the Sun. Construction of the theory 
of the Moon's orbital and rotational motions and its improvement using lunar 
laser ranging (LLR) observations are the most difficult tasks in creating modern 
ephemerides of planets and the Moon. This work was carried out at the IAA under 
the direction of G. A. Krasinsky and is described in a series of papers 
(Aleshkina et al., 1997; Krasinsky, 2002; Yagudina et al., 2012). The lunar 
theory takes into account the effects associated with elasticity, tidal 
dissipation of energy, and the frictional interaction between the Moon's liquid 
core and its mantle, and cites selenodynamical parameters obtained through the 
analysis of LLR observations made from 1970 to 2010.

The influence of solar oblateness on planetary motion was established 
theoretically a long time ago, and some researchers even tried to attribute to 
it the anomalous motion of Mercury's perihelion which was discovered by Le 
Verrier in the late 19th century. The solar oblateness causes secular variations 
of the orbital elements of planets, with the exception of semimajor axes and 
eccentricities, and has to be taken into account when constructing the model of 
planetary motion. The problem lies in the fact that the solar oblateness is 
determined indirectly from some complex astrophysical measurements that are 
subject to various systematic errors caused by equipment imperfection and the 
solar atmosphere and activity. The use of modern equipment made it possible to 
give a more reliable estimate $J_2 = 2\cdot 10^{-7}$. This value is used for the
construction of ephemerides starting with DE 405 (Standish, 1998) and EPM2000 
(Pitjeva, 2001). Recently, it became possible to determine the dynamical solar 
oblateness while processing of high-precision radar observations when 
constructing planetary ephemerides (see Pitjeva, 2005b).

A serious problem arises in the construction of modern high-precision planetary 
ephemerides due to the necessity of taking into account the perturbations caused 
by asteroids. The DE200 and EPM87 ephemerides considered the perturbations only 
from the 3-5 largest asteroids; the experiments revealed that this was impossible 
to attane a proper representation of high-precision observations of the {\it 
Viking 1} and {\it Viking 2} landers, i.e., a representation which would match 
the a priori errors (6-12 meters) of these observations. Amplitudes of the 
perturbations from asteroids were determined analytically by Williams (1984) 
considering commensurability between the orbital periods of the asteroids and 
Mars. The perturbations from 300 asteroids that were selected by Williams due to 
the significant perturbations of the orbit of Mars caused by them (Williams, 1989) 
are taken into account starting with the DE 403 (Standish et al., 1995) and EPM98 
(Pitjeva, 1998) ephemerides. However, the masses of the majority of these 
asteroids are either unknown or known with insufficient accuracy, and Standish 
and Fienga (2002) showed that the accuracy of planetary ephemerides deteriorated 
substantially with time due to this factor. Direct dynamical estimates of the 
masses of asteroids may be obtained by analyzing their perturbations to other 
celestial bodies caused by them. This technique may be applied when examining 
spacecraft near asteroids, binary asteroids or asteroids with satellites, 
perturbations on the Mars and the Earth caused by asteroids and revealed through 
the processing of radar observations of Martian spacecraft and landers, and close 
encounters of asteroids. Applying the latter (classical) method requires great 
caution, since optical observations may produce large errors (Krasinsky et al., 
2002). These techniques were used to measure the masses of several dozen 
asteroids, but the construction of high-precision planetary ephemerides demands 
taking into account the perturbations from about 300 large asteroids. If the 
estimates of the diameters and densities of these asteroids are available, one 
may also estimate their masses. The diameters of hundreds of asteroids were 
determined by processing the infrared data from the {\it Infrared Astronomical 
Satellite} (IRAS) and {\it Midcourse Space Experiment} (MSX) satellites. When 
constructing the DE and EPM ephemerides, these asteroids were divided into the C 
(Carbonic), S (Sillicum), and M (Metallic) taxonomic types according to their 
spectral classes, and the estimates of their densities were derived from radar 
observations while improving the ephemerides. Apart from the sufficiently large 
asteroids, thousands of small asteroids, many of which are too small to be ever 
discovered from the Earth, produce a substantial cumulative effect on the orbits 
of the inner planets. The majority of these bodies travel within the main 
asteroid belt, and the distribution of their instantaneous positions in the main 
belt may be considered uniform. Thus, the perturbations from the small asteroids 
that were not considered individually in the integration may be modeled by 
additional perturbations from a massive ring in the plane of the ecliptic with a 
uniform mass distribution. Starting with EPM2004 (Pitjeva, 2005a), the two 
parameters characterizing the ring (its mass $M_r$ and radius $R_r$) are included 
in the set of parameters that are improved from observations.

Hundreds of trans-Neptunian objects (TNOs) that were discovered lately also exert 
influence on the motion of planets, especially the outer planets. The updated 
dynamical model of the EPM ephemerides includes Eris (a dwarf planet discovered 
in 2003, which is more massive than Pluto) and 20 of the largest TNOs into 
simultaneous integration. The perturbations from the other TNOs were modeled by 
a homogeneous TNO ring lying in the plane of the ecliptic and having a radius of 
43 AU and an estimated mass (Pitjeva, 2010a).

Thus, the dynamical model created at the IAA RAS, takes into account (besides the 
mutual perturbations of large planets and the Moon) a number of relatively weak 
gravitational effects that contribute appreciably while processing modern 
high-precision observations:

\begin{itemize}
\itemsep -3mm
\item perturbations from 301 of the most massive asteroids;
\item perturbations from other minor planets in the main asteroid belt, modeled 
by a homogeneous ring;
\item perturbations from the 21 largest TNOs;
\item perturbations from the other trans-Neptunian planets, modeled by a 
homogeneous ring at a mean distance of 43 AU;
\item perturbations from the solar oblateness $(2\cdot10^{-7})$;
\item relativistic perturbations from the Sun, the Moon, planets (including 
Pluto), and five largest asteroids.
\end{itemize}

When constructing the EPM ephemerides, the equations of motion of n bodies with 
masses $m_1, \ldots m_n$ in a non-rotating barycentric coordinate system were 
used. These equations take the form of

$$ {\bf\ddot r}_i =  A + B + C + D + E,  $$

where $A$ stands for the Newtonian gravitational accelerations:

$$  A = \sum_{j \not = i}{\mu_j({\bf r}_j-{\bf r}_i) \over r^3_{ij}}; $$

$B$ stands for the relativistic terms:

$$  B = \sum_{j \not = i}{\mu_j({\bf r}_j-{\bf r}_i) \over r^3_{ij}}
\left \{-{2(\beta+\gamma) \over c^2}\sum_{k \not = i}{\mu_k \over r_{ik}}
-{2\beta-1 \over c^2}\sum_{k \not = j}{\mu_k \over r_{jk}}
+ \gamma\left({v_i \over c}\right)^2 + \right.$$
$$
\left.
+(1+\gamma)\left({v_j \over c}\right)^2 - {2(1+\gamma) \over c^2}{\bf\dot r}_i \cdot {\bf\dot r}_j
-{3 \over 2c^2} \left[{({\bf r}_i-{\bf r}_j)\cdot{\bf\dot r}_j \over r_{ij}}\right]^2
+{1 \over 2c^2}({\bf r}_j-{\bf r}_i) \cdot {\bf\ddot r}_j \right \} + $$
$$ + {1 \over c^2} \sum_{j \not = i}{\mu_j \over r^3_{ij}}
\left \{ \left[{\bf r}_i-{\bf r}_j \right] \cdot \left[(2+2\gamma){\bf\dot r}_i
-(1+2\gamma){\bf\dot r}_j \right] \right \}({\bf\dot r}_i-{\bf\dot r}_j)
+ {3+4\gamma \over 2c^2}  \sum_{j \not = i}{\mu_j {\bf\ddot r}_j  \over r_{ij}};
$$

$C$ stands for the terms caused by the solar oblateness (the solar quadrupole 
moment):

$$ C = 3  J_2 \mu_S {R^2 \over  r^4_{iS}}
\left \{ \left[ {5 \over 2}\left( {({\bf r}_i-{\bf r}_S) \over r_{iS}} \cdot {\bf p}  \right)^2
-{1 \over 2} \right]{({\bf r}_i-{\bf r}_S) \over r_{iS}}
-\left({({\bf r}_i-{\bf r}_S)  \over r_{iS}} \cdot {\bf p} \right) {\bf p} \right \};
$$

$D$ stands for the terms caused by the asteroid and TNO rings to the inner 
planets:

$$ D= {1 \over2}{M_r \over R^3_r} F \left(1.5,1.5,2,{{r_i}^2 \over {R_r}^2} \right){\bf r}_{i};
$$

and $E$ stands for the terms caused by the asteroid ring to the outer planets:

$$ E= -{M_r \over {r_i}^3} \left[F \left(0.5,0.5,1,{{R_r}^2 \over {r_i}^2} \right)
+ {1 \over 2}{{R_r}^2 \over {r_i}^2}  F \left(1.5,1.5,2,{{R_r}^2 \over {r_i}^2} \right)
\right]{\bf r}_{i}.
$$

\noindent Here the following designations were introduced: ${\bf r}_i$, ${\bf\dot r}_i$, 
${\bf\ddot r}_i$ (barycentric vectors) are the coordinate, velocity, and 
acceleration vectors of the $i$th body; $\mu_j=Gm_j$, where $G$ is the 
gravitational constant and $m_j$ is the mass of the $j$th body; $r_{ij} = 
|{r_j - r_i}| $; $\beta, \gamma$ are the parameters of the PPN (parameterized 
post-Newtonian) formalism; $v_i= |{\bf\dot r}_i| $; $c$ is the speed of light; 
$J_2$ is the second zonal harmonic of the Sun; $R$ is the equatorial solar 
radius; ${\bf p}$ is the unit vector pointing to the Sun's north pole; 
${M_r}=Gm_r$, $m_r$, $R_r$ are the masses and radii of the rings; and $F$ is the 
hypergeometric function.

The summation in the equation that pertains to the Newtonian gravitational 
accelerations ($A$) includes (besides planets, the Sun, and the Moon) 301 
asteroids and 21 TNOs. The five main asteroids (Ceres, Pallas, Vesta, Iris, and 
Bamberga) are entered not only in $A$, but also in the equations $B$ (the 
relativistic terms) and $C$ (the terms caused by the solar oblateness). Thus, the
equations of motion for the 16 main objects incorporate all the mutual 
perturbations, including relativistic ones and the perturbations due to the solar 
oblateness.

The variable ${\bf\ddot r}_j$, that appears in two terms in the right side of the 
equations stands for the barycentric acceleration of the $j$th body due to the 
Newtonian acceleration of other bodies.

It should be noted that only the equations of motion of planets, asteroids, TNOs, 
and the Moon are actually integrated. The barycentric coordinates and velocities 
of the Sun are derived from the following equation:
$$  \sum_{i} \mu_i^*{\bf r}_i=0,$$

\noindent{where}
$$\mu_i^* = \mu_i \left \{1+ {1 \over 2c^2}v_i^2 - { 1 \over 2c^2}
\sum_{j \not = i}{\mu_j \over r_{ij}} \right \}.
$$

All modern high-precision ephemerides are based on relativistic time scales and 
relativistic equations of motion of celestial bodies and radio and light rays.
The main common feature of the DE, EPM, and INPOP series of ephemerides is the 
simultaneous numerical integration of the equations of motion of nine major planets, the 
Sun, the Moon, and the lunar physical libration carried out in the post-Newtonian 
approximation for GR ($\beta = \gamma = 1$) in a harmonic coordinate system ($\alpha = 0$).

Thus, the terms $A$, $B$, and $C$ are identical in all those major planetary 
ephemerides. Various versions of ephemerides differ in modeling the lunar 
libration, reference frames in which the ephemerides are computed, adopted values 
of the solar oblateness and other parameters, modeling of perturbations from
asteroids, and used sets of observations and estimated parameters. The main 
distinction of the latest EPM ephemerides (starting with EPM2008, as described in
Pitjeva, 2009) from the DE and INPOP ephemerides is the inclusion of the 
perturbations from TNOs that are actually present in the Solar System. The 
inclusion of any additional objects into the simultaneous integration leads to 
the shift 
of the barycenter of the Solar System. Since TNOs are located beyond the orbit of 
Neptune, and there are many large objects (for example, Eris) among them, the 
said shift becomes significant. In the process of calculations, the barycenter 
remains in its place, while the coordinates of all objects involved in the 
integration change. Therefore, comparing the EPM ephemerides with the DE and 
INPOP ephemerides requires using relative (heliocentric, geocentric, etc.) 
coordinates of objects, but not barycentric ones. Such a comparison was carried 
out for DE421, EPM2008, and INPOP08 by Hilton and Hohenkerk (2011). Since any 
observations are relative (are usually made from the Earth), the shift of the 
barycenter does not influence the representation of observations.

In recent years, a large number of high-precision radiometric observations of 
spacecraft, revolving around or passing close to planets, and optical 
observations of the satellites of planets carried out by both terrestrial 
observatories and the {\it Hubble Space Telescope} became available. This enabled 
the researchers to derive new masses of planets and other bodies of the Solar 
System. These values were adopted as the current best values of the constants of 
dynamical astronomy by XXVII IAU GA in 2009 (Luzum et al., 2011) and are used in 
updated versions of the EPM ephemerides (starting with EPM2008).

The integration in the barycentric coordinate system at the J2000.0  epoch was 
done using Everhart's method over a 400-year interval (from 1800 to 2200) by a 
lunar and planetary integrator of the ERA-7 software system.

\bigskip
\centerline{OBSERVATIONAL DATA, THEIR REDUCTION, AND TT - TDB}
\smallskip

The observations that were used to improve the accuracy of the EPM2011 
ephemerides included 677670 positional measurements of various types (from 
classical meridian observations to modern radio observations of planets and 
spacecraft) obtained from 1913 to 2011. Optical observations dating from 1913, 
when an improved micrometer was installed at the United States Naval Observatory 
and the measurements became more accurate ($\sim$0$\rlap.''5$), and all the 
available radio observations (up to the year 2011) were used.
It should be noted that the accuracies of modern CCD observations approach a few 
hundredths of an arcsecond. A real revolution in dynamical astronomy started in 
1961 when the first successful radiolocation of Venus was carried out 
simultaneously in the United States (at the California Institute of Technology 
and the Massachusetts Institute of Technology), the USSR (at the Institute of 
Radio Engineering and Electronics), and England (at the Jodrell Bank Observatory). 
The significance of astronomical radar observations stems from two factors.
Firstly, they added two new types of measurements, namely, the measurement of the 
delay time (ranging) that could be converted to distance using the known speed of 
light and the measurement of the Doppler frequency shift that gives the relative 
radial velocity of the reflecting surface. Secondly, radar observations are 
highly accurate. Nowadays the relative accuracy that ranges from $10^{-11}$ to 
$10^{-12}$ has become ordinary for trajectory measurements of spacecraft. These 
values are five orders of magnitude better than the accuracy of classical optical 
measurements.
However, only the terrestrial planets are fully provided by with radio observations. 
Fewer observations of this type are made for Jupiter and Saturn, and there exists 
only one three-dimensional normal point provided by {\it Voyager 2} for Uranus (and 
Neptune). Therefore, optical observations still retain their 
significance for the outer planets. The main factors that limit the accuracy of 
photographic and CCD observations of planets are the brightness of planets 
compared to reference stars (the equalization of brightness); the distortion of 
photographic images due to meteorological, instrumental, and astronomical (the 
phase effect) causes; and the difficulty of measuring an extended object of a 
non-uniform density. This applies especially to bright planets (Jupiter and
Saturn) with large visible disks. Positional observations of planetary satellites are 
not prone to any of these restrictions. Since the position of a satellites relative to
the stars is determined both by the planetary motion and the satellite's own motion 
around the planet, the measurements of the positions of satellites may be used to 
define the planetary orbits more accurately. The astrometric photographic 
observations of the satellites of Jupiter and Saturn were started in the Nikolaev
Observatory in 1962. In 1998, astronomers in Flagstaff began observing the satellites 
of the outer planets (in addition to the observations of the outer planets 
themselves), and all their measurements are referred to the ICRF system with the 
use of reference stars from the AST and TYCHO2 catalogues. Observations of satellites
are also carried out at a number of other observatories. Theories of the motion 
of satellites are required to process such observations.
Analytical theories of the motion of the satellites of Jupiter (Lieske), Saturn 
(Vienn and Duriez), and Uranus (Lascar and Jacobson) are incorporated in the 
ERA-7 software system. The drawback of these analytical theories lies in the fact 
that they do not provide an opportunity to correctly introduce the parameters of 
the satellite' motion when improved from observations. Therefore, the researchers 
at the IAA RAS, construct their own numerical theories of the motion of the 
satellites of Mars and the outer planets (Poroshina et al., 2012). These theories 
are successfully used to improve the ephemerides of satellites and planets alike. 
Lately, the previous observations (prior to 2005) were supplemented with the new 
data from spacecraft, namely, measurements of ranging made using {\it Odyssey}, 
{\it Mars Reconnaissance Orbiter} (MRO), {\it Mars Express} (MEX), and {\it Venus 
Express} (VEX); VLBI observations of {\it Odyssey} and MRO; and three-dimensional 
normal point observations of {\it Cassini} and {\it Messenger}. These measurements 
were complemented by CCD 
observations of the outer planets and their satellites made at the Flagstaff and 
Table Mountain observatories. The observations used are shown on the page 12 
(1 mas = $0\rlap.''001$); the numbers in the headings (57560, 58112, and 561998) 
indicate the number of observations.

The majority of these observations were taken from the Jet Propulsion Laboratory 
(JPL) database (http://iau-comm4.jpl.nasa.gov/plan-eph-data/index.html) 
which was created by E. M. Standish and is now maintained and expanded by W.M. 
Folkner. This data set was supplemented by Russian radar observations of planets
made from 1961 to 1995 (http://www.ipa.nw.ru/PAGE/DEPFUND/LEA/ENG/rrr.html) and 
data from {\it Venus Express} and {\it Mars Express} obtained through the courtesy of A. 
Fienga.

\bigskip
\centerline{ \large{\bf Astrometric observations of planets and spacecraft}}
\smallskip
\leftline{Optical observations of the outer planets and their satellites made 
from 1913 to 2011 (57560)}
\vspace{-4mm}
\begin{figure}[h!]
\parbox[b]{0.2\textwidth}{
$ \left.
\begin{array}{l}
USNO \\[-2pt]
Pulkovo \\[-2pt]
Nikolaev \\[-2pt]
Tokyo   \\[-2pt]
Bordeaux \\[-2pt]
La Palma \\[-2pt]
Flagstaff \\[-2pt]
TMO
\end{array}
\right\}
$
}
\parbox[b]{0.5\textwidth}{
\begin{tabular}{l|l|l|}
\hline
\noalign{\smallskip}
 Observation type & Interval & A priori accuracy\\
\noalign{\smallskip}
\hline
\noalign{\smallskip}
Transits        & 1913--1994 & 1$'' \to $0$\rlap.''$5 \\
Photoelectric transits & 1963--1998 & 0$\rlap.''$8 $ \to $0$\rlap.''$25 \\
Photographic        & 1913--1998 & 1$'' \to $0$\rlap.''$2 \\
CCD                    & 1995--2011 & 0$\rlap.''$2 $ \to $0$\rlap.''$05 \\
\end{tabular}
}
\vspace{-4mm}
\end{figure}
\leftline{Radar observations of Mercury, Venus, and Mars (58112)}
\vspace{-4mm}
\begin{figure}[h!]
\parbox[b]{0.2\textwidth}{
$ \left.
\begin{array}{l}
Millstone \\[-2pt]
Haystack \\[-2pt]
Arecibo \\[-2pt]
Goldstone \\[-2pt]
Crimea
\end{array}
\right\}
$
}
\hspace{-5pt}
\parbox[b]{0.2\textwidth}{
\begin{tabular}{l|l|l|}
\hline
\noalign{\smallskip}
 Observation type & Interval & A priori accuracy\\
\noalign{\smallskip}
\hline
\noalign{\smallskip}
Ranging        & 1961--1997 & 100 km $\to$ 150 m \\
\end{tabular}
}
\end{figure}
\centerline{Radio data provided by spacecraft from 1971 to 2010 (561998)}
\vspace{-4mm}
\begin{figure}[h!]
\parbox[b]{0.4\textwidth}{
$ \left.
\begin{array}{ll}
Mariner-9&\hbox{Venus} \\[-2pt]
Pioneer-10,-11&\hbox{Jupiter} \\[-2pt]
Voyager&\hbox{Jupiter} \\[-2pt]
Phobos&\hbox{Mars} \\[-2pt]
Ulysses&\hbox{Jupiter} \\[-2pt]
Magellan&\hbox{Venus} \\[-2pt]
Galileo&\hbox{Jupiter} \\[-2pt]
Viking-1,-2&\hbox{Mars} \\[-2pt]
Pathfinder&\hbox{Mars} \\[-2pt]
MGS&\hbox{Mars} \\[-2pt]
Odyssey&\hbox{Mars} \\[-2pt]
MRO&\hbox{Mars} \\[-2pt]
Cassini&\hbox{Saturn} \\[-2pt]
VEX&\hbox{Venus} \\[-2pt]
Messenger&\hbox{Mercury} \\[-2pt]
MEX&\hbox{Mars} \\
\end{array}
\right\}
$
}
\hspace{-5pt}
\parbox[b]{0.2\textwidth}{
\begin{tabular}{l|l|l|}
\hline
\noalign{\smallskip}
Observation type&Interval&A priori accuracy\\
\noalign{\smallskip}
\hline
\noalign{\smallskip}
Ranging       & 1971--2009 & 6 km $\to$ 1 m \\
Dif.range & 1976--1997 & 1.3 $\to$ 0.1 mm/s \\
Rad.velos. & 1992--1994 & 0.1 $\to$ 0.002 mm/s \\
Flybys  & 1973--2010 & 400 mas $\to$ 0.4 mas \\
$\Delta$VLBI & 1990--2010 & 12 mas $\to$ 0.2 mas \\
\end{tabular}
}
\end{figure}

The processing of observational data was done using proven and reliable 
techniques with due account for all the needed reductions (Pitjeva, 2005a). The 
following reductions were applied to radar data:
\begin{itemize}
\itemsep -3mm
\item 	reduction of time moments to a uniform scale;;
\item 	relativistic corrections, namely, the delay of radio signals in the 
gravitational field of the Sun, Jupiter, and Saturn (the Shapiro effect) and the 
transition from the coordinate time (the argument of ephemerides) to the proper 
time of the observer;
\item 	the delay of radio signals in the Earth's troposphere;
\item 	the delay of radio signals in the plasma of the solar corona;
\item 	correction for topography of the surfaces of planets (Mercury, Venus, and Mars).
\end{itemize}

The following reductions were applied to optical data:
\begin{itemize}
\itemsep -3mm
\item 	reduction to the ICRF system: from reference catalogues to FK4, then to 
the FK5 catalog, and at last to the ICRF frame;
\item 	correction for additional phase effect; 
\item   correction for gravitationa deflection of light by the Sun.
\end{itemize}

The transition from the observing time (UTC = TAI + an integer number of seconds) 
to the barycentric dynamic time (TDB) of the ephemerides requires knowing the 
differences between the terrestrial time (TT = TAI $+32.184$ s) and TDB. Until 
recently, these differences were computed by applying the analytical expansions 
for the DE405 ephemerides. However, the differences TT - TDB depend on the 
coordinates of all bodies that are involved in the integration of the 
corresponding ephemerides. Therefore, the construction of these differences by 
numerical integration using the corresponding ephemerides is more correct.

\begin{figure}[h!]
\begin{center}
\includegraphics[scale=0.4]{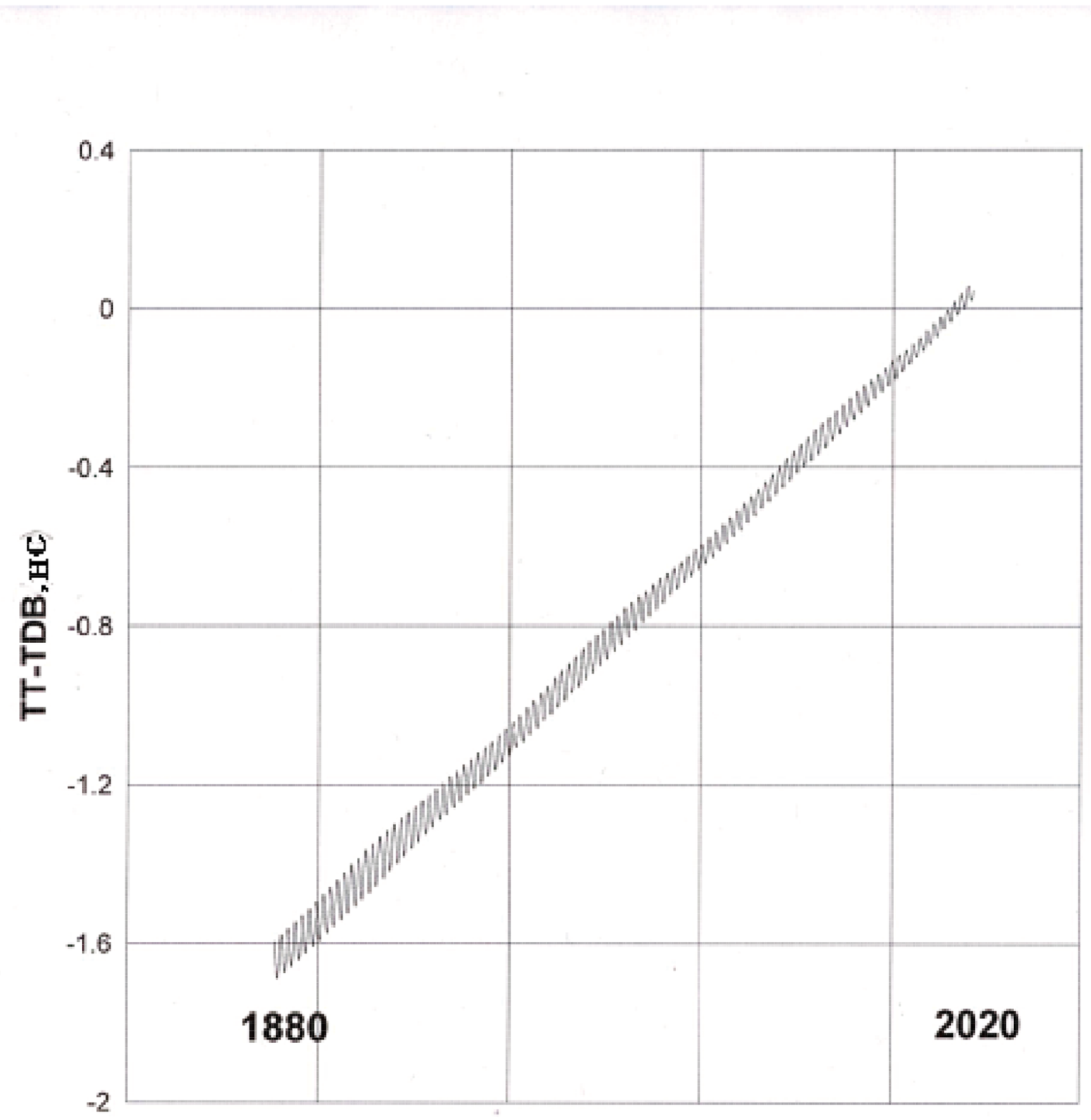}
\end{center}
\centerline{ {\bf Fig. 1.} Differences in (TT--TDB) for the EPM2004 and EPM2008} 
\centerline{ ephemerides expressed in nanoseconds.}
\end{figure}

\noindent The following differential equation taken from the paper by Klioner 
(2010) was used for connection between TT and TDB:
$$ \frac{\mbox{d}{(TT-TDB)}}{\mbox{d}TDB}=
\frac{L_B-L_G}{1-L_B}+\frac{1-L_G}{1-L_B}\left(\frac{1}{c^{2}}\alpha'
+\frac{1}{c^4}\beta'\right),$$

\noindent where $L_B=1.550519768\cdot10^{-8}$, $L_G=6.969290134\cdot10^{-10}$, $c$ is
the speed of light,
$$\alpha'=-\frac{1}{2}v^{2}_{E}-\sum_{A \neq E}\frac{GM_{A}}{r_{EA}},$$
\begin{eqnarray}
\beta'=-\frac{1}{8}v^{4}_{E}+\frac{1}{2}\left[\sum_{A \neq E}\frac{GM_{A}}{r_{EA}}\right]^{2}+\sum_{A \neq  E}\frac{GM_{A}}{r_{EA}}\left\{4\mathbf{v}_A\cdot\mathbf{v}_E-\phantom{\frac12}\right. \nonumber\\
\left.-\frac{3}{2}v_E^2-2v_A^2+\frac{1}{2}{\bf a}_A\cdot {\bf r}_{EA}+
\frac{1}{2}\left(\frac{{\bf v}_A\cdot{\bf r}_{EA}}{r_{EA}}\right)^2+
\sum_{B \neq A}\frac{GM_{B}}{r_{AB}}\right\}. \nonumber
\end{eqnarray}

\noindent Figure 1 shows, as an example, differences in TT - TDB for the EPM2004 
and EPM2008 ephemerides expressed in nanoseconds.

\bigskip
\centerline{EPM2011 PARAMETERS AND REPRESENTATION OF OBSERVATIONS}
\smallskip

About 270 parameters were determined in the process of improving the planetary 
part of the EPM2011 ephemerides:
\begin{itemize}
\itemsep -3mm
\item the orbital elements of planets and 18 satellites of the outer planets; 
\item the value of the astronomical unit or $GM_{\odot}$;
\item the angles of orientation of the ephemerides with respect to the ICRF;
\item parameters of the rotation of Mars and the coordinates of three Martian 
landers;
\item the masses of 21 asteroids and the mean densities of three taxonomic 
classes (C, S, and M) of asteroids;
\item the mass and radius of the asteroid ring and the mass of the TNO ring;
\item the ratio of the Earth and Moon masses; 
\item the Sun's quadrupole moment and parameters of the solar corona for 
different conjunctions of planets with the Sun;
\item the coefficients of Mercury's topography and corrections to the level 
surfaces of Venus and Mars; 
\item the coefficients for additional phase effect of the outer planets;
\item the constant shifts for the series of observations of Venus in Goldstone 
(1964) and Venus (1969) and Mercury (from 1986 to 1989) in Crimea, as well as the
shifts (and, in certain cases, their derivatives) for all spacecraft that were 
interpreted as the calibration errors;
\item post model parameters, such as the PPN parameters ($\beta, \ \gamma), \  
\dot{\pi_i}, \dot{GM_{\odot}}/GM_{\odot}, \ \dot{a_i}/a_i $.
\end{itemize}

Mean values and rms's of the residuals of observations are shown in the tables 
1, 2, where ``n. p.'' stands for normal points (with the exception of {\it Viking} 
and {\it Pathfinder} for which the total number of observations is given).

\bigskip
\noindent{\bf Table 1.} Mean values and rms's of the residuals of radio observations
\bigskip

\begin{center}
\vskip -0.4cm
\begin{tabular}{|@{\hspace{1pt}}c@{\hspace{1pt}}|c|@{\hspace{3pt}}c@{\hspace{3pt}}|c|@{\hspace{1pt}}c@{\hspace{1pt}}|c@{\hspace{0pt}}|}
\noalign{\smallskip}
\hline
\noalign{\smallskip}
\hskip 0.1cm Planet & \hskip 0.1cmObservation type & Interval & Number of n. p. &  $<O-C>$ & $\sigma$ \\
\noalign{\smallskip}
\hline
\noalign{\smallskip}
Mercury & $\tau$ [¬] & 1964--1997 & 746 & 0.0 & 610 \\
& ЉЂ $\tau$ [¬] & 1974--2009 & 5 & 1.3 & 18.9 \\
Venus  & $\tau$ [¬]  & 1961--1995 & 1354 & 0.0 & 594 \\
& Magellan d$r$ [¬¬/б] &  1992--1994 & 195 & 0.0 & 0.007 \\
& MGN,VEX  VLBI [mas] & 1990--2010 & 47 & 0.0 & 2.7 \\
& Cassini  $\tau$ [¬] & 1998--1999 & 2 & -2.6 & 2.4 \\
& VEX  $\tau$ [¬] & 2006--2010 & 1721 & 0.0 & 2.8 \\
Mars   & $\tau$ [¬]  & 1965--1995 & 403 & 0.0 & 745\\
& ЉЂ $\tau$ [¬] & 1971--1989 & 644 & -13.7 & 43.9 \\
&  Viking $\tau$ [¬] &  1976--1982  & 1258 & 0.0 & 9.5  \\
&  Viking d$\tau$ [¬¬/б] &  1976--1978  & 14978 & 0.0 & 0.89  \\
&  Pathfinder $\tau$ [¬] &  1997  & 90 & 0.0 & 2.7  \\
&  Pathfinder d$\tau$ [¬¬/б] &  1997  & 7574 & 0.0 & 0.09 \\
&  MGS $\tau$ [¬] & 1998--2006 & 7341 & 0.0 & 1.3  \\
&  Odyssey $\tau$ [¬] & 2002--2009 & 8187 & 0.0 & 1.1  \\
&  MRO $\tau$ [¬] & 2006--2009 & 930 & 0.0 & 1.2  \\
&  MEX $\tau$ [¬] & 2009--2010 & 970 & 0.0 & 1.5  \\
&  ЉЂ VLBI [mas] & 1989--2010 & 144 & 0.0 & 0.8  \\
Jupiter & ЉЂ $\tau$ [¬] &  1973--2000  & 7 & 0.0 & 12.4  \\
&  ЉЂ VLBI [mas] & 1996--1997 & 24 & -1.0 & 11.4  \\
Saturn & ЉЂ $\tau$ [¬] &  1979--2006  & 34 & 0.0 & 2.8  \\
Uranus &  $\tau$ [¬] &  1986  & 1 & 1.7 & 105  \\
Neptune & Voyager-2 $\tau$ [¬] &  1989  & 1 & 0.0 & 14  \\
\noalign{\smallskip}
\hline
\end{tabular}
\end{center}
\bigskip

\noindent{\bf Table 2.} Mean values and rms's of the residuals of optical 
observations and data from spacecraft near planets (marked by $^*$) obtained from 
1913 to 2011 for $\alpha$ and $\delta$ expressed in mas
\smallskip

\begin{center}
\vskip -0.4cm
\begin{tabular}{|l|c|@{\hspace{1pt}}c@{\hspace{1pt}}|c|@{\hspace{1pt}}c@{\hspace{1pt}}|c|}
\noalign{\smallskip}
\hline
\noalign{\smallskip}
\hskip 0.7cmPlanet& Number of observations & $<O-C>_{ \alpha}$ & 
$ \sigma_{ \alpha}$ & $<O-C>_{ \delta}$ & $\sigma_{ \delta}$ \\
\noalign{\smallskip}
\hline
\noalign{\smallskip}
Mercury$^*$ & 6 & 0.0 & 0.7 & 1.0 & 1.8 \\
Venus$^*$ & 4 & 0.3 & 1.7 & 1.8 & 6.5 \\
Jupiter & 13364 & 12 & 181 & -28 & 194 \\
Jupiter$^*$ & 16 & -1.0 & 2.2 & -4.9 & 7.9 \\
Saturn &  15056 & -1.0 & 160 & -1.0 & 157 \\
Saturn$^*$ &  92 & 0.1 & 0.3 & 0.0 & 0.8 \\
Uranus &  11846 & 3.0 & 171 & 0.4 & 203 \\
Uranus$^*$ &  2 & -45 & 8.0 & -27 & 12 \\
Neptune &  11634 & 4.9 & 152 & 6.4 & 195 \\
Neptune$^*$ &  2 & -12 & 3.5 & -13 & 4.0 \\
Pluto  &  5660  & 0.4 & 138 & 3.0 & 140 \\
\noalign{\smallskip}
\hline
\noalign{\smallskip}
\end{tabular}
\end{center}

The residuals of ranging for {\it Odyssey}, MRO, MEX, and VEX are shown in 
Fig. 2. In ranging the increase of the dispersion O--C is evident during solar 
conjuctions when the signal passes through the solar corona. The delay in the 
solar corona was taken into 
account with the improvement of the coefficients of the corona model from 
observations, but getting rid of the solar corona noise completely requires 
the two frequencies measurements. The rms errors of the residuals amount to 
1.1 m ({\it Odyssey}), 1.2 m (MRO), 1.5 m (MEX), and 2.8 m (VEX).

\smallskip
\begin{figure}[h!]
\begin{center}
\hbox{
\includegraphics[scale=0.34]{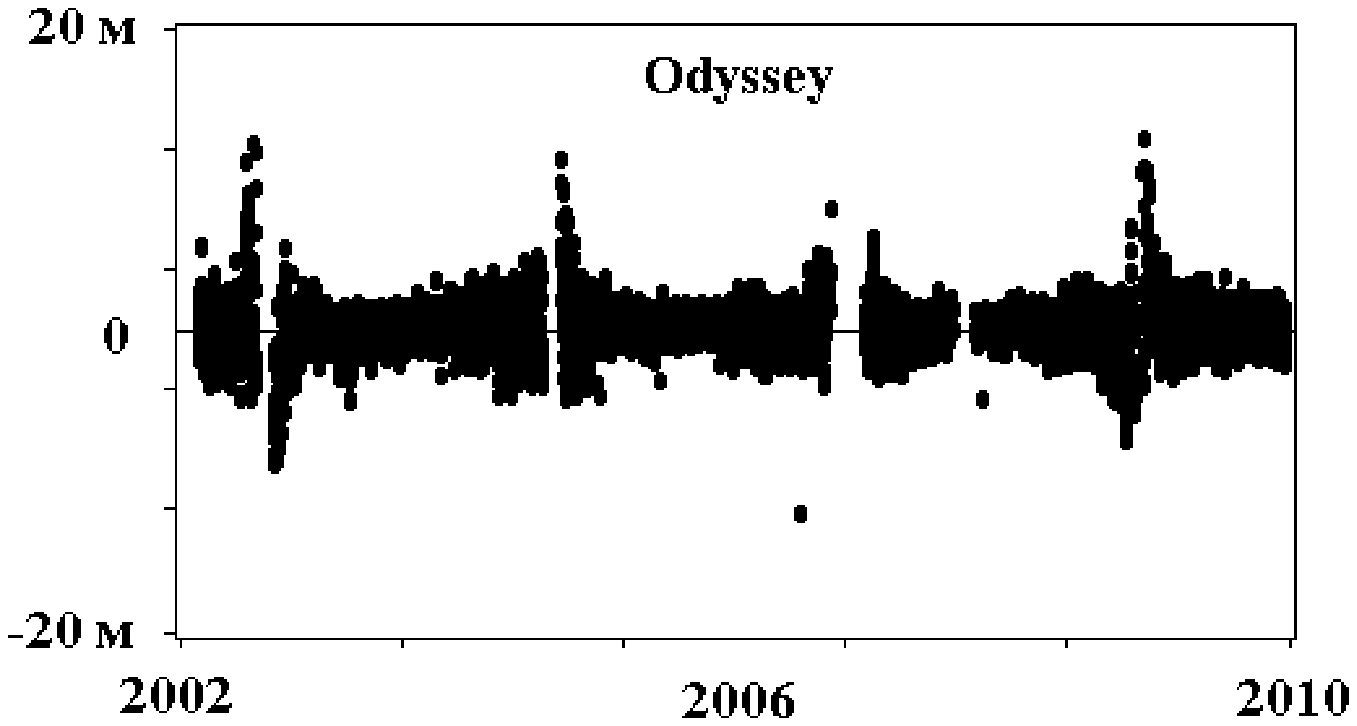}
\hskip 1.1truecm
\includegraphics[scale=0.34]{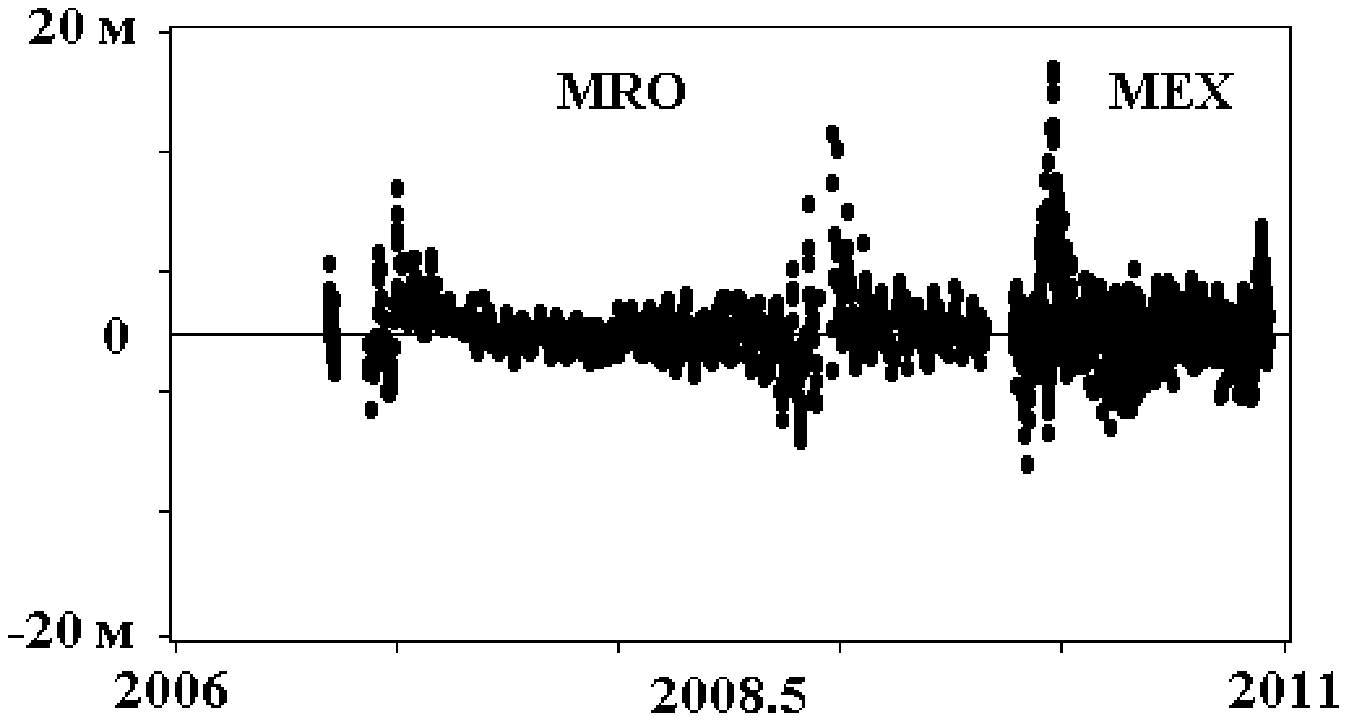}
\hskip 0.8truecm
\includegraphics[scale=0.34]{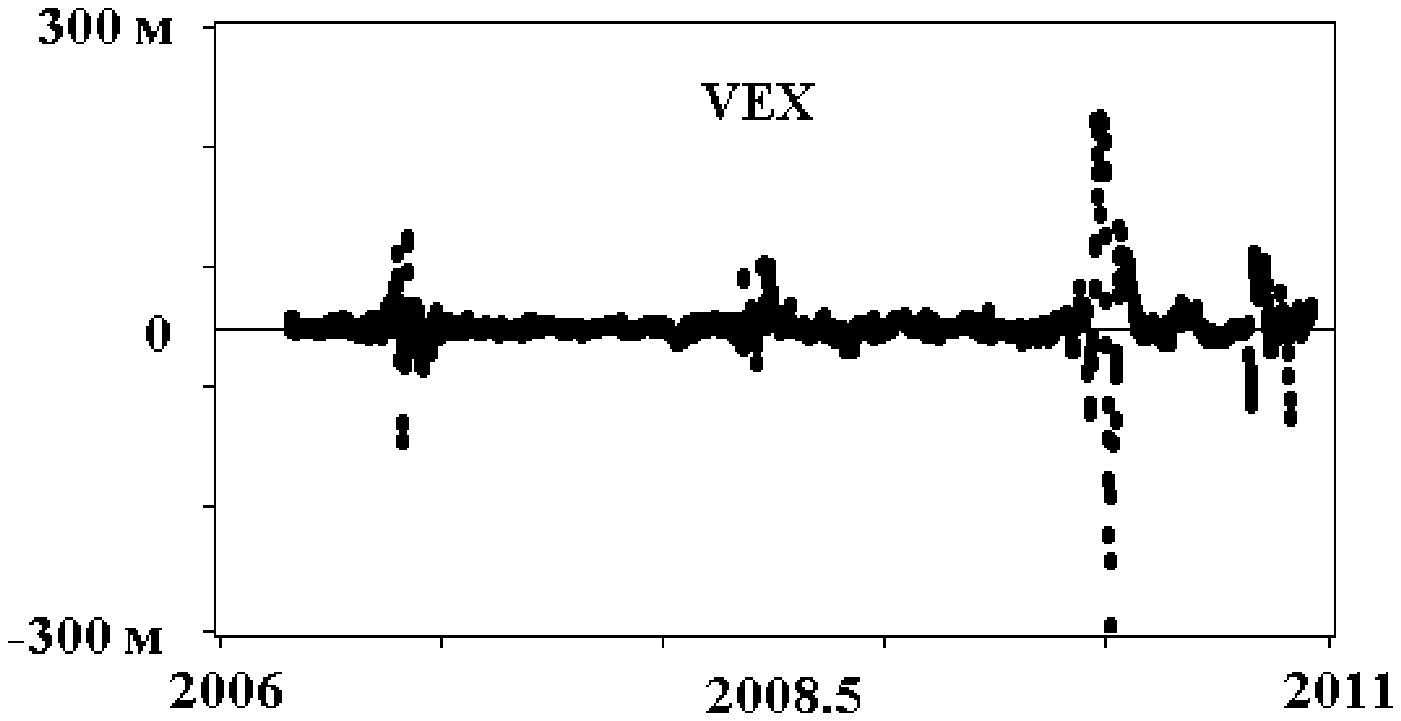}
}
\end{center}
\centerline{ {\bf Fig. 2.} Residuals of ranging (expressed in meters) for 
observations made by Odyssey,} 
\centerline{{\it Mars Reconnaissance Orbiter} (MRO), {\it Mars Express} 
(MEX), and {\it Venus Express} (VEX).}
\end{figure}

\noindent The residuals of observations of right ascensions (or, to be more 
precise, $\alpha cos \delta$) and declinations for the outer planets and their 
satellites are presented in Fig.~3.

\begin{figure}[h!]
\begin{center}
\hbox{
\includegraphics[scale=0.43]{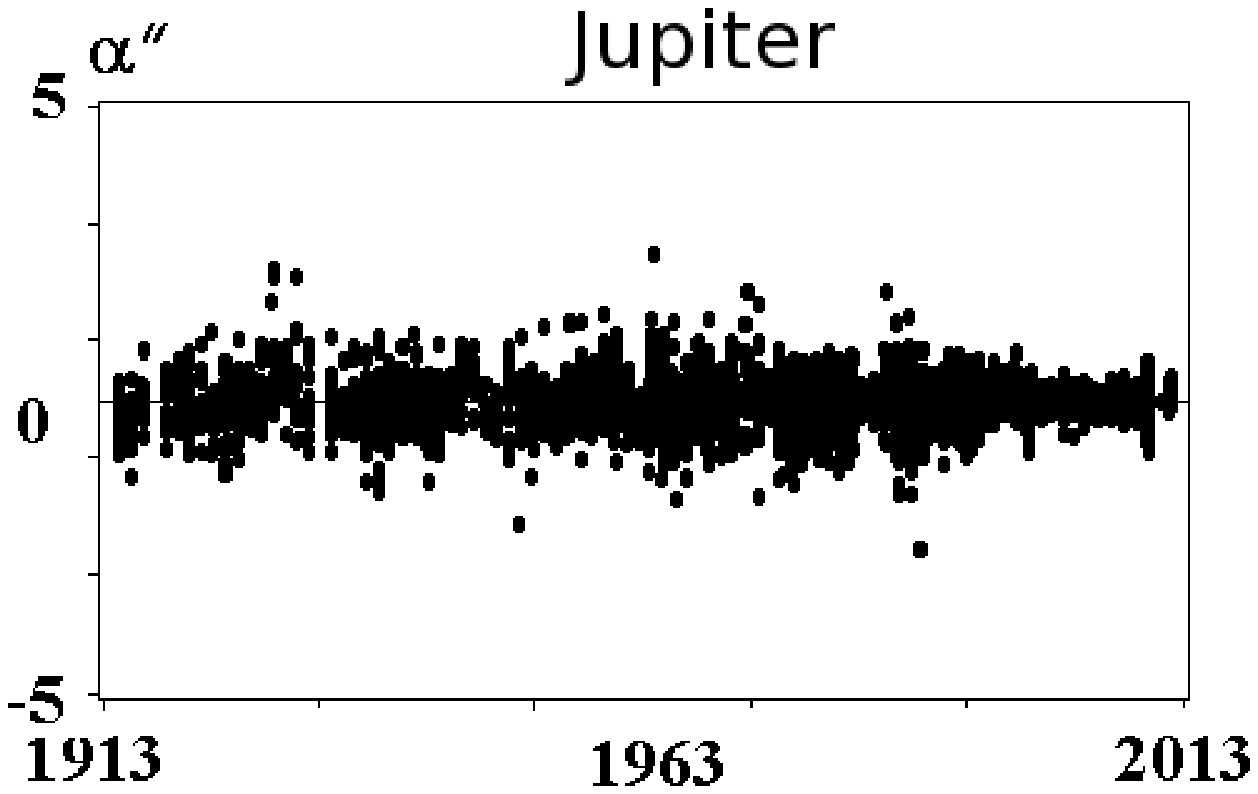}
\hskip 2.0truecm
\includegraphics[scale=0.43]{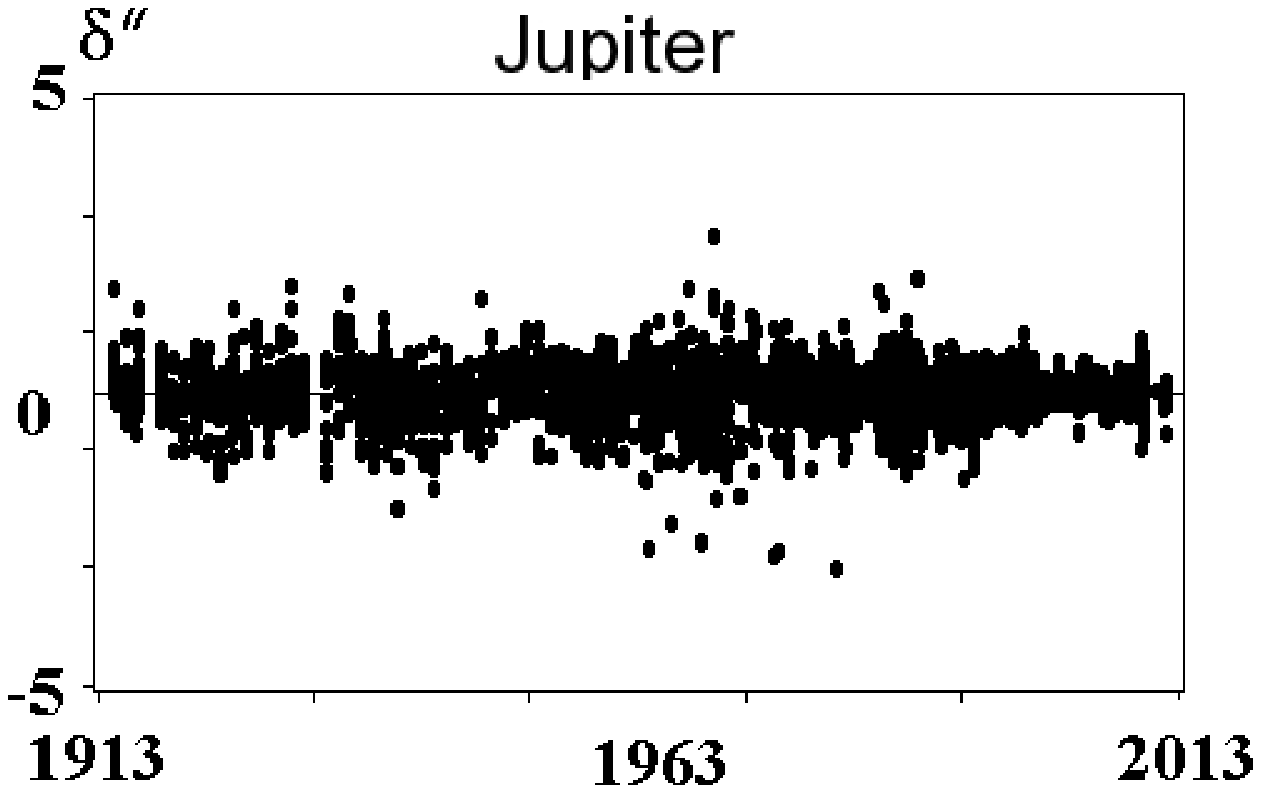}
}
\vskip 0.85truecm
\hbox{
\hskip 0.15truecm
\includegraphics[scale=0.43]{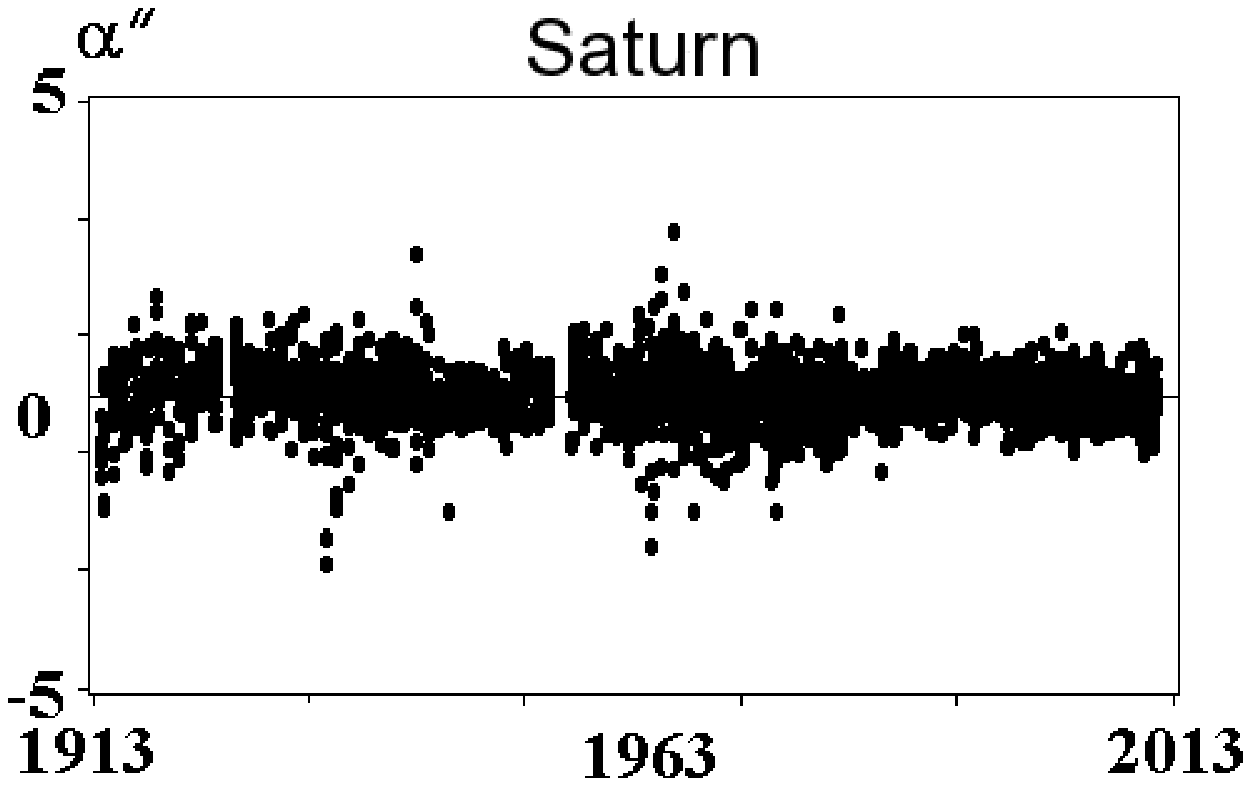}
\hskip 2.0truecm
\includegraphics[scale=0.43]{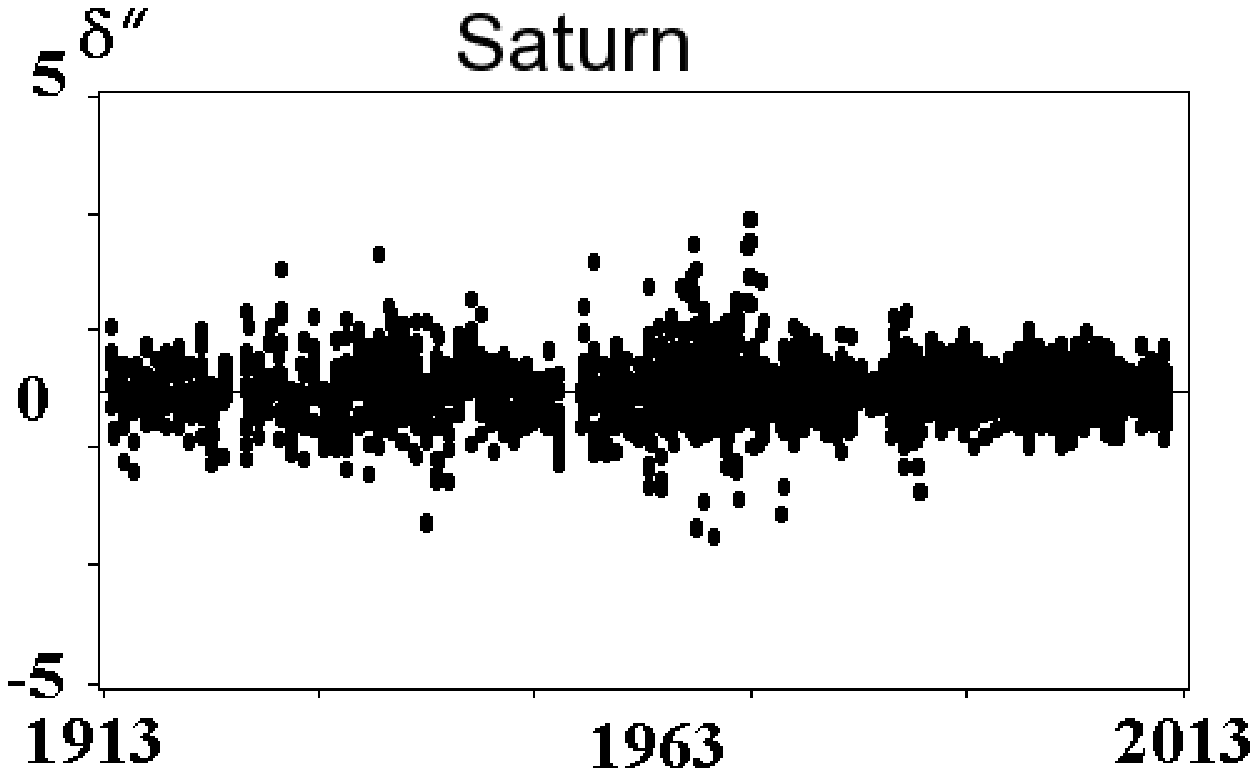}
}
\vskip 0.85truecm
\hbox{
\includegraphics[scale=0.43]{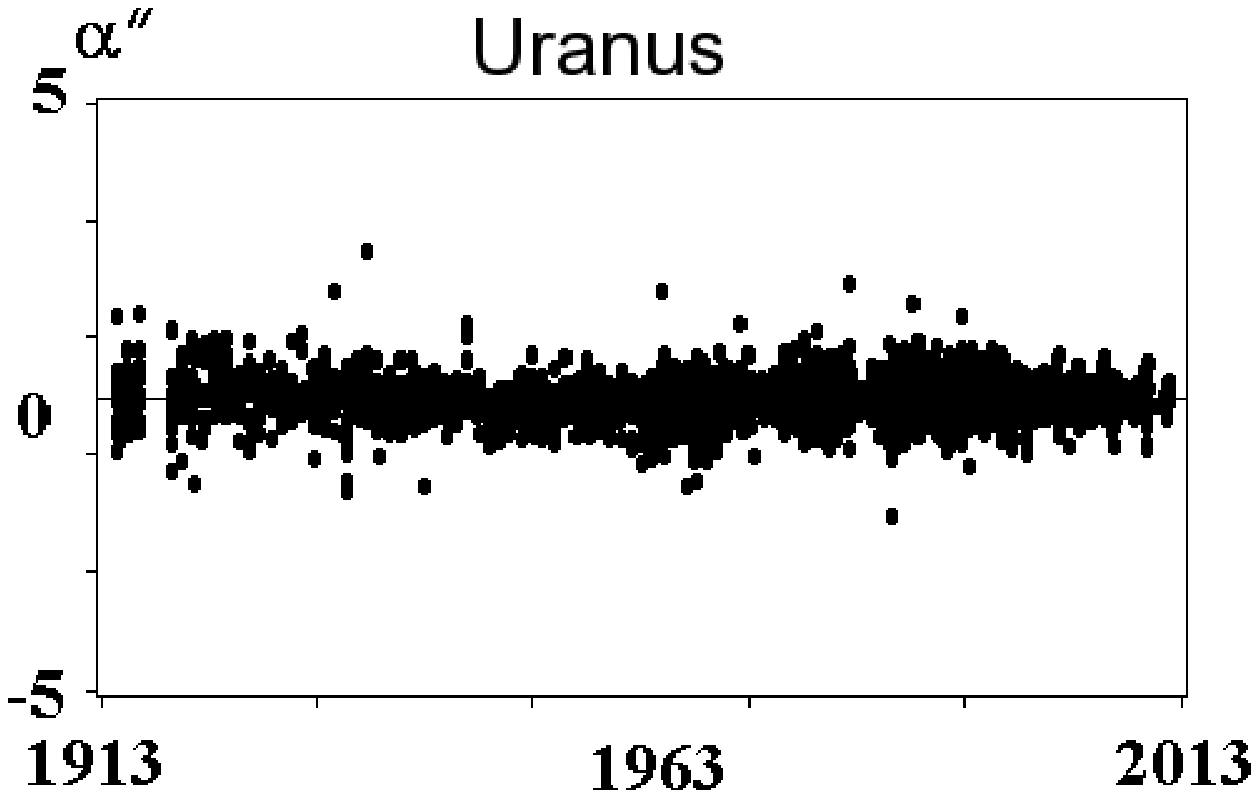}
\hskip 2.0truecm
\includegraphics[scale=0.43]{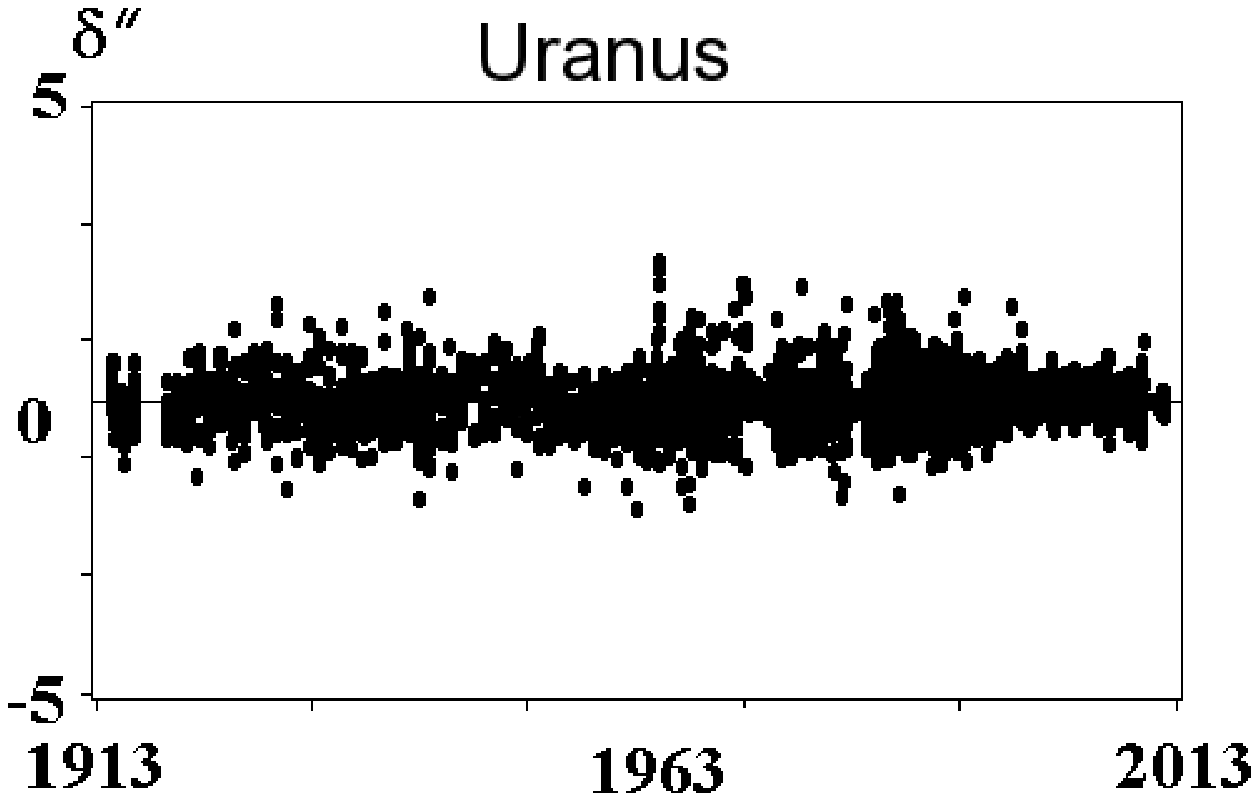}
}
\vskip 0.85truecm
\hbox{
\includegraphics[scale=0.43]{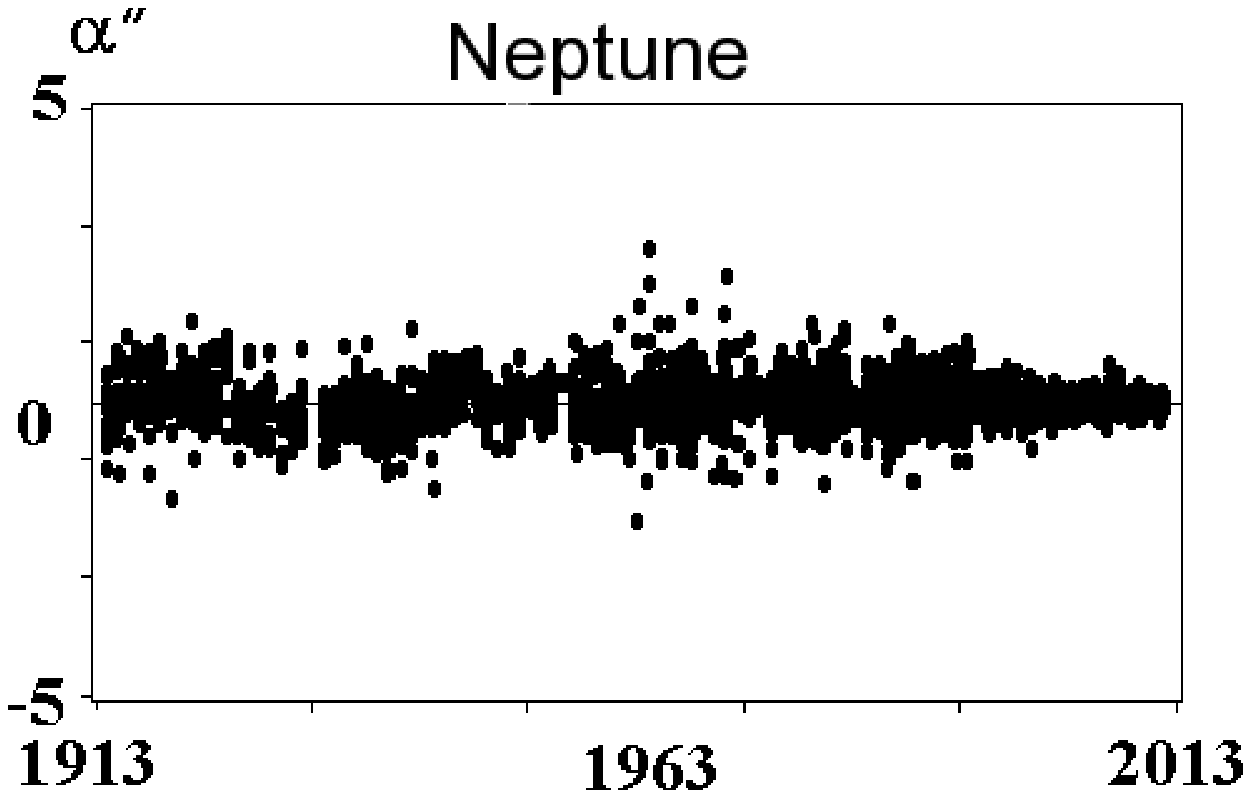}
\hskip 2.0truecm
\includegraphics[scale=0.43]{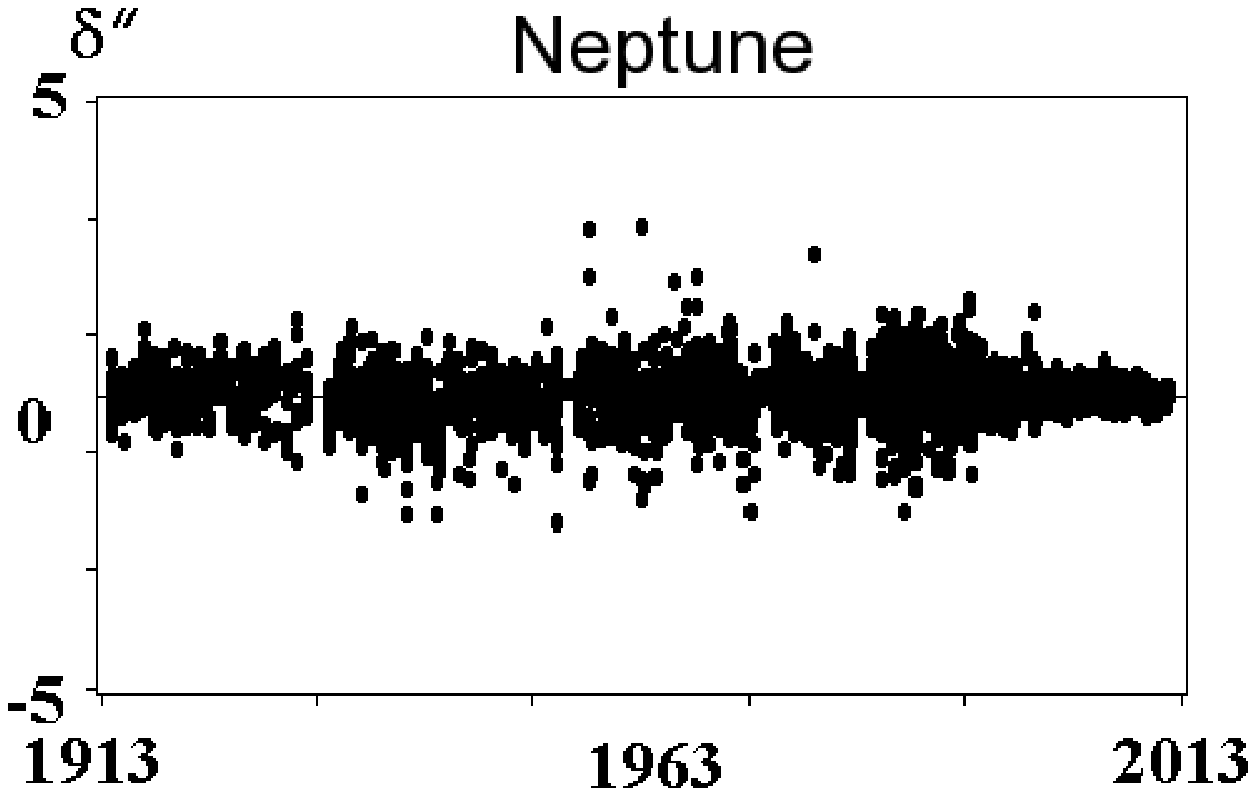}
}
\vskip 0.85truecm
\hbox{
\includegraphics[scale=0.43]{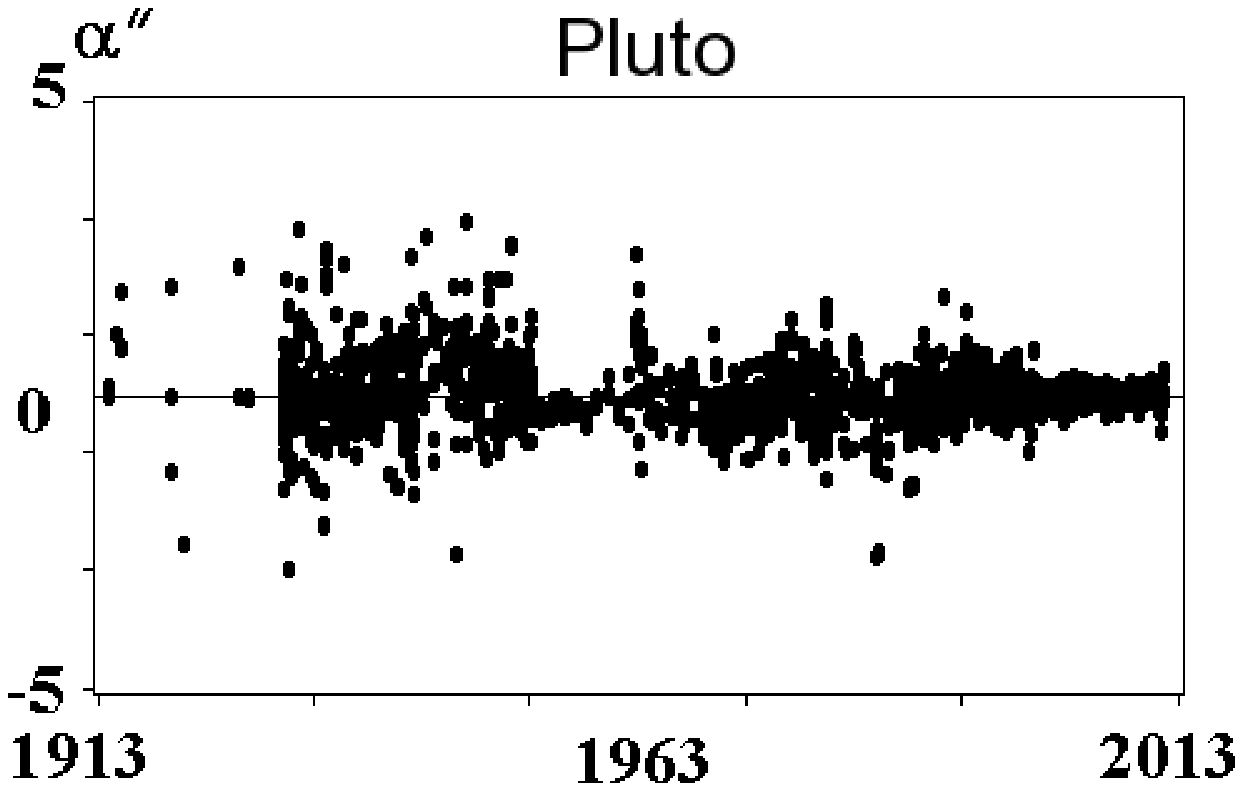}
\hskip 2.0truecm
\includegraphics[scale=0.43]{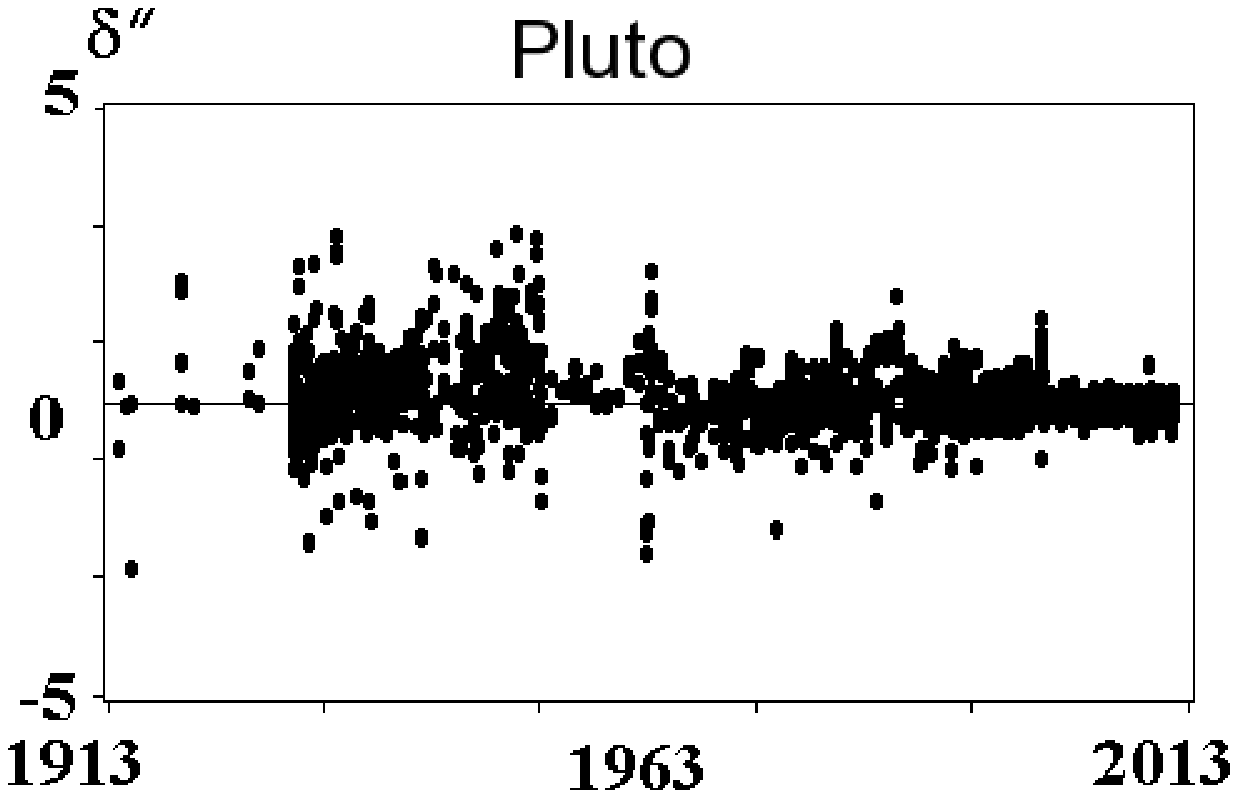}
}
\end{center}
\centerline{ {\bf Fig. 3.} Residuals of observations of $\alpha\cos\delta$ and 
$\delta$} 
\centerline{ (1913-2011) for the outer planets on a scale of $\pm5''$.}
\end{figure}

\hskip -0.4cm

Tables 1 and 2 show that the majority of observations that form the basis of the 
ephemerides are classified as radio observations, mostly ranging obtained with 
the use of spacecraft. These measurements allow us to obtain all the orbital 
elements of planets with the exception of the three angles of the Earth's 
orientation, which is equivalent to the orientation of the whole system of the 
ephemerides (angles $\varepsilon_x, \varepsilon_y,$ and $\varepsilon_z$). The 
earliest numerical planetary ephemerides (DE118 and EPM87) were referred to the 
FK4 catalogue system, while the DE200 ephemerides were referred to the system of 
the dynamical equator and equinox. At present, planetary ephemerides are oriented 
with respect to the international ICRF system through the use of VLBI 
observations of various spacecraft near planets with background quasars, the 
coordinates of which are given in the ICRF system. The accuracy of these 
observations has improved considerably from 2001 to 2010 and reached several 
tenths of a milliarcsecond for Saturn and Mars (Jones et al., 2011). This made it 
possible to significantly improve the orientation of the EPM2011 ephemerides. The 
angles of rotation between the EPM ephemerides and the ICRF system and their 
errors obtained at present and previously are presented in Table 3. Figure 4 
shows the residuals of VLBI observations of various spacecraft near Mars and of
{\it Cassini} near Saturn.

\noindent{{\bf Table 3.} Rotation angles for orientation of the EPM ephemerides into the ICRF system
\smallskip

\begin{tabular}{|c|c|c|c|c|}
\noalign{\smallskip}
\hline
\noalign{\smallskip}
Observation  &Number of & $\varepsilon_x$ & $\varepsilon_y$ & $\varepsilon_z$ \\
interval & observations & mas & mas & mas \\
\noalign{\smallskip}
\hline
\noalign{\smallskip}
1989-1994 & 20 & $4.5 \pm 0.8$ & $-0.8 \pm 0.6$ & $-0.6 \pm 0.4$ \\[-2pt]
\hline
\noalign{\smallskip}
1989-2003 & 62 & $1.9 \pm 0.1$ & $-0.5 \pm 0.2$ & $-1.5 \pm 0.1$ \\[-2pt]
\noalign{\smallskip}
\hline
\noalign{\smallskip}
1989-2007 & 118 & $-1.528\pm0.062$ & $1.025\pm0.060$ & $1.271\pm0.046$ \\[-2pt]
\noalign{\smallskip}
\hline
\noalign{\smallskip}
1989-2010 & 213 & $-0.000 \pm 0.042$ & $-0.025 \pm 0.048$ & $0.004 \pm 0.028$ \\[-2pt]
\hline
\noalign{\smallskip}
\end{tabular}

\bigskip
\begin{figure}[h!]
\begin{center}
\hbox{
\hskip -0.15truecm
\includegraphics[scale=0.34]{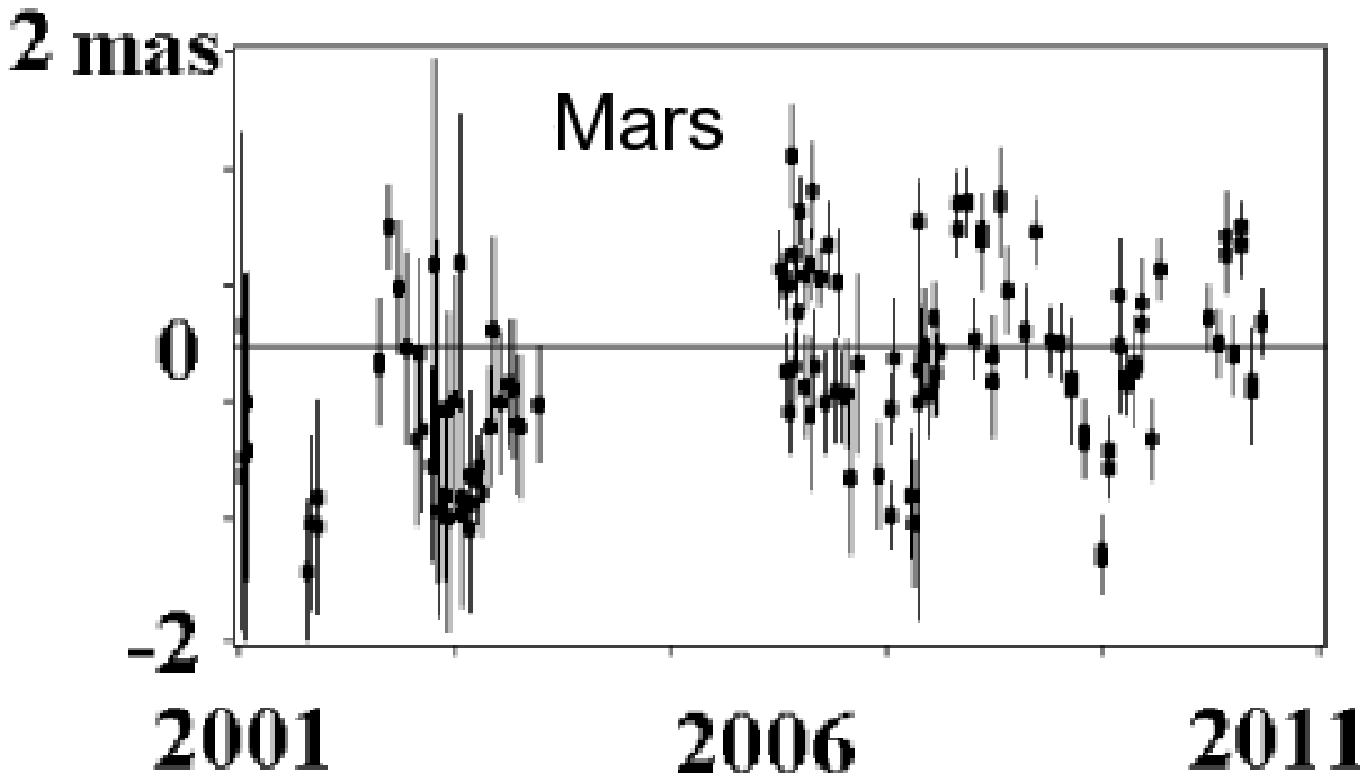}
\hskip 0.75truecm
\includegraphics[scale=0.34]{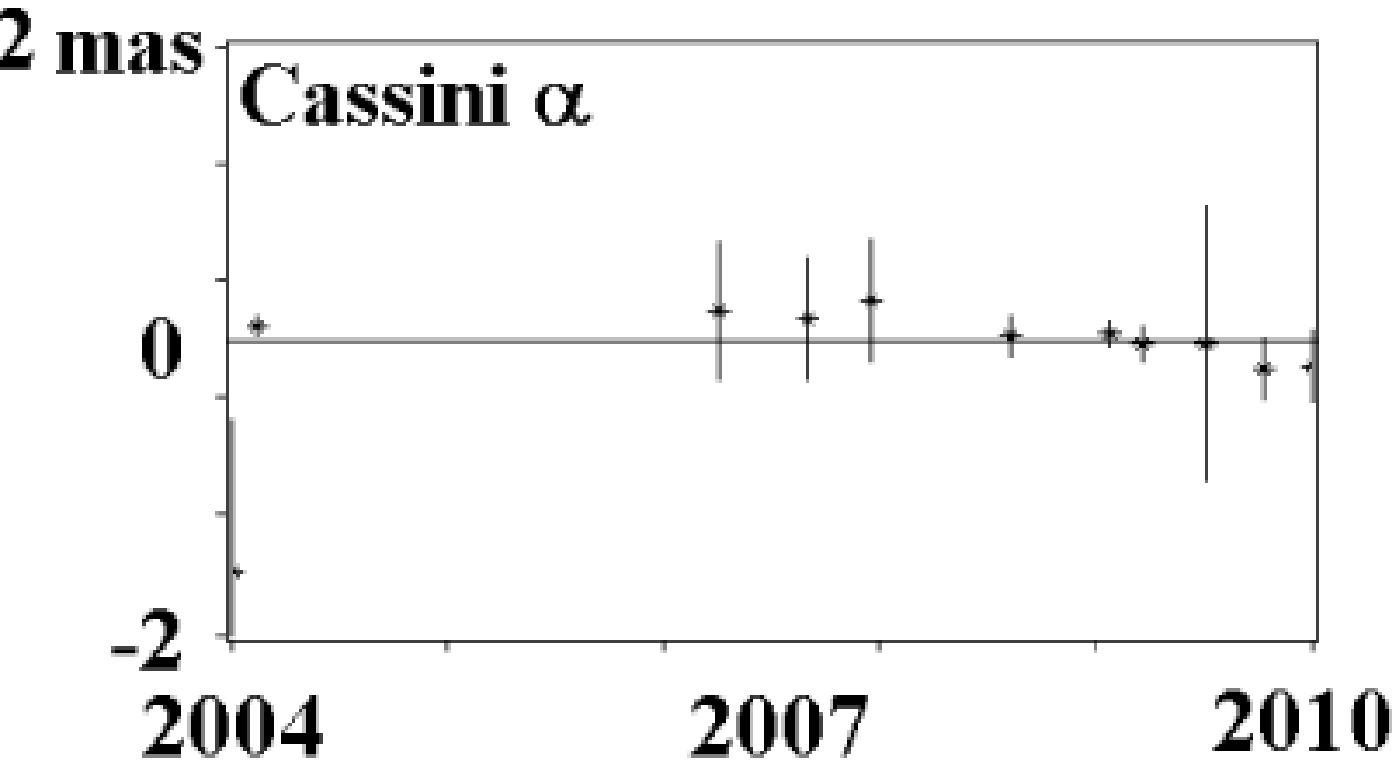}
\hskip 0.75truecm
\includegraphics[scale=0.34]{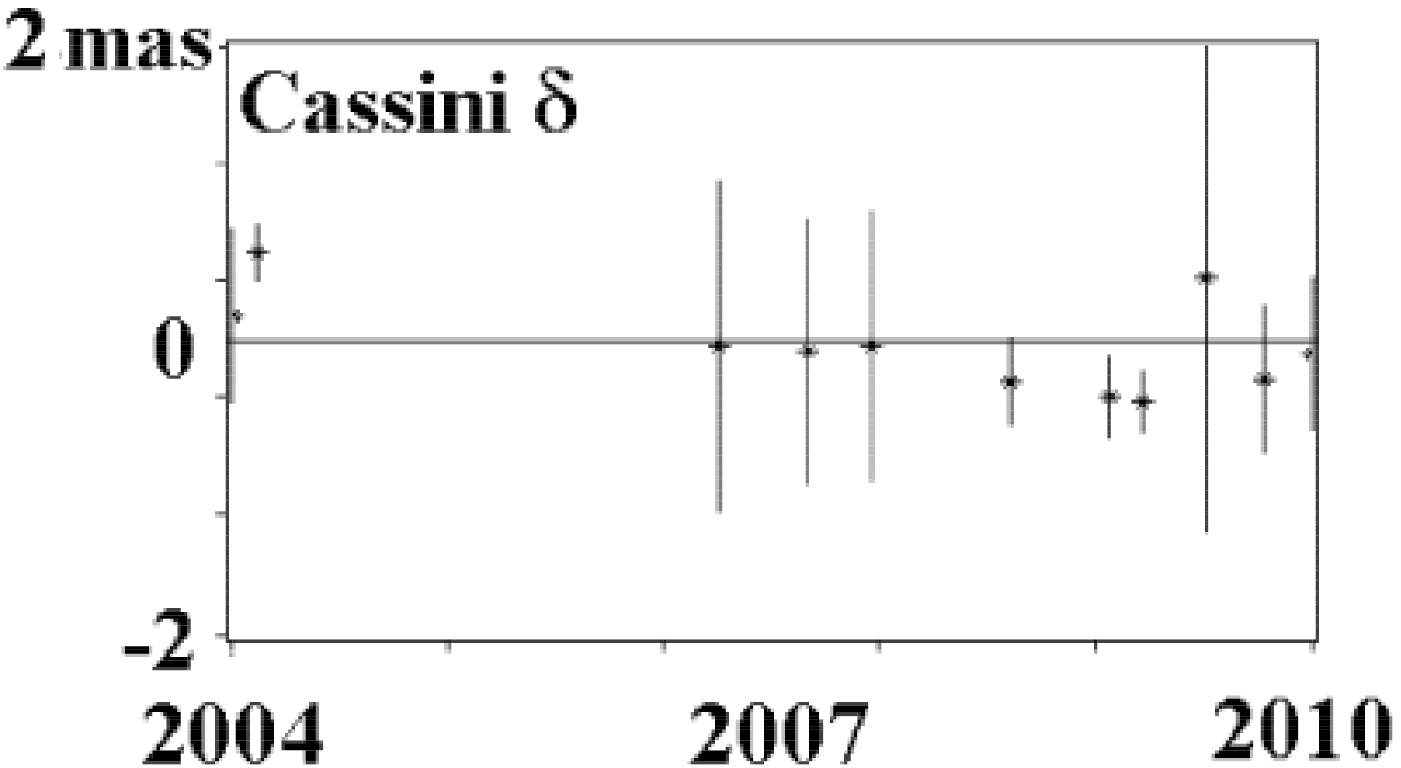}
}
\end{center}
\centerline{ {\bf Fig. 4.} Residuals of VLBI observations of various spacecraft 
near Mars} 
\centerline{ and of Cassini near Saturn expressed in mas.}
\end{figure}

The improvement of the orientation of the EPM ephemerides made it possible to 
reach the accuracy of the Earth's heliocentric coordinates ($X, Y, Z$) over a
100-year interval (from 1950 to 2050) of at least 250 m and the accuracy of 
velocities ($\dot X, \dot Y, \dot Z $) of at least 0.05 mm/s (see Fig.~5). The 
knowledge of the Earth's accurate heliocentric coordinates is particularly
important when studying pulsars, variable stars, and exoplanets. However, the 
comparison of the EPM2011 and JPL DE424 ephemerides showed that differences of 
heliocentric distances of the Earth, determined by ranging, for these ephemerides
over the same interval are much smaller and do not exceed 6 m (see left side of 
Fig.~6 for the geocentric Sun).
\bigskip
\begin{figure}[h!]
\begin{center}
\hbox{
\includegraphics[scale=0.43]{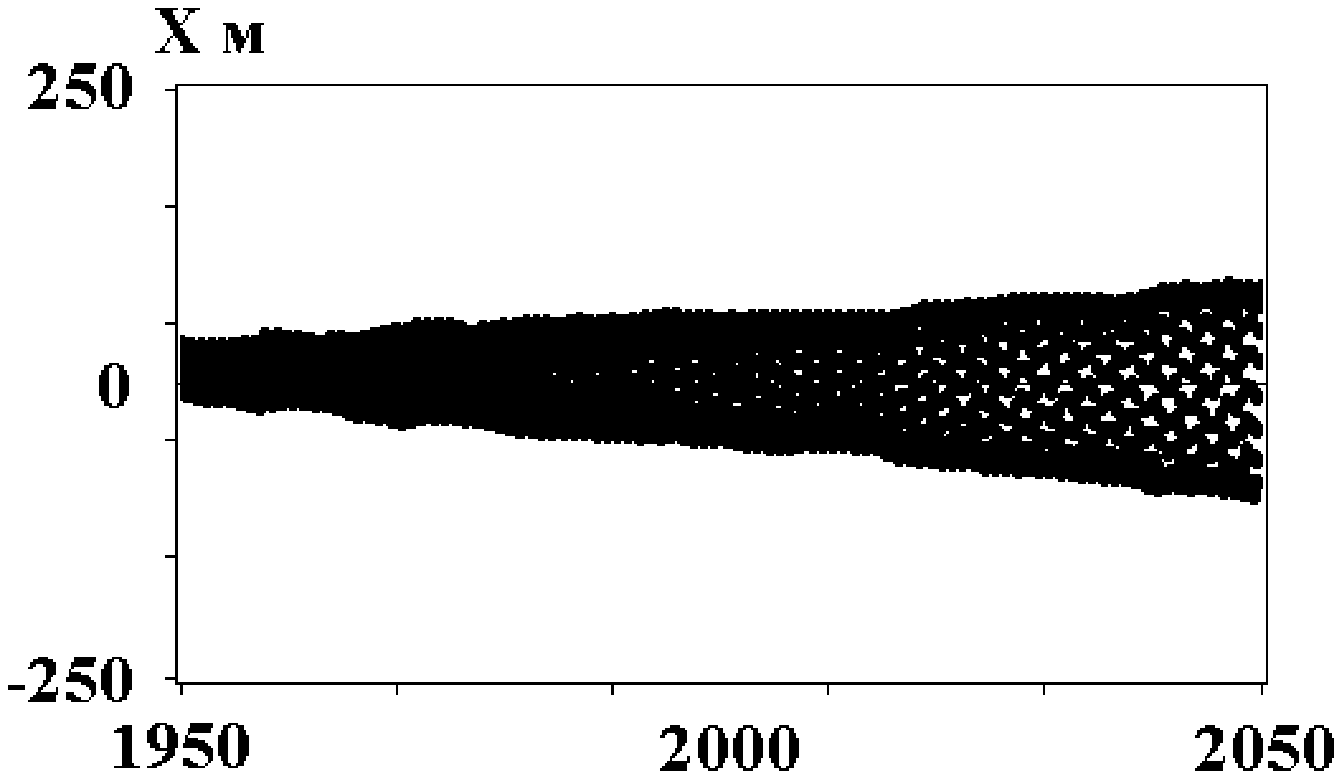}
\hskip 2.0truecm
\includegraphics[scale=0.43]{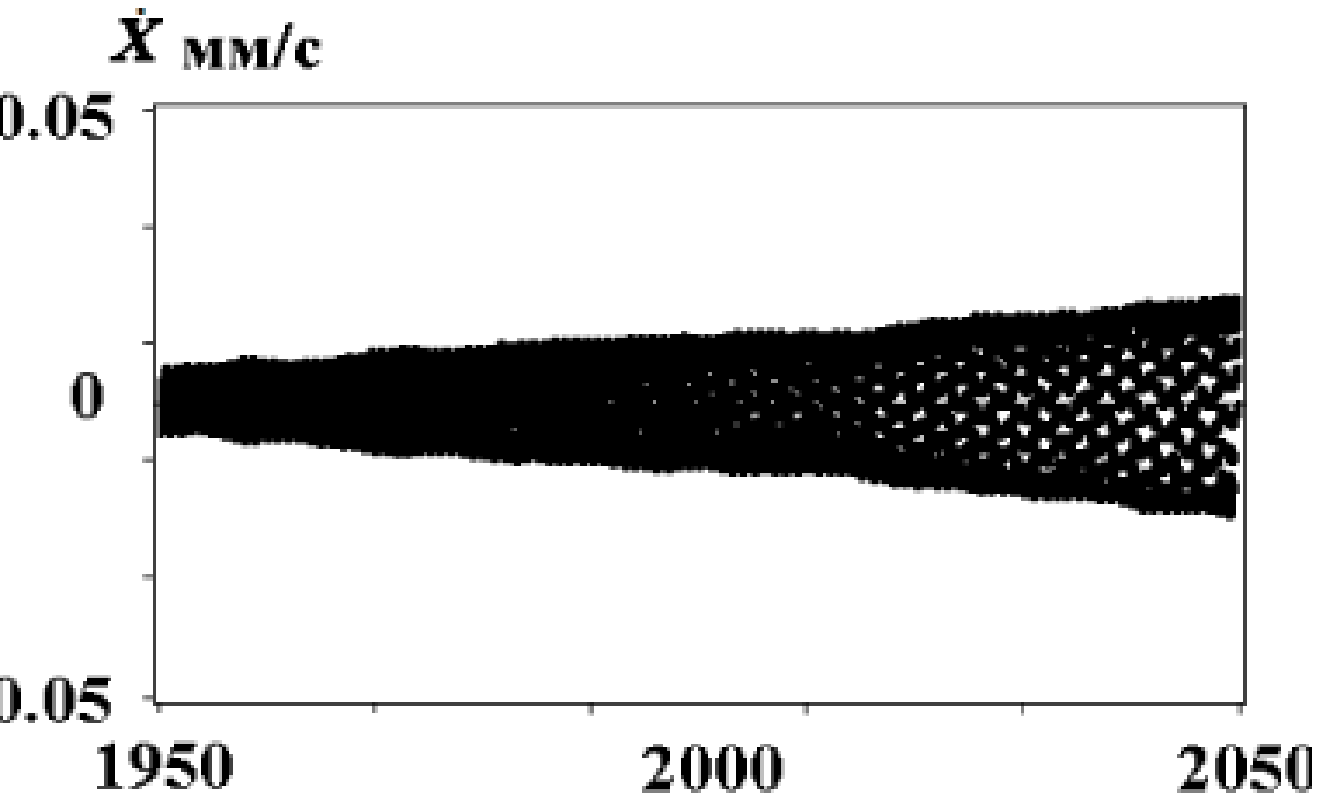}
}
\vskip 0.85truecm
\hbox{
\includegraphics[scale=0.43]{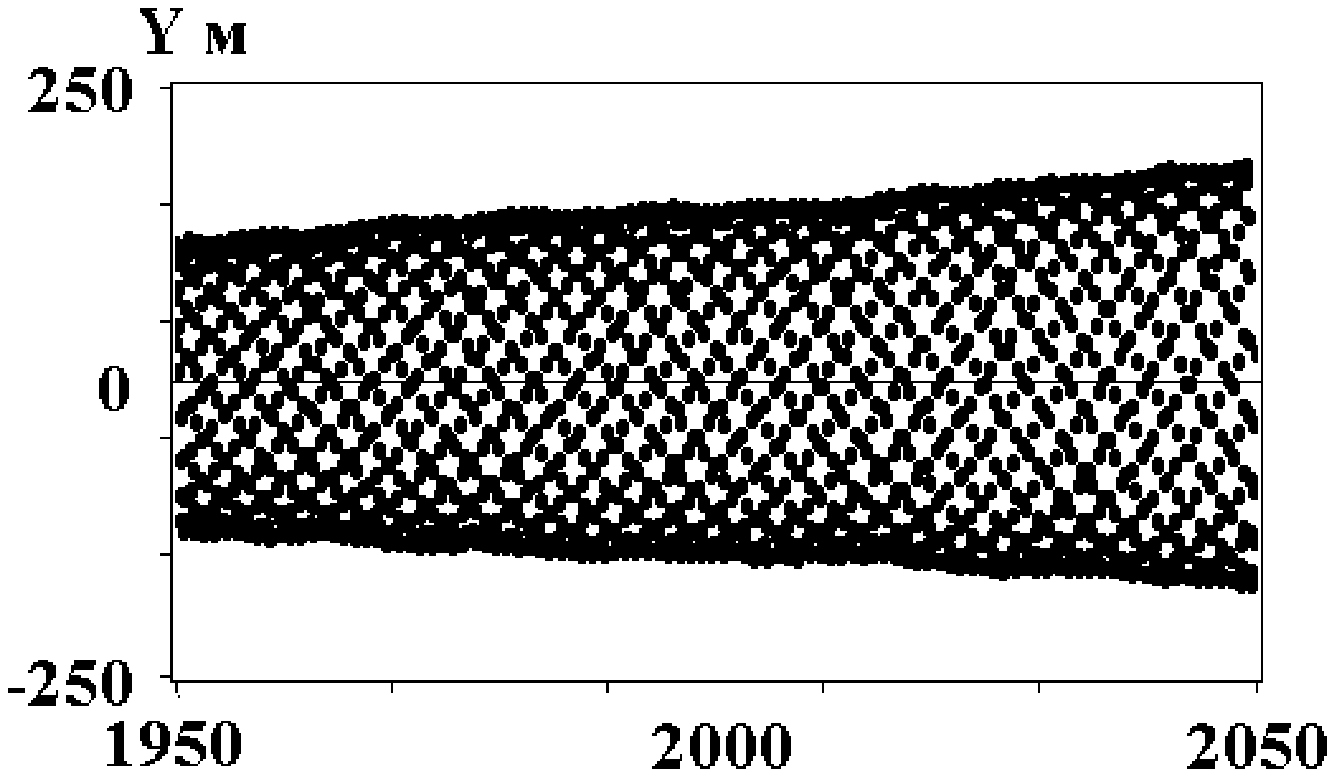}
\hskip 2.0truecm
\includegraphics[scale=0.43]{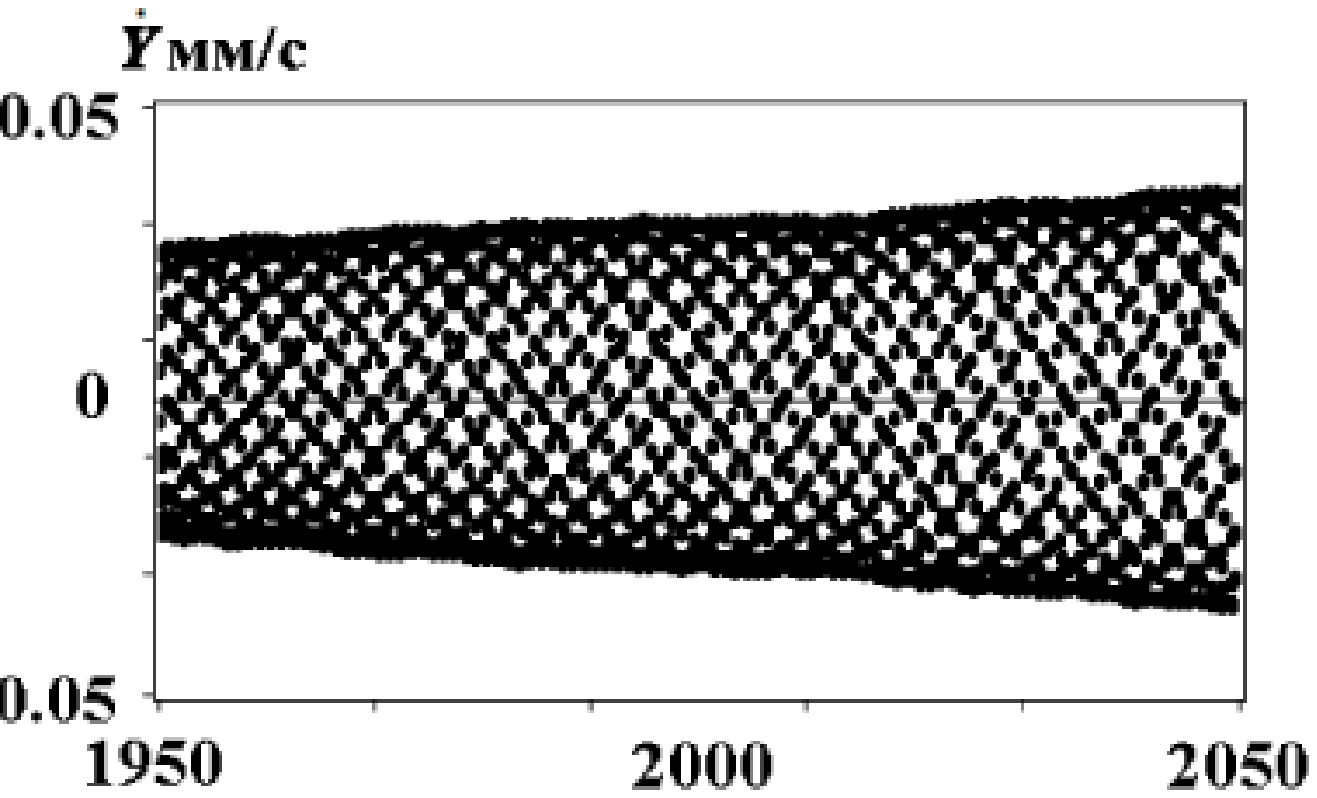}
}
\vskip 0.85truecm
\hbox{
\includegraphics[scale=0.43]{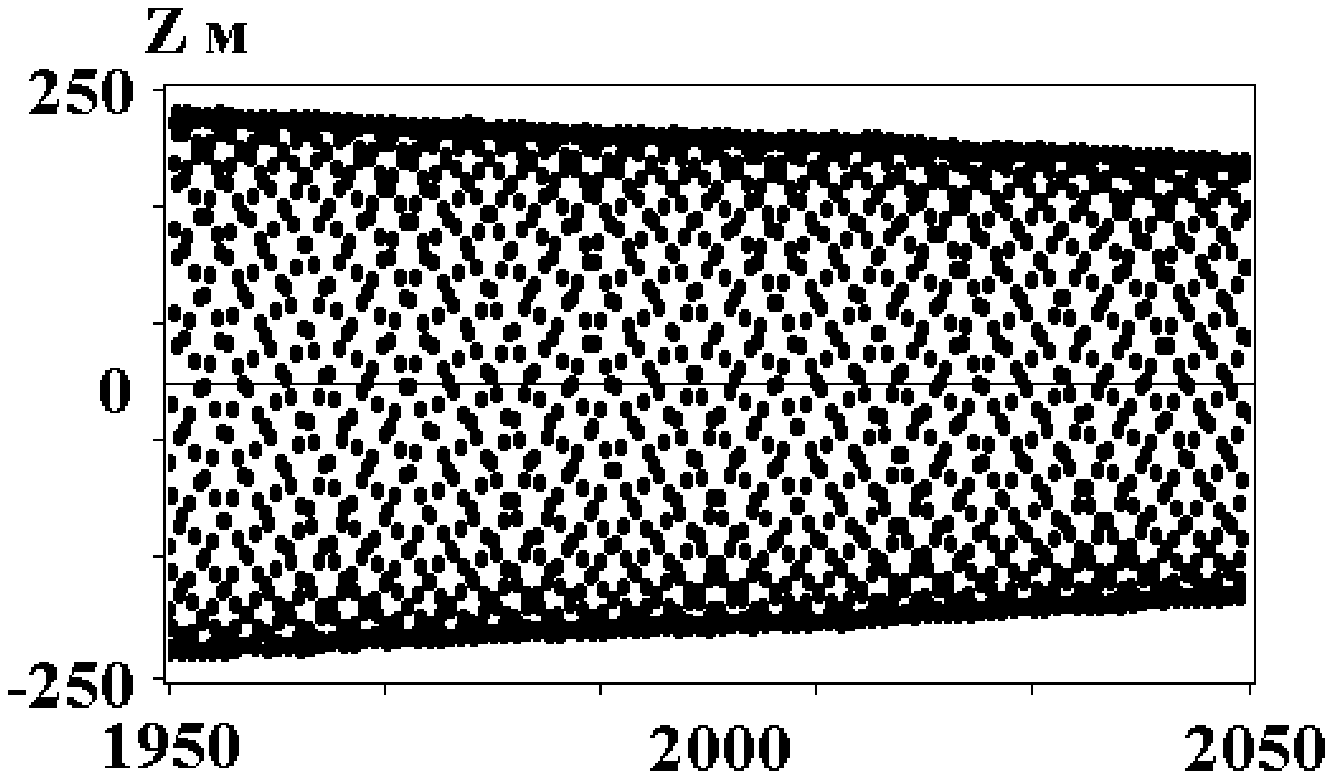}
\hskip 2.0truecm
\includegraphics[scale=0.43]{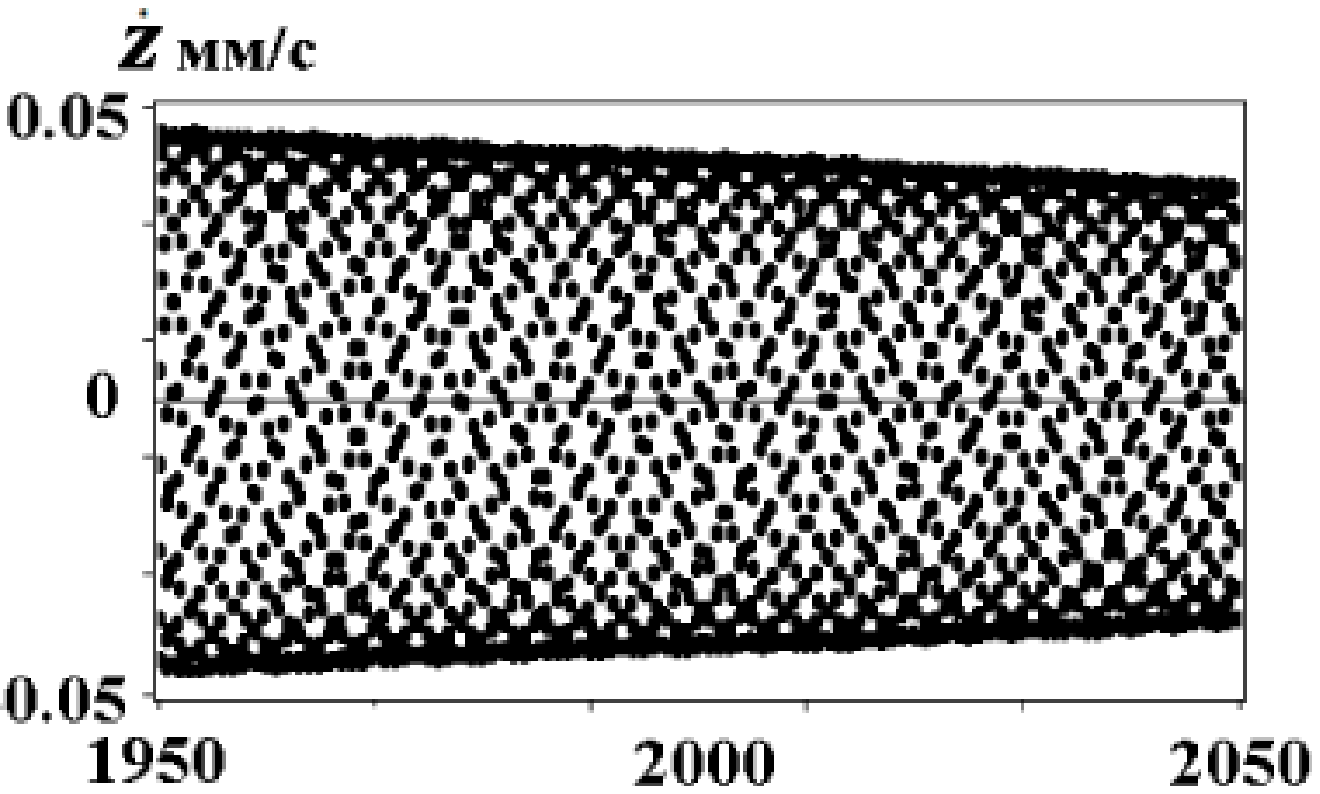}
}
\end{center}
\centerline{ {\bf Fig. 5.} Differences EPM2011 -- DE424 in heliocentric 
coordinates ($X, Y, Z$)}
\centerline{and velocities ($\dot X, \dot Y, \dot Z $) of the Earth from 1950 to 
2050.}
\end{figure}


Some parameters determined in the process of improving the DE and EPM ephemerides 
(Pitjeva and Standish, 2009) and adopted as the current best values for ephemeris 
astronomy by XXVII IAU GA in 2009 (Luzum et al., 2011) were taken as initial in 
EPM2011 and were then improved from all observations. Among them are such 
parameters as the ratio of masses of the Earth and the Moon 
$M_{\hbox{Earth}}/M_{\hbox{Moon}} = 81.30056763 \pm 0.00000005 $ and the masses 
of largest asteroids (Ceres, Pallas, and Vesta) and 18 other asteroids. Table 4 
gives the masses and the estimates of these masses with ones taken from papers by 
Konopliv et al. (2011) and Fienga et al. (2011), where they were obtained in the 
same way using the DE423 and INPOP10a ephemerides. All parameters obtained in the 
present work and 
mentioned in this section are given with uncertainties that correspond to 
$3\sigma$ (formal standard error of the least squares method). Experience shows 
that formal standard errors are overly optimistic. Uncertainties given by 
Konopliv et al. (2011) are obtained with a special method that is characterized 
by the fact that the uncertainties of the masses of asteroids that are not 
estimated are taken into account while calculating all the adjusted parameters. 
The uncertainties obtained in this way are probably close to the actual errors.
Uncertainties specified in a paper by Fienga et al. (2011) are larger than the 
ones obtained here due to the large quantity (145) of the estimated masses of 
asteroids. The data presented in Table 4 point to the fact that the estimates of 
masses of asteroids largely agree with each other within the limits of their 
errors. The two exceptions are the masses of (52) Europa and (511) Davida for the 
INPOP10a ephemerides. On August 16, 2011, the {\it Dawn} spacecraft approached 
Vesta, one of the largest asteroids. The spacecraft studied the asteroid for a 
year and determined its mass to be $(130.2927 \pm 0.0005) \cdot 10^{-12} GM_{\odot}$ 
(Russel et al., 2012). This value virtually coincides with the estimate of the 
mass of Vesta obtained in the present work.

Special effort was given to producing an accurate estimate of the total influence 
of asteroids on the motion of planets, the majority of which lie in the main 
asteroid belt. In EPM the main belt is modeled by the motion of the 301 largest 
asteroids and a homogeneous material ring that represents the influence of 
numerous other small asteroids. Parameters $M_{\hbox{ring}}$ and 
$R_{\hbox{ring}}$ that characterize the ring of small asteroids were determined 
through the processing of observations:
$$ M_{\hbox{ring}}=(1.06\pm1.12)\cdot10^{-10} M_{\odot},\quad
R_{\hbox{ring}}=(3.57\pm0.26)\, \hbox{AU} $$

\smallskip
\noindent {\bf Table 4.} Estimates of the masses of asteroids obtained by using 
the observations of ranging for the EPM 2011, DE423 (Konopliv et al., 2011), and 
INPOP10a (Fienga et al., 2011) ephemerides and expressed in 
$(GM_i/GM_{\odot})\times 10^{-12}$
\bigskip

\begin{tabular}{|l|c|c|c|}
\noalign{\smallskip}
\hline
\noalign{\smallskip}
Asteroid  & EPM2011 & DE423 & INPOP10a \\
\noalign{\smallskip}
\hline
\noalign{\smallskip}
(1) Ceres & $472.17 \pm 0.79$ & $467.90 \pm 3.25$ & $475.8 \pm 2.8$ \\
(2) Pallas & $104.72 \pm 0.92$ & $103.44 \pm 2.55$ & $111.4 \pm 2.8$ \\
(3) Juno & $14.67 \pm 0.25$ & $12.10 \pm 0.91$ & $11.6 \pm 1.3$ \\
(4) Vesta & $129.70 \pm 0.45$ & $130.97 \pm 2.06$ & $133.1 \pm 1.7$ \\
(6) Hebe & $4.05 \pm 0.46$ & $6.73 \pm 1.64$ & $7.1 \pm 1.2$ \\
(7) Iris & $6.54 \pm 0.30$ & $5.53 \pm 1.32$ & $7.7 \pm 1.1$ \\
(8) Flora & $2.05 \pm 0.18$ & $2.01 \pm 0.42$ & $4.07 \pm 0.63$ \\
(9) Metis & $1.64 \pm 0.25$ & $3.28 \pm 1.08$ & --- \\
(10) Hygiea & $41.61 \pm 1.34$ & $44.97 \pm 7.36$ & --- \\
(14) Irene & $3.61 \pm 0.28$ & $1.91 \pm 0.81$ &  --- \\
(15) Eunomia & $14.45 \pm 0.55$ & $14.18 \pm 1.49$ & $18.8 \pm 1.6$ \\
(16) Psyche & $12.75 \pm 1.03$ & $12.41 \pm 3.44$ & $11.2 \pm 5.2$ \\
(19) Fortuna & $4.36 \pm 0.13$ & $3.20 \pm 0.53$ & --- \\
(23) Thalia & $1.24 \pm 0.21$ & $1.11 \pm 0.71$ &  --- \\
(29) Amphitrite & $5.39 \pm 0.50$ & $7.42 \pm 1.49$ & --- \\
(41) Daphne & $4.17 \pm 0.44$ & $4.24 \pm 1.77$ & $9.2 \pm 2.6$ \\
(52) Europa & $9.06 \pm 1.32$ & $11.17 \pm 8.40$ & $42.3 \pm 8.0$ \\
(324) Bamberga & $5.10 \pm 0.14$ & $5.34 \pm 0.99$ & $4.67 \pm 0.38$ \\
(511) Davida & $6.11 \pm 1.74$ & $8.58 \pm 5.93$ & $19.9 \pm 4.1$ \\
(532) Herculina & $7.07 \pm 0.62$ & $4.97 \pm 2.81$ & $2.89 \pm 0.76$ \\
(704) Interamnia & $12.22 \pm 0.96$ & $19.97 \pm 6.57$ & --- \\
\hline
\noalign{\smallskip}
\end{tabular}
\bigskip

The total mass $M_{belt}$ of the main belt asteroids is expressed as the sum of the 
masses of 301 largest asteroids and the asteroid ring and is equal to
$ M_{belt} = (12.3 \pm 2.1)\cdot10^{-10} M_{\odot}$ (about 3 times the 
mass of Ceres).
The gravitational attraction of trans-Neptunian objects is modeled in much the 
same way by summing the influences of 21 known TNOs and an additional homogeneous
ring with a radius of 43 AU that represents numerous other smaller objects. The 
mass of the TNO ring $M_{TNOring}$ was determined to be equal to
 $ M_{TNOring} = (501 \pm 249)\cdot10^{-10} M_{\odot} $ while processing 
observations.

The total TNO mass $M_{TNO}$ that includes the masses of Pluto, the 21 largest TNOs, 
and the TNO ring is equal to
$M_{TNO} = 790\cdot10^{-10} M_{\odot}$, (about 164 times the mass of
Ceres or 2 times the mass of the Moon).

\bigskip
\centerline{ACCURACY OF EPHEMERIDES AND COMPARISON BETWEEN} 
\centerline{THE EPM2011 AND DE424 EPHEMERIDES}
\smallskip

Firstly, the accuracy of the constructed ephemerides may be estimated from the 
representation of observations, i.e., from comparison of the observal values (O)
with the computed values (C) of observations. Tables 1--3 and Figures 2--4 
present the residuals, their mean values, and their errors ($\sigma$) that do not
exceed their a priori errors. Secondly, the accuracy of ephemerides may be 
evaluated by comparing them with other ephemerides constructed by independent
research teams. Starting from the 1970s, the EPM ephemerides computed at the IAA 
RAS, were regularly compared with the DE ephemerides created at JPL. In a paper 
by Pitjeva (2005a), the differences in heliocentric distances of planets for the 
EPM2004 and DE410 ephemerides over a 40-year interval (from 1970 to 2010) are 
presented. In the present work, the differences in three coordinates over a 
100-year interval are presented in Figures 6 and 7. These coordinates --
geocentric distances ($D$), right ascensions ($\alpha$), and declinations 
($\delta$) -- fully characterize the accuracy of the ephemerides determined by 
comparing the EPM2011 and DE424 ephemerides.

\smallskip
\begin{figure}[h!]
\begin{center}
\hbox{
\hskip -0.15truecm
\includegraphics[scale=0.35]{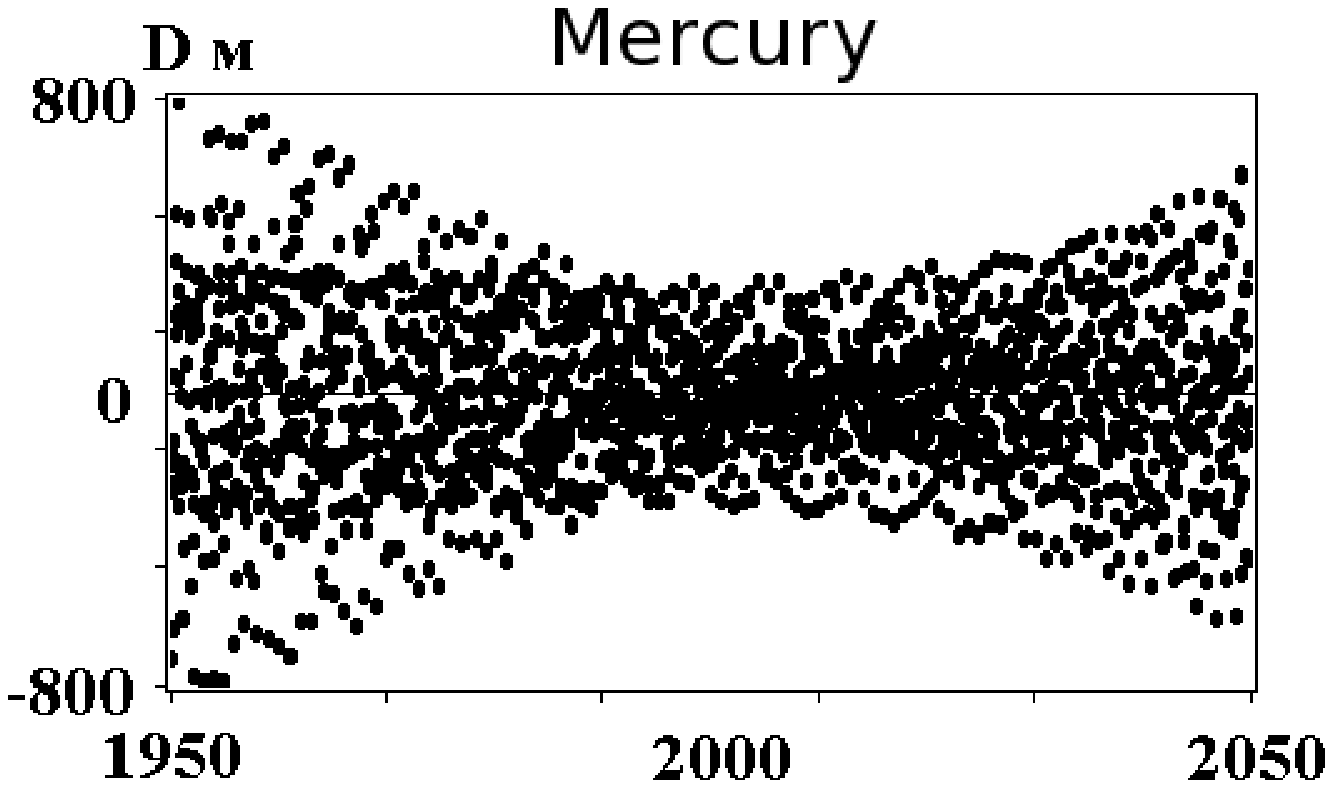}
\hskip 1.4truecm
\includegraphics[scale=0.35]{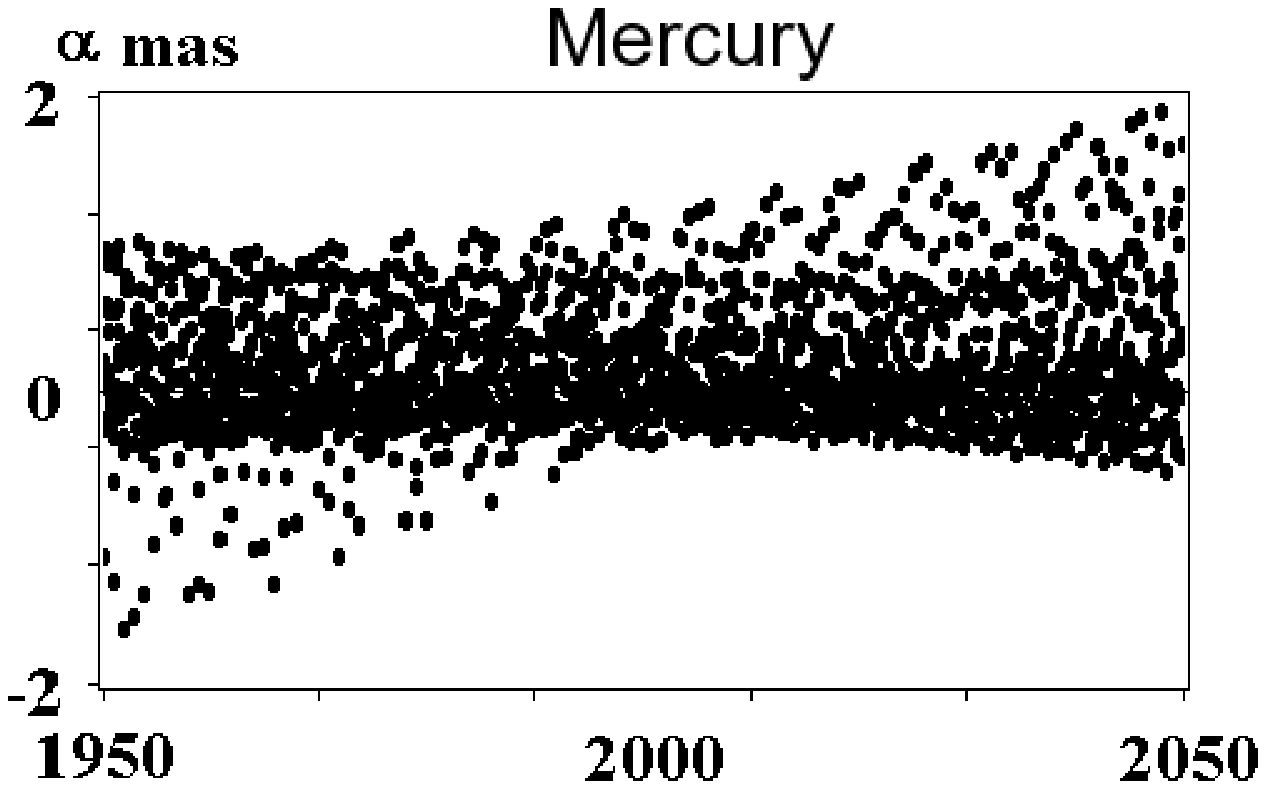}
\hskip 1.4truecm
\includegraphics[scale=0.35]{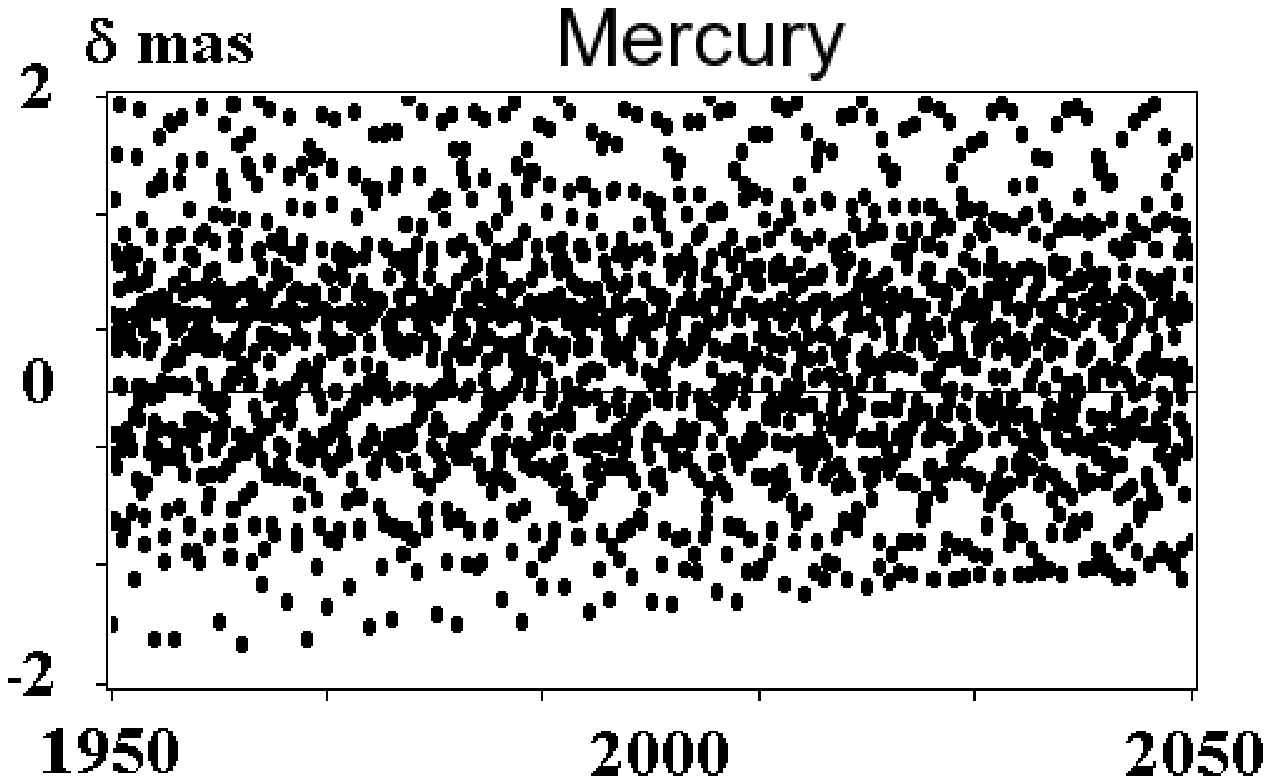}
}
\vskip 0.65truecm
\hbox{
\hskip -0.15truecm
\includegraphics[scale=0.35]{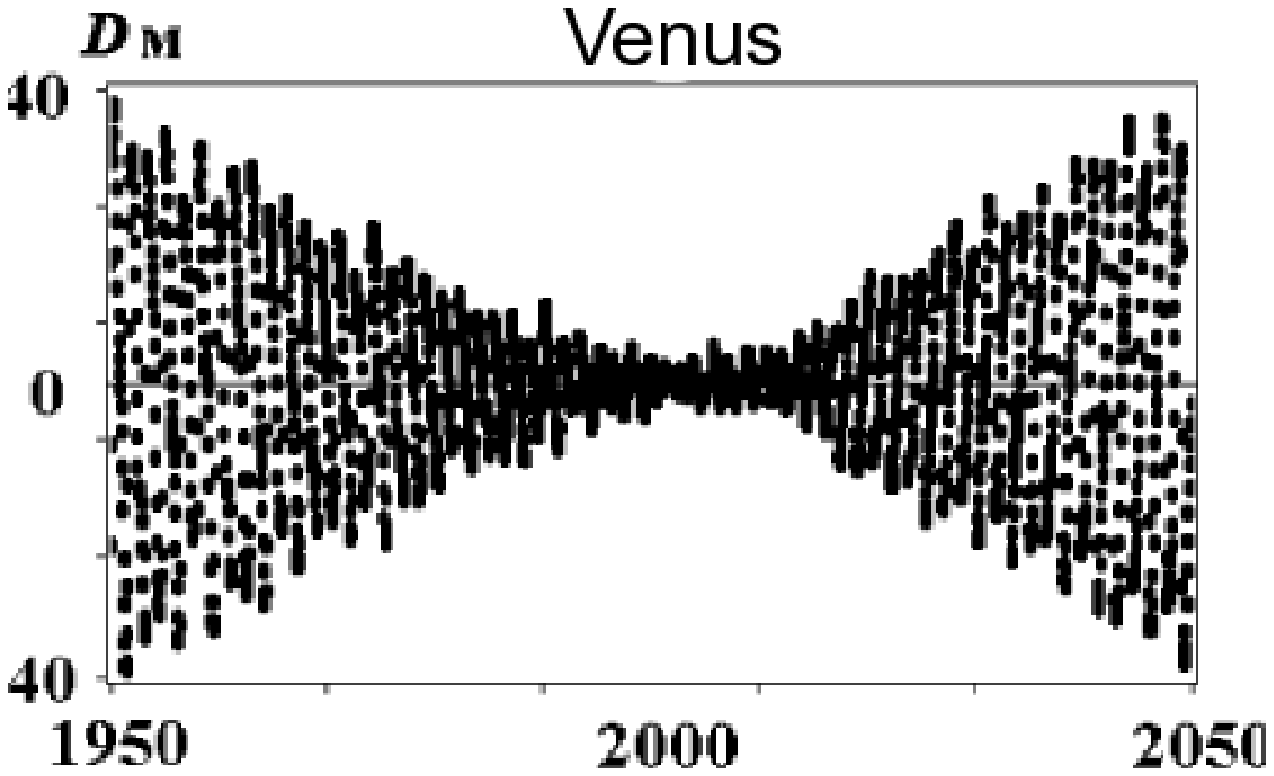}
\hskip 1.4truecm
\includegraphics[scale=0.35]{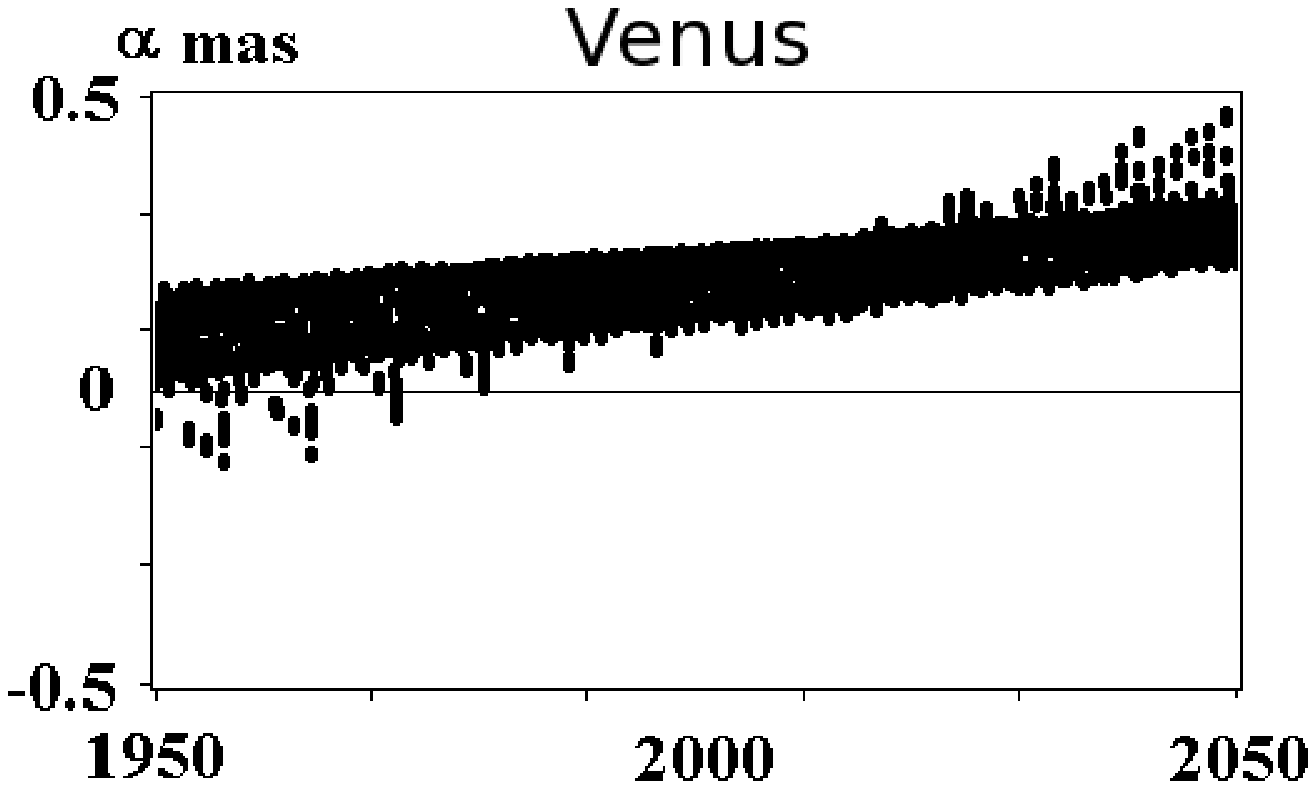}
\hskip 1.4truecm
\includegraphics[scale=0.35]{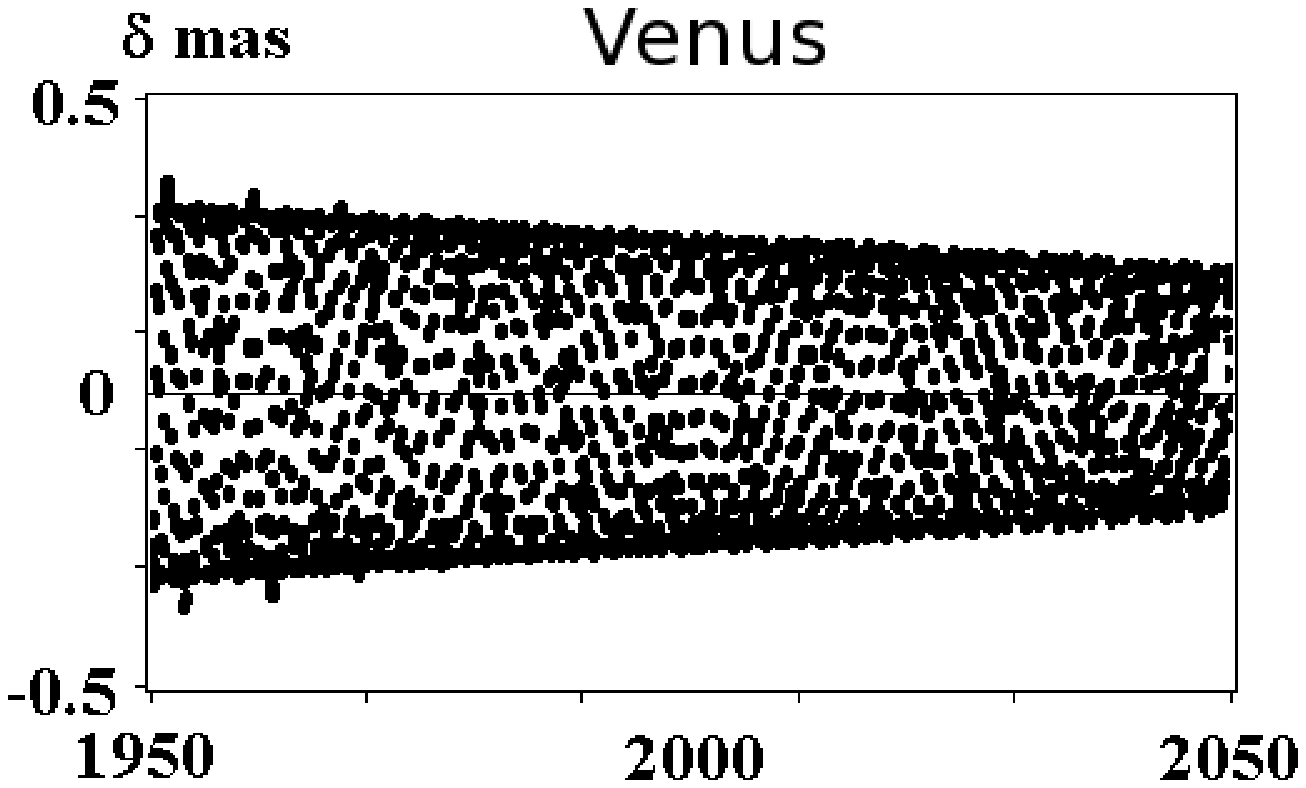}
}
\vskip 0.65truecm
\hbox{
\hskip -0.15truecm
\includegraphics[scale=0.35]{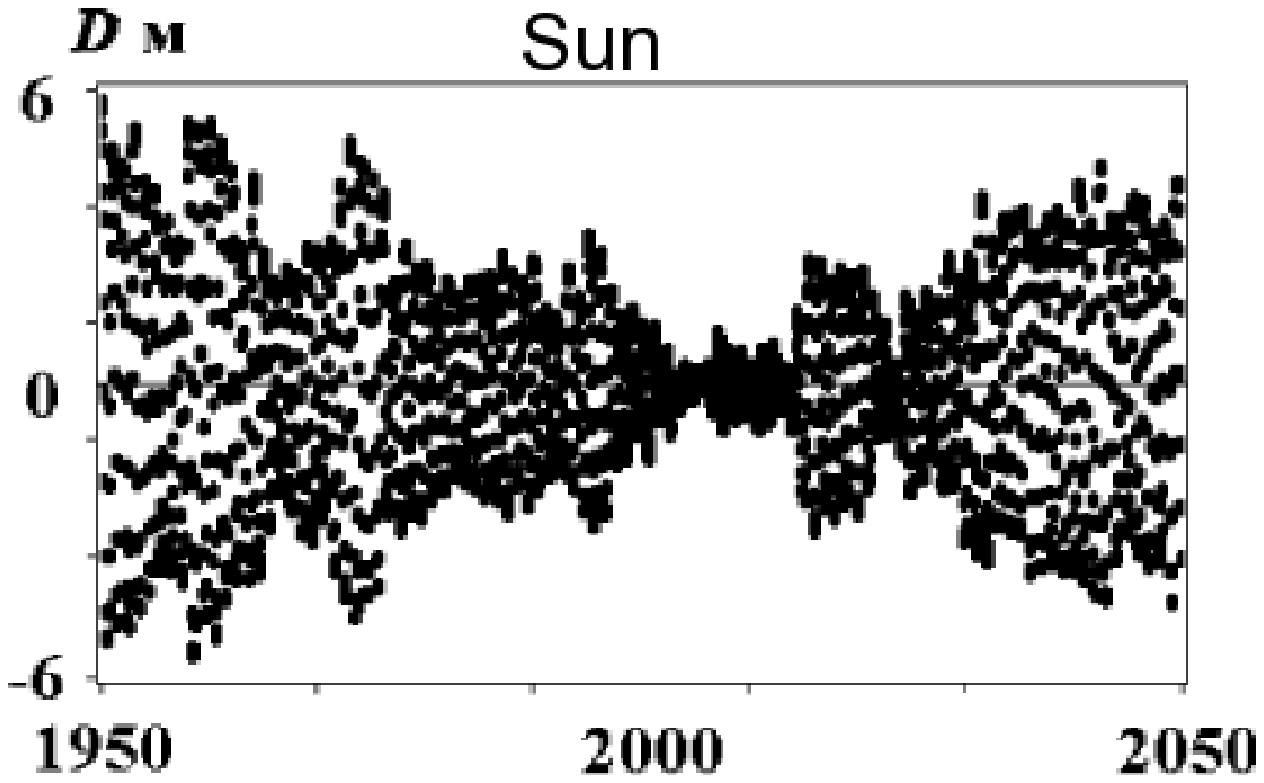}
\hskip 1.4truecm
\includegraphics[scale=0.35]{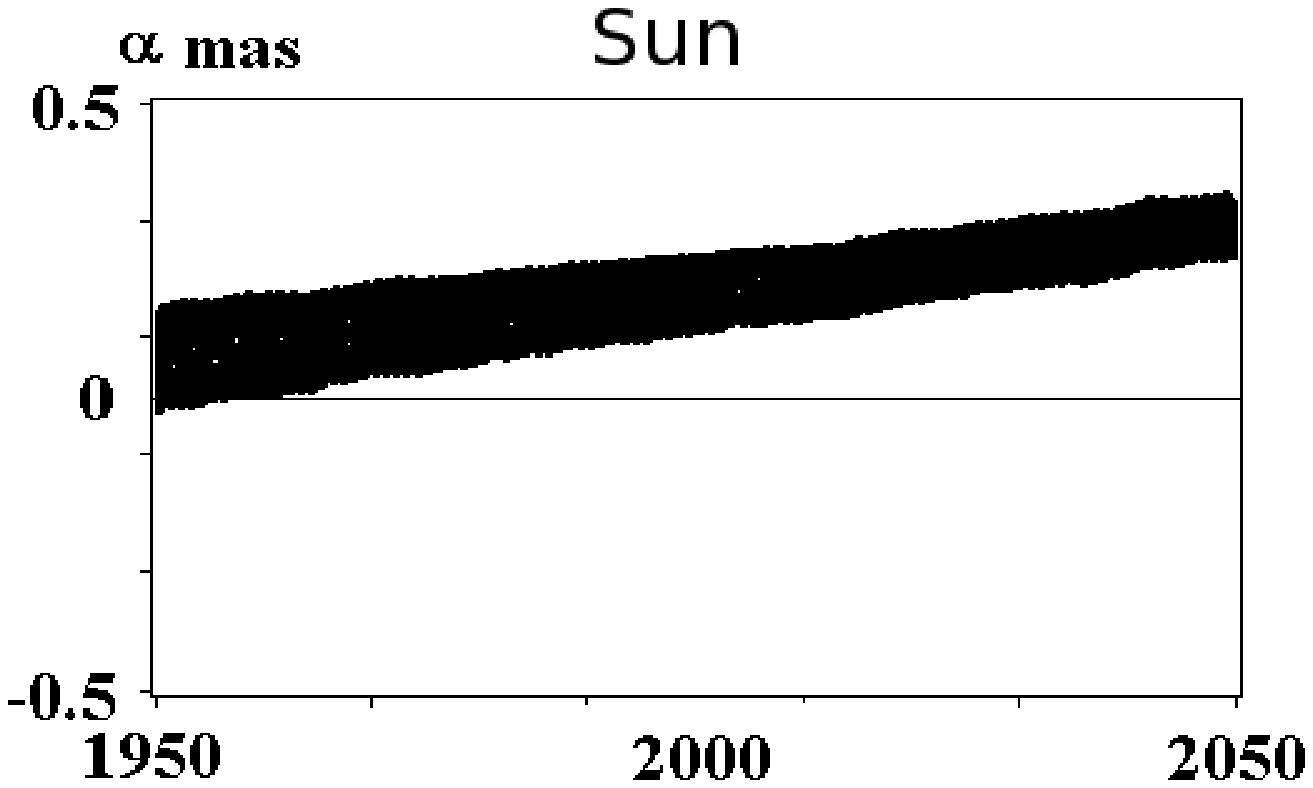}
\hskip 1.4truecm
\includegraphics[scale=0.35]{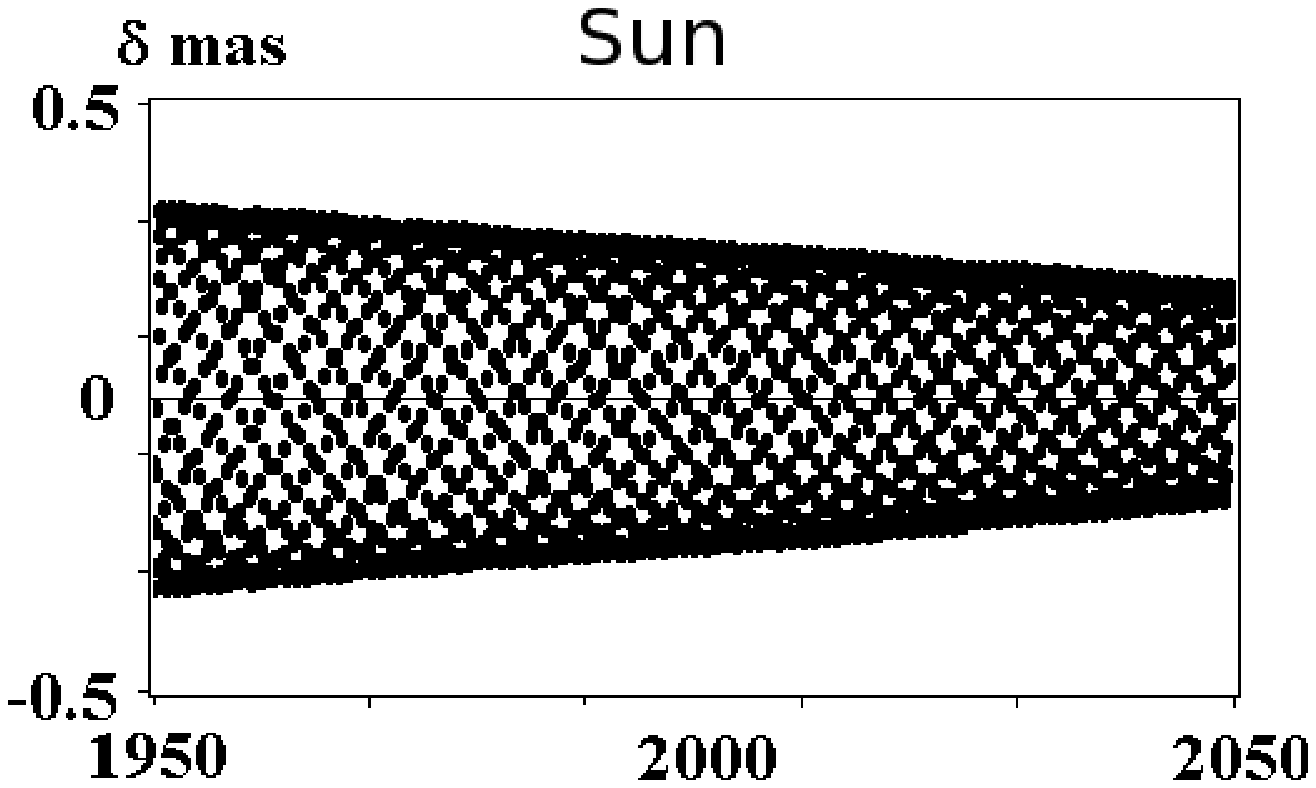}
}
\vskip 0.65truecm
\hbox{
\hskip -0.15truecm
\includegraphics[scale=0.35]{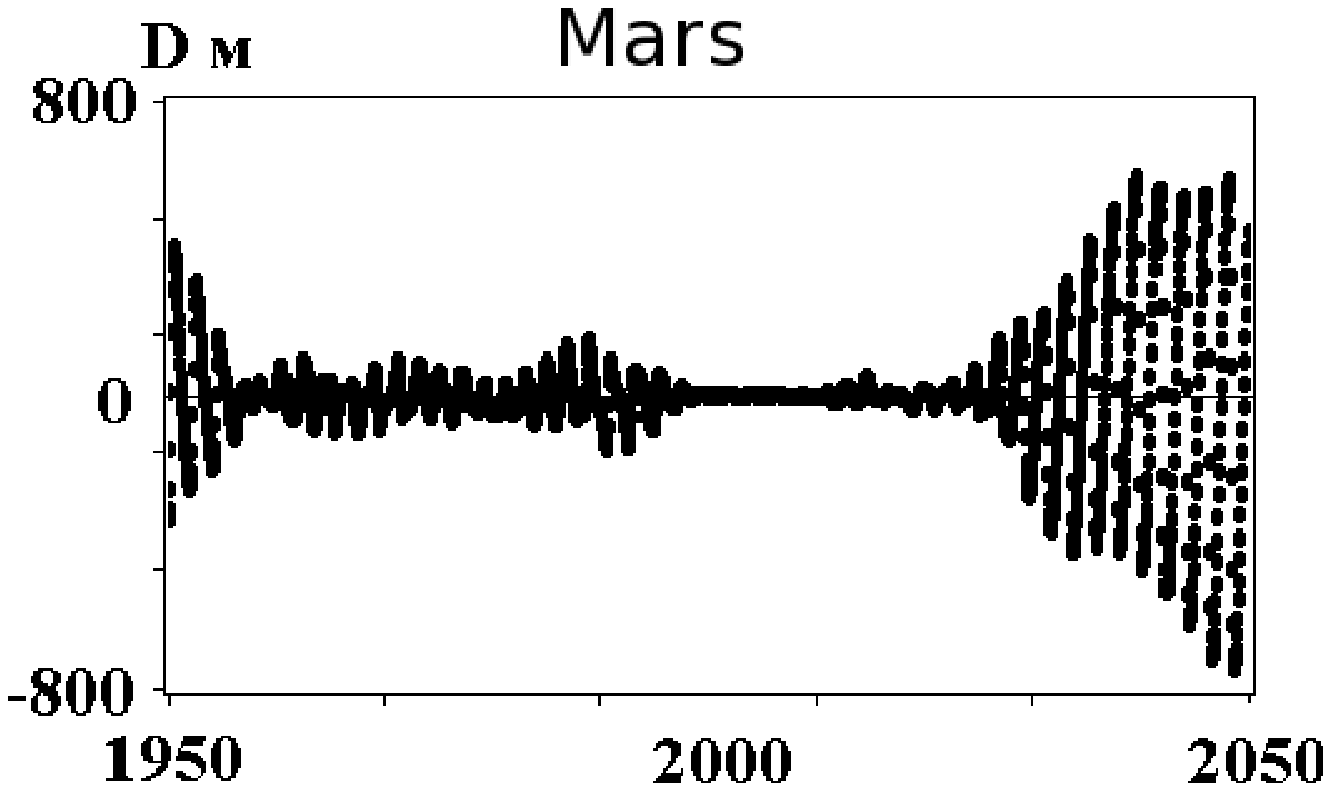}
\hskip 1.4truecm
\includegraphics[scale=0.35]{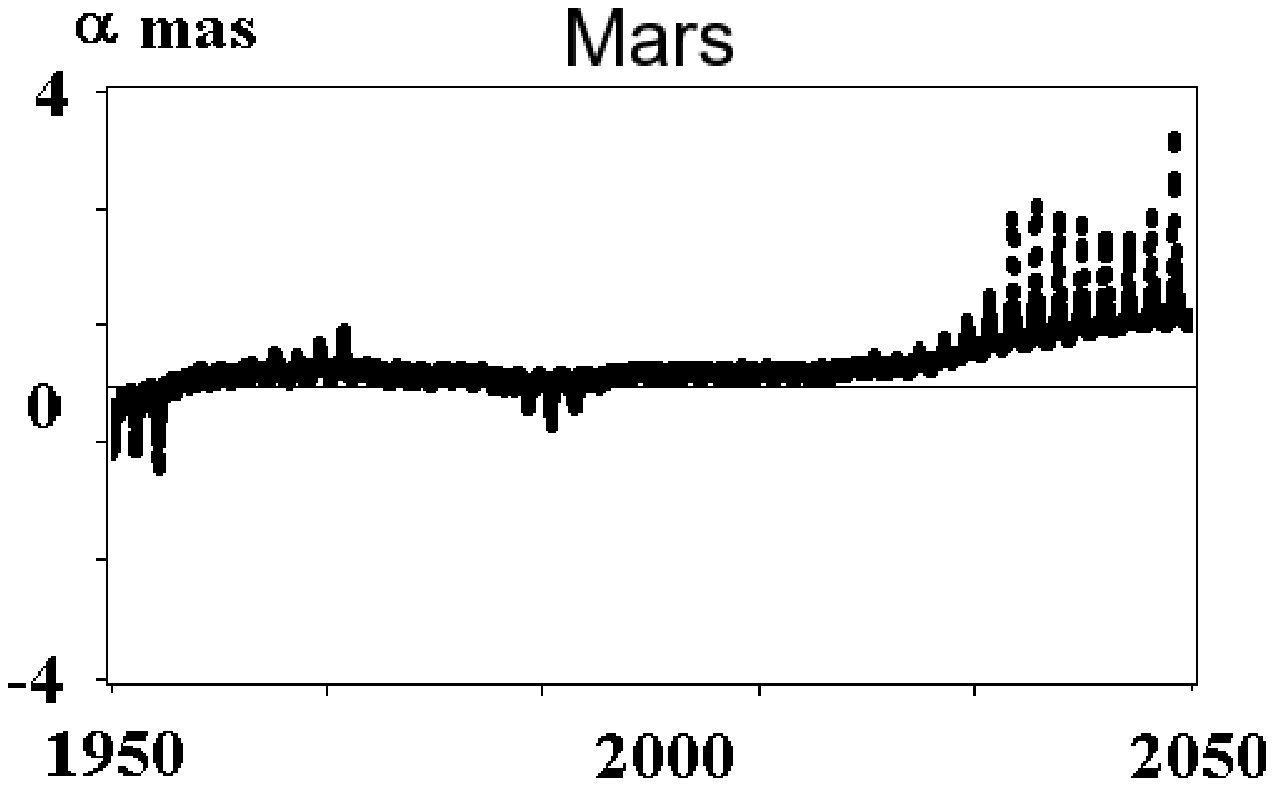}
\hskip 1.4truecm
\includegraphics[scale=0.35]{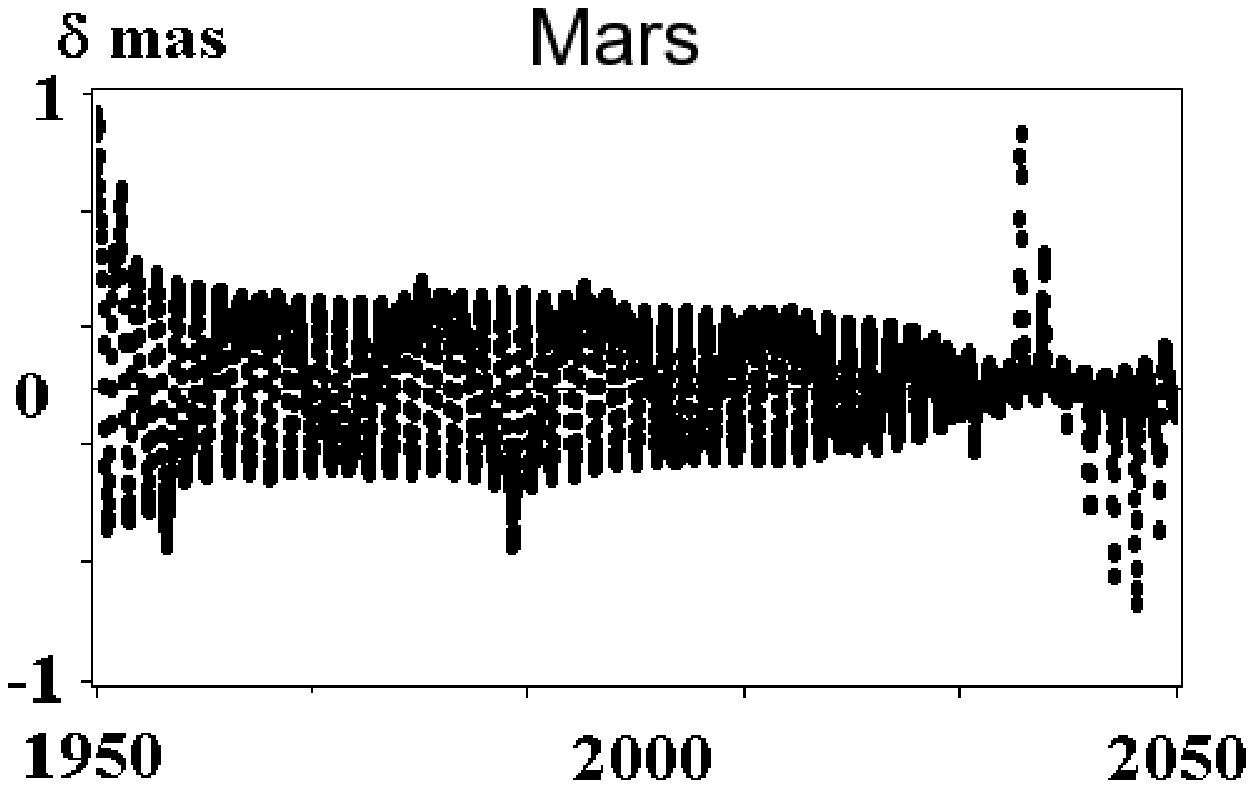}
}
\end{center}
\centerline{ {\bf Fig. 6.} Differences EPM2011--DE424 in geocentric distances 
($D$), right ascensions ($\alpha$),} 
\centerline{ and declinations ($\delta$) of Mercury, Venus, the Sun, and Mars 
over a 100-year interval}
\centerline{ (from 1950 to 2050).}
\end{figure}

Since the coordinates of the inner planets were obtained through high-precision 
radio observations, the differences calculated for them are much smaller for all 
the coordinates ($D, \ \alpha, \ \delta$) than the differences for the outer 
planets (the geocentric position of the Sun may be viewed as the heliocentric 
position of the Earth with an opposite sign). The fact that the difference in 
Mercury distances is slightly larger than the one given in a paper by Pitjeva 
(2005a) is explained by the use of new {\it Messenger} data, so far inaccessible 
to us, in DE424. The differences in distances (over the interval considered in the 
2005 paper) for all the other planets have become less. In the case of Mars, the 
differences remain minor over an interval which is somewhat wider than the one 
covered by observations. More precisely, the differences in distance, $\alpha$, 
and $\delta$ do not exceed 150 m, 0.7 mas, and 0.5 mas, respectively, over a 
58-year interval (from 1970 to 2028).

The availability of some radio observations of Jupiter and particularly Saturn 
(studied by the {\it Cassini} spacecraft) allowed us to reconstruct their orbits with
an accuracy greater than that achievable for the other outer planets' orbits 
defined virtually only by optical observations. There exists only one 
three-dimensional point ($D, \ \alpha, \ \delta$) provided by {\it Voyager~-~2} for 
Uranus and Neptune. Besides that, not even one period of orbital rotation of 
Neptune and Pluto is covered with more or less accurate observations. The 
uncertainty of the Pluto's distance, which was specified by Folkner in his talk 
at XXVIII IAU GA, changes from 1100 to 3000 km over a 18-year interval (from 2000 
to 2018). These values are roughly correspondent to the uncertainty obtained in 
the present work (3300 km) by comparing the EPM2011 and DE424 ephemerides (see 
left bottom part of Fig. 7).

\bigskip
\begin{figure}[h!]
\begin{center}
\hbox{
\hskip -0.15truecm
\includegraphics[scale=0.35]{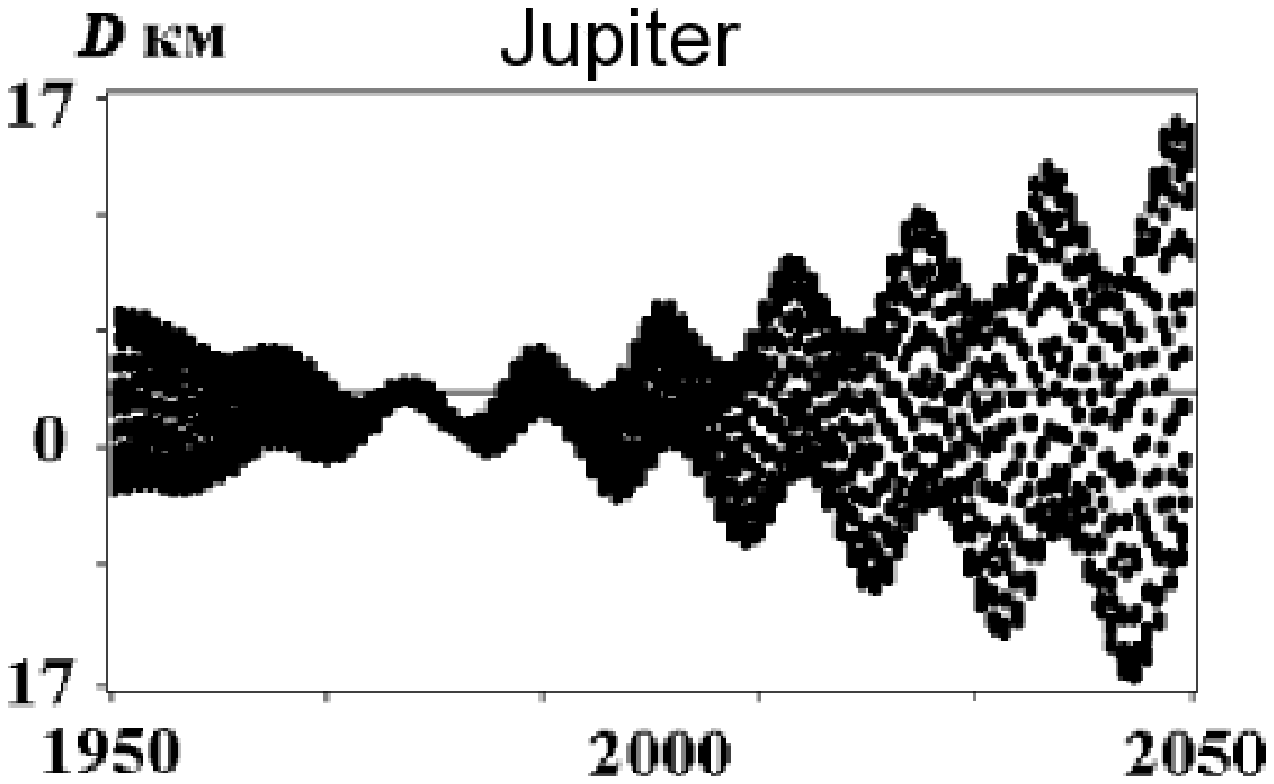}
\hskip 1.4truecm
\includegraphics[scale=0.35]{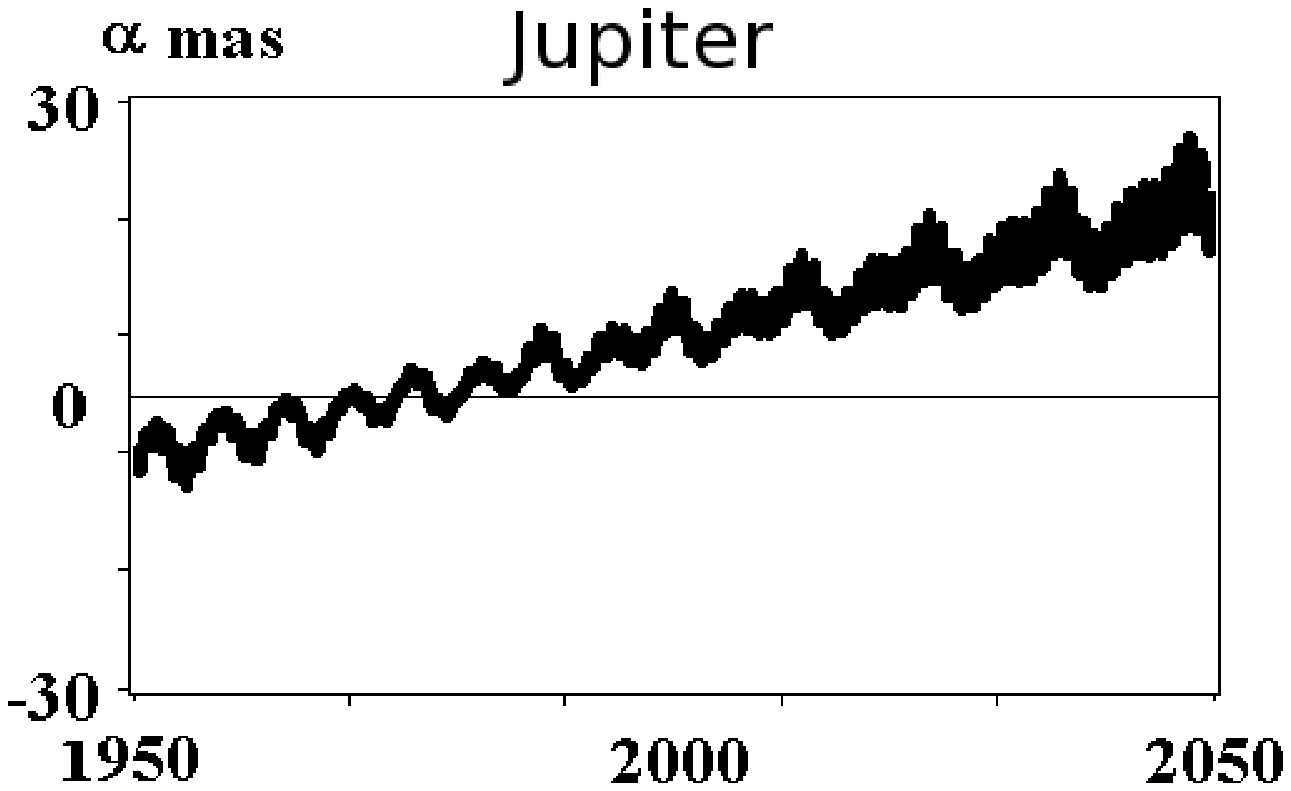}
\hskip 1.4truecm
\includegraphics[scale=0.35]{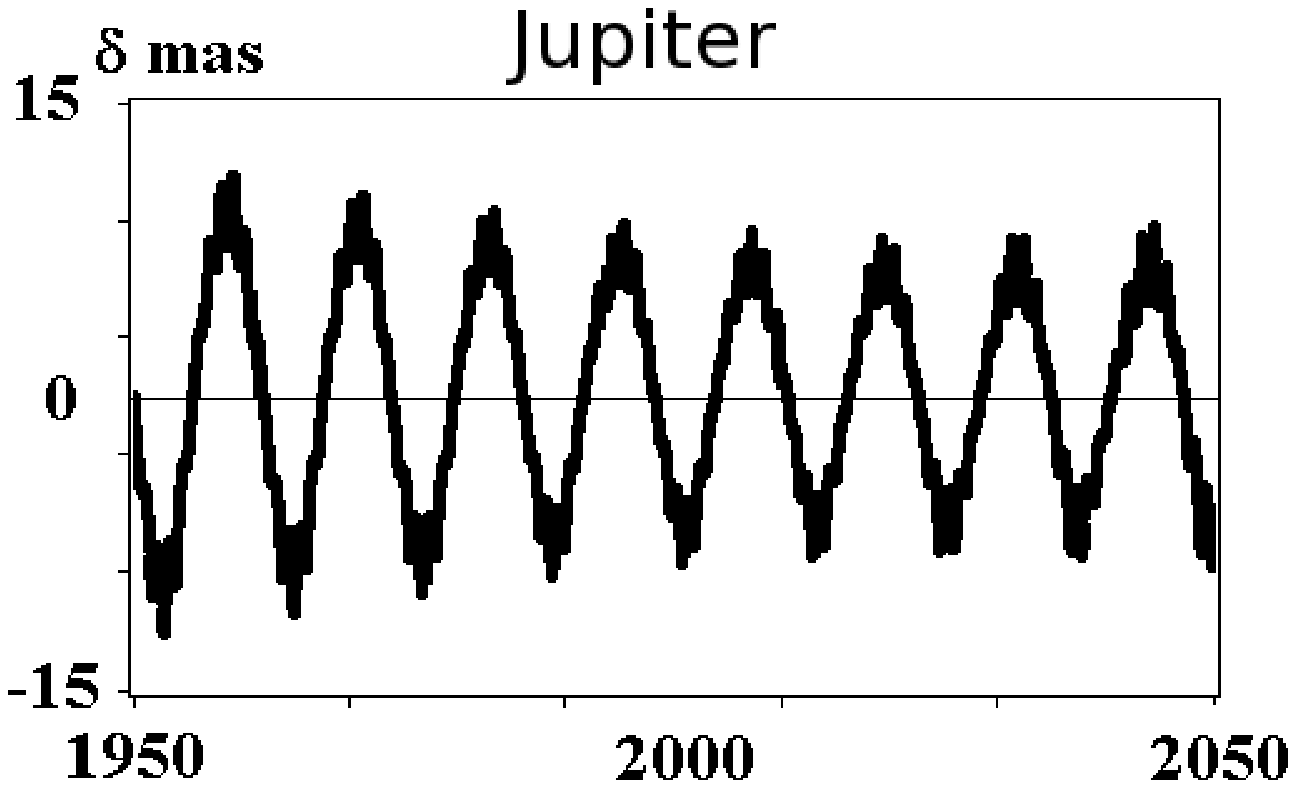}
}
\vskip 0.65truecm
\hbox{
\hskip -0.15truecm
\includegraphics[scale=0.35]{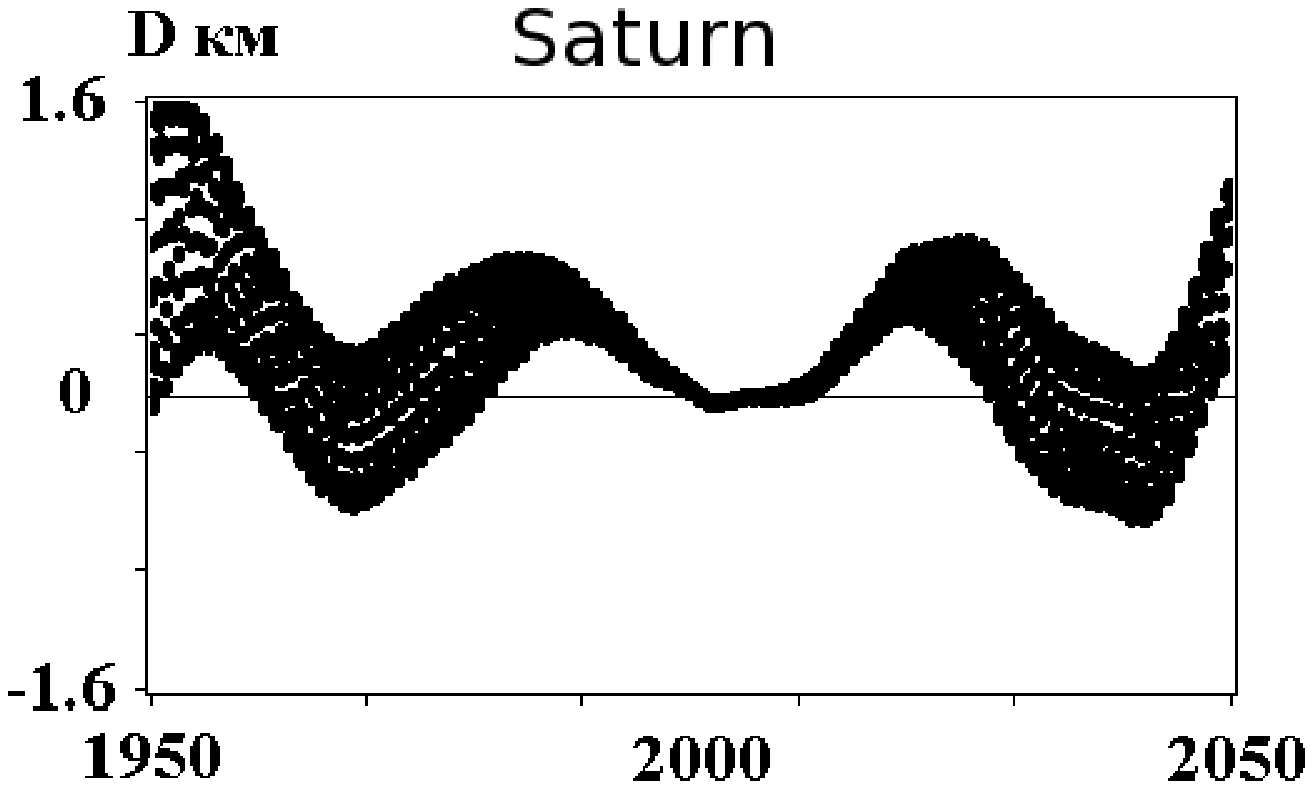}
\hskip 1.4truecm
\includegraphics[scale=0.35]{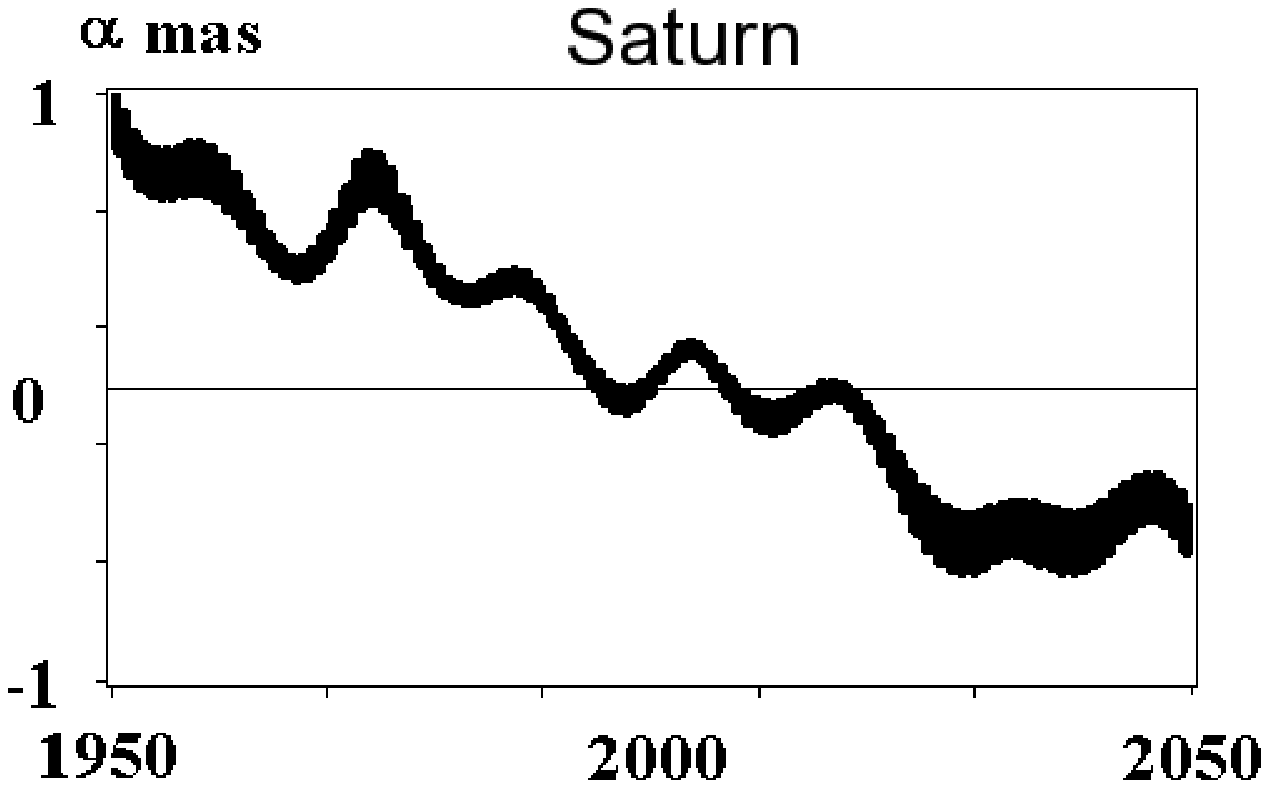}
\hskip 1.4truecm
\includegraphics[scale=0.35]{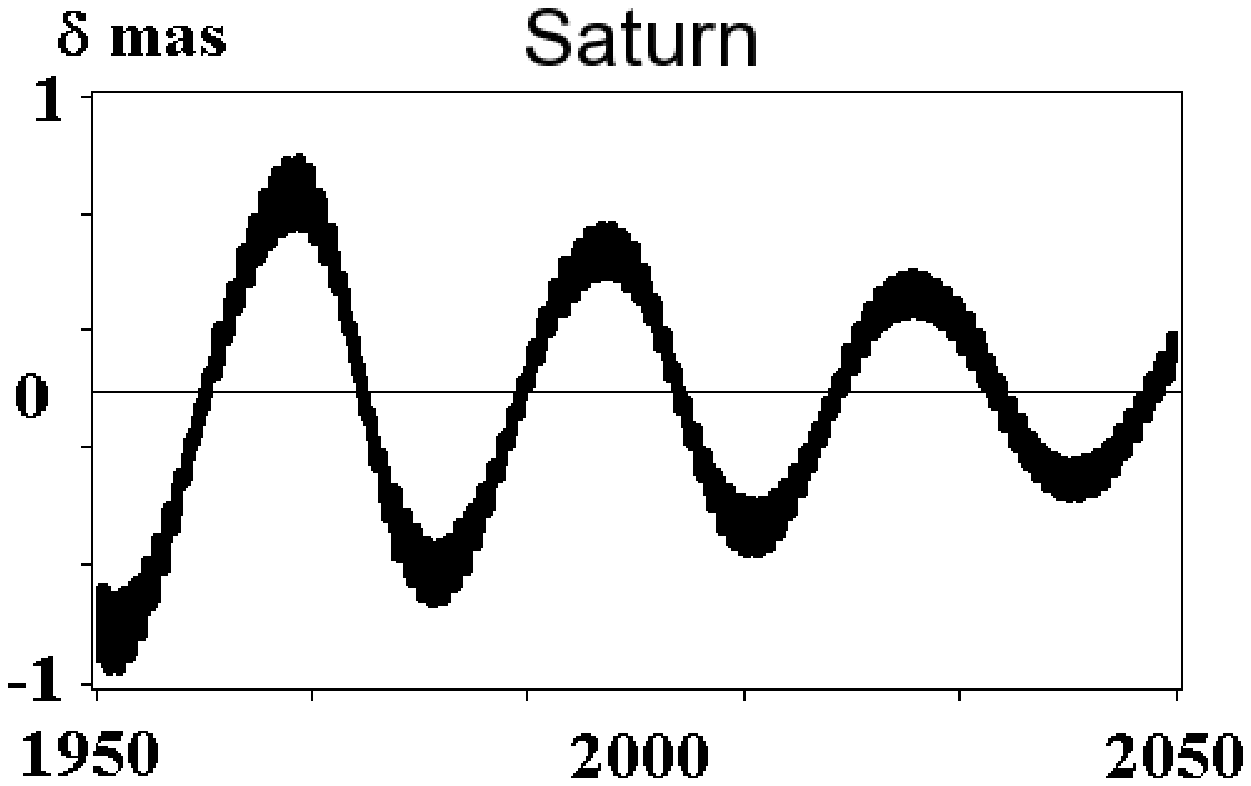}
}
\vskip 0.65truecm
\hbox{
\hskip -0.15truecm
\includegraphics[scale=0.35]{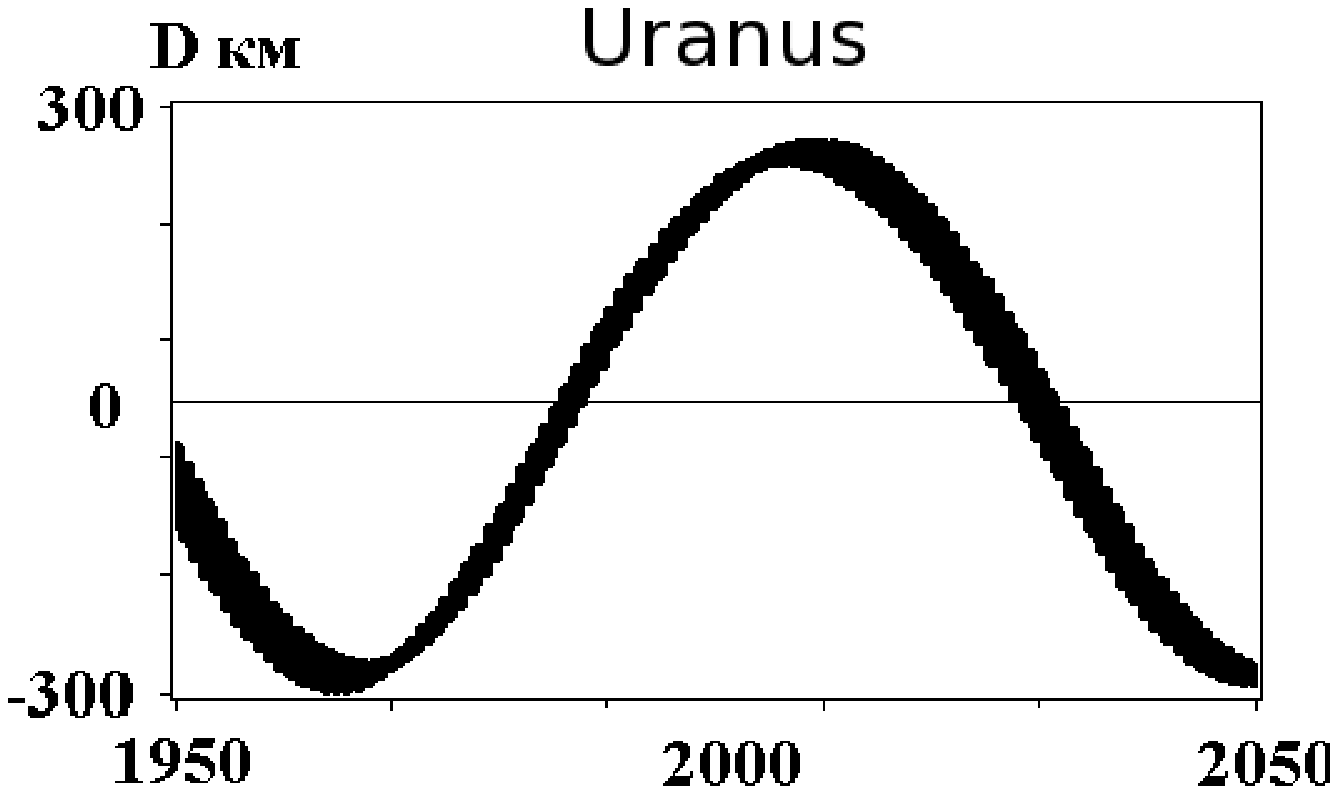}
\hskip 1.4truecm
\includegraphics[scale=0.35]{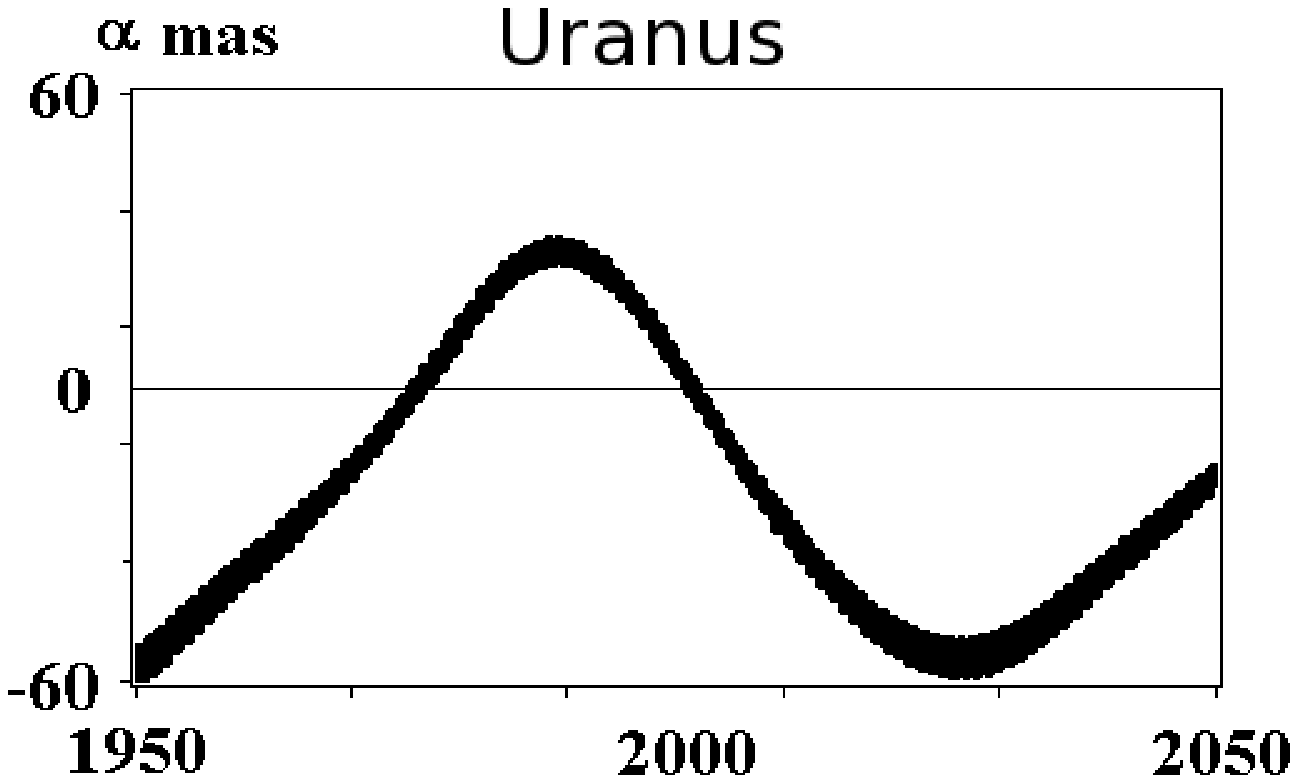}
\hskip 1.4truecm
\includegraphics[scale=0.35]{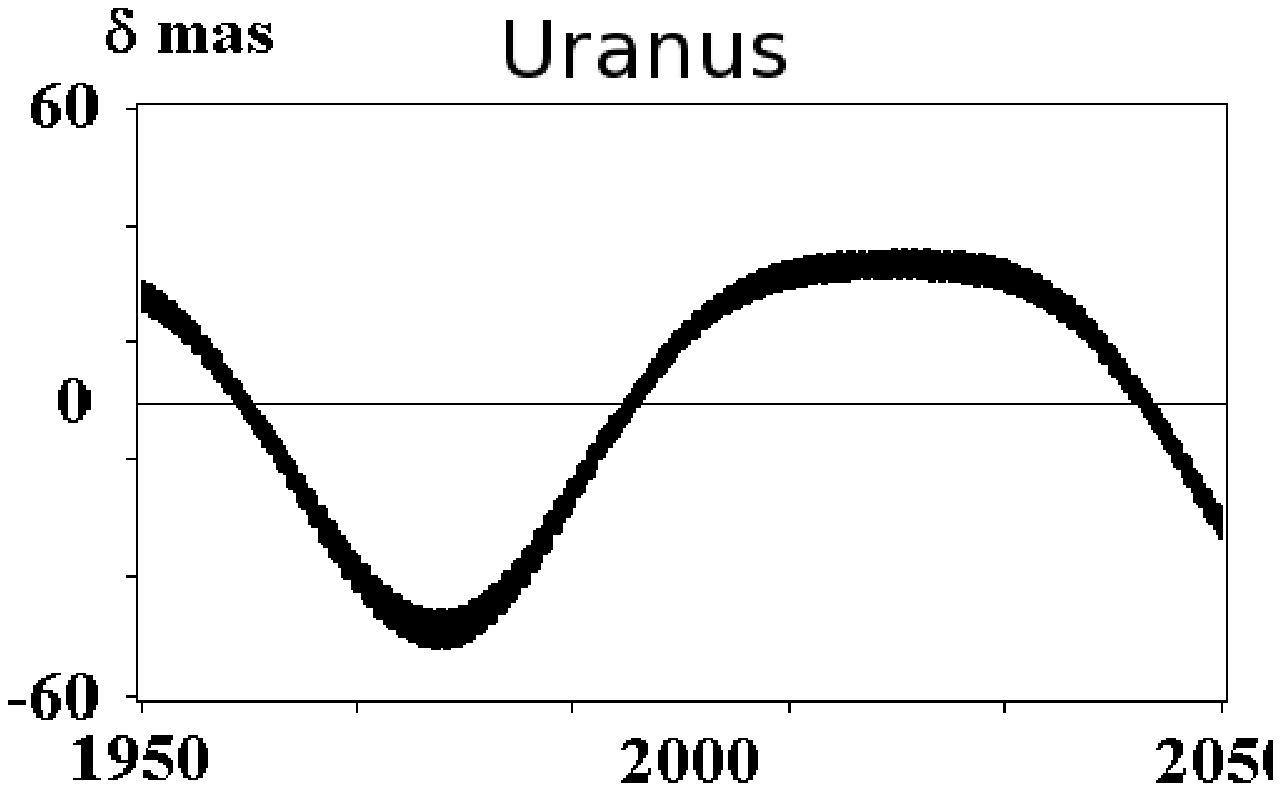}
}
\vskip 0.65truecm
\hbox{
\hskip -0.15truecm
\includegraphics[scale=0.35]{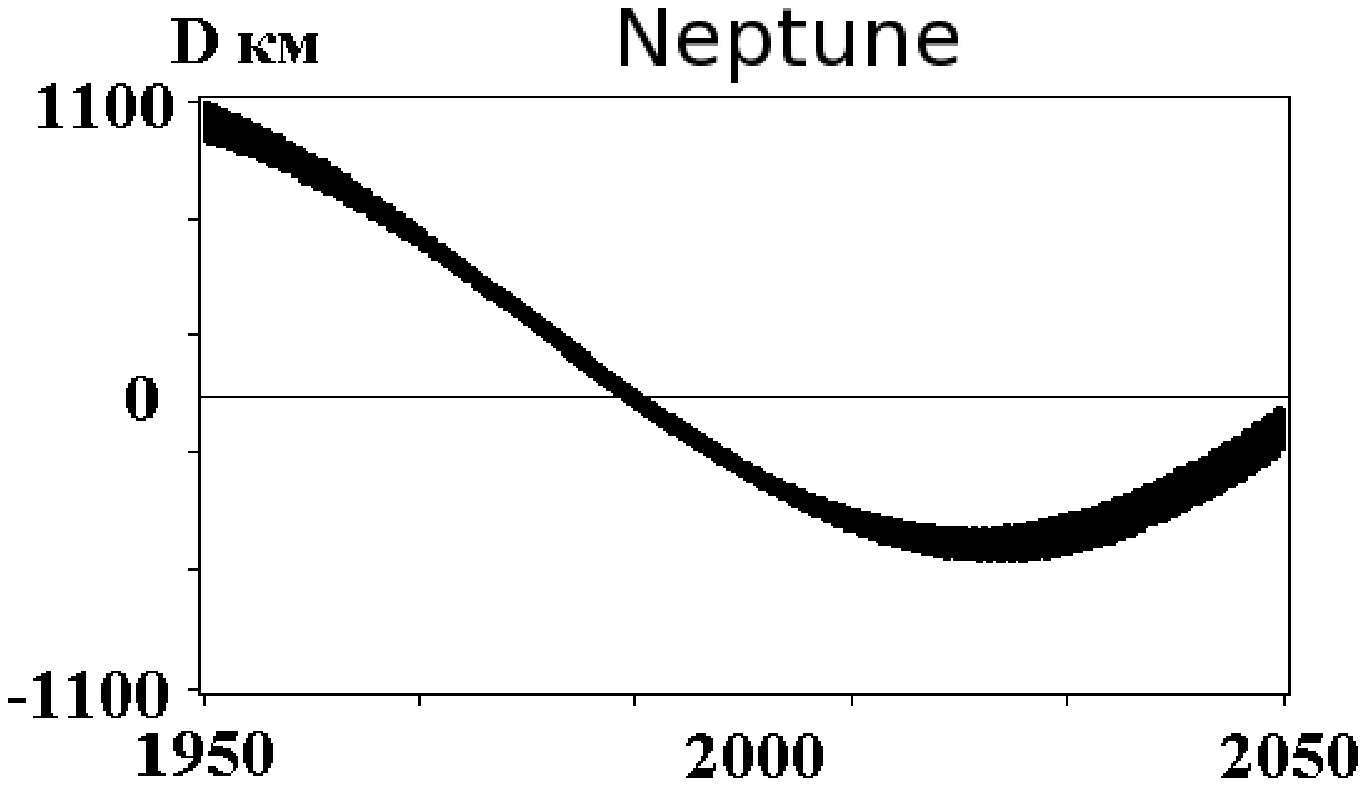}
\hskip 1.4truecm
\includegraphics[scale=0.35]{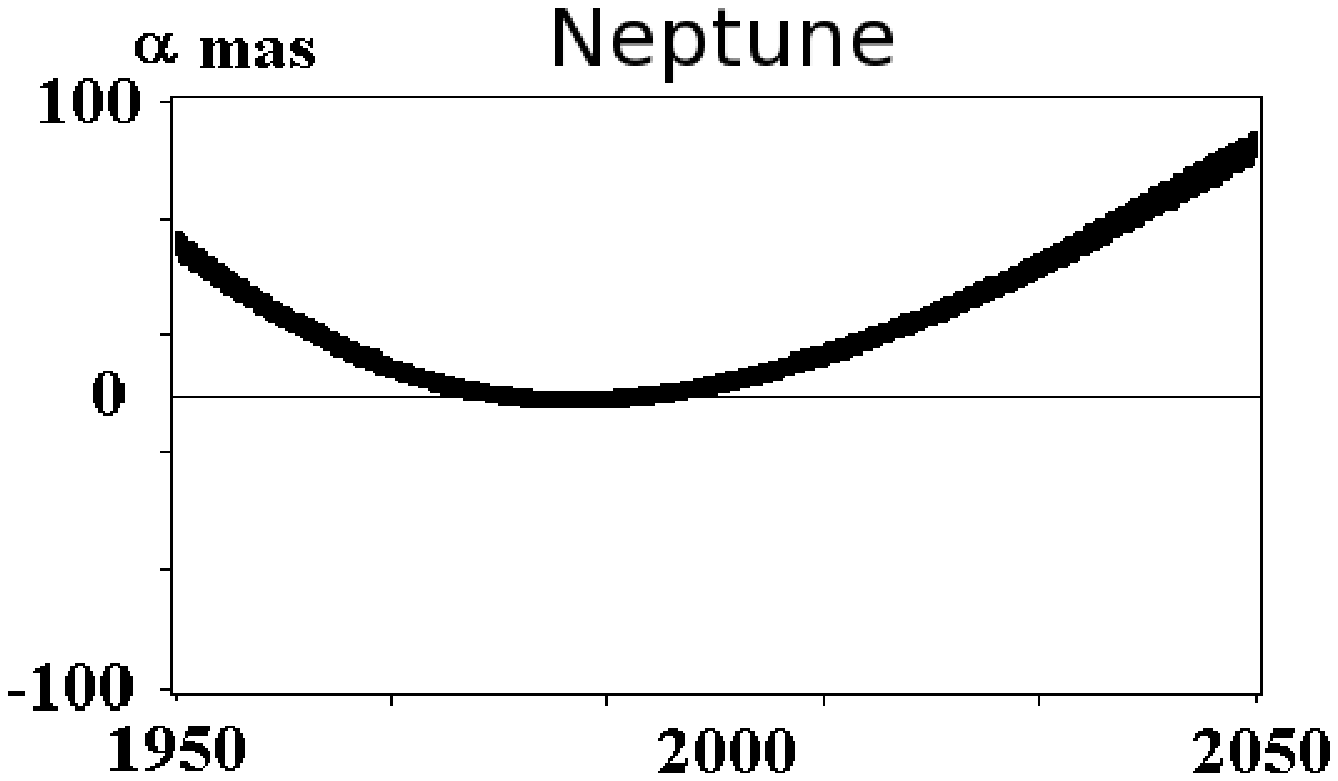}
\hskip 1.4truecm
\includegraphics[scale=0.35]{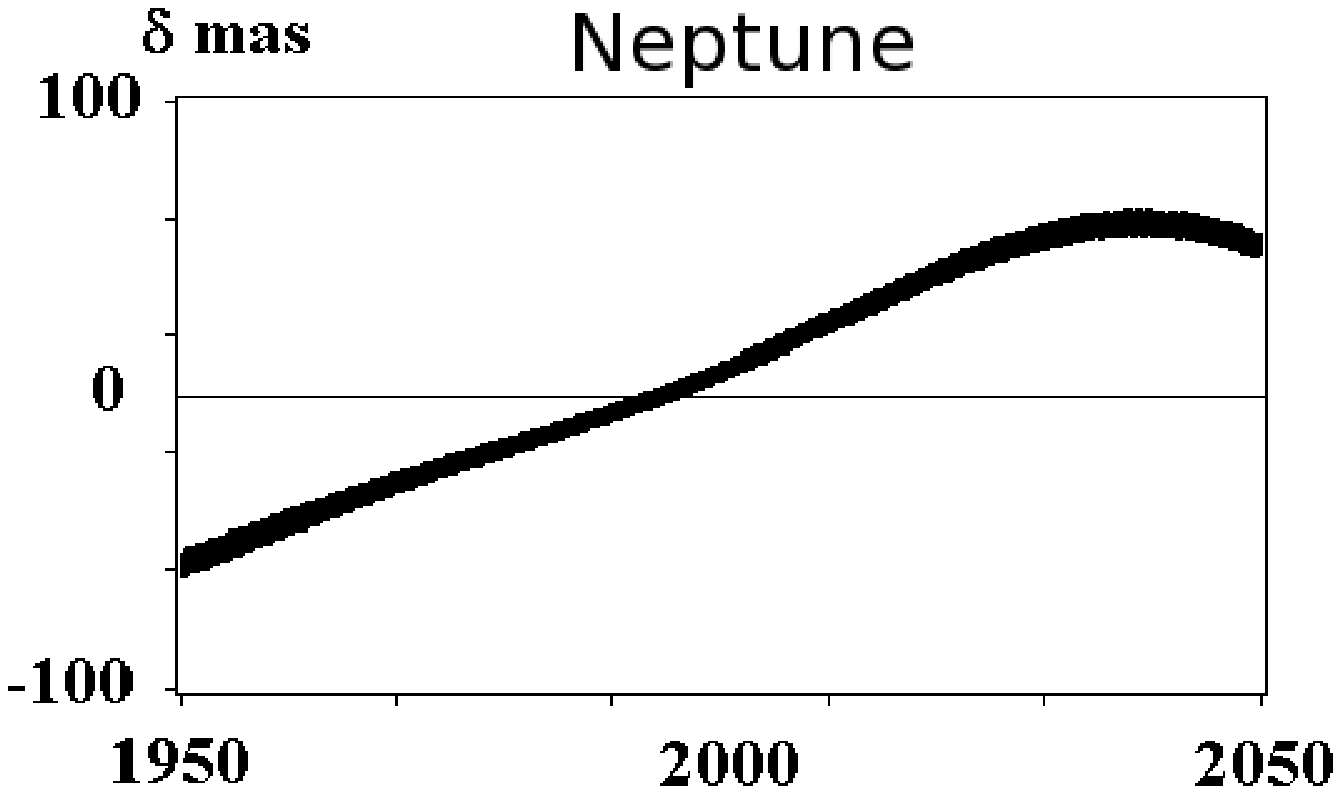}
}
\vskip 0.65truecm
\hbox{
\hskip -0.15truecm
\includegraphics[scale=0.35]{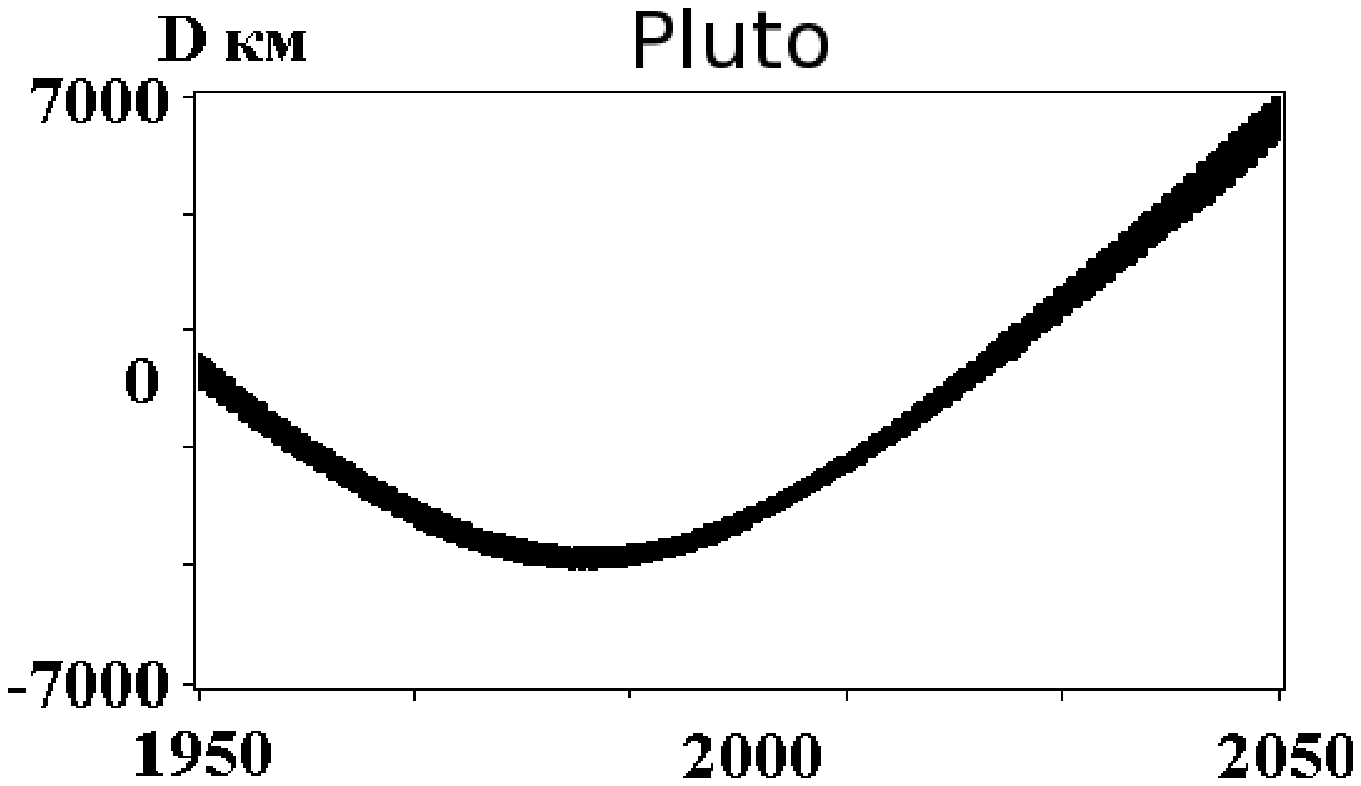}
\hskip 1.4truecm
\includegraphics[scale=0.35]{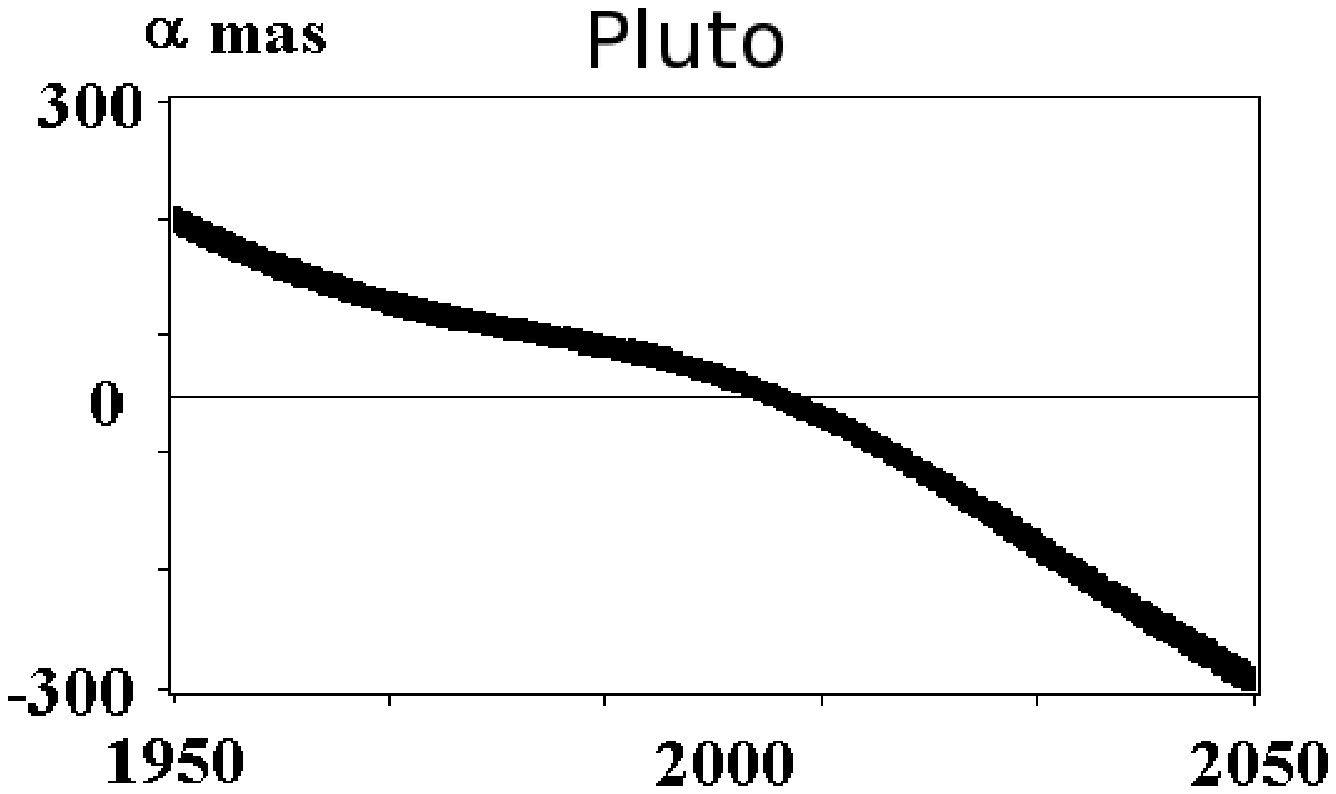}
\hskip 1.4truecm
\includegraphics[scale=0.35]{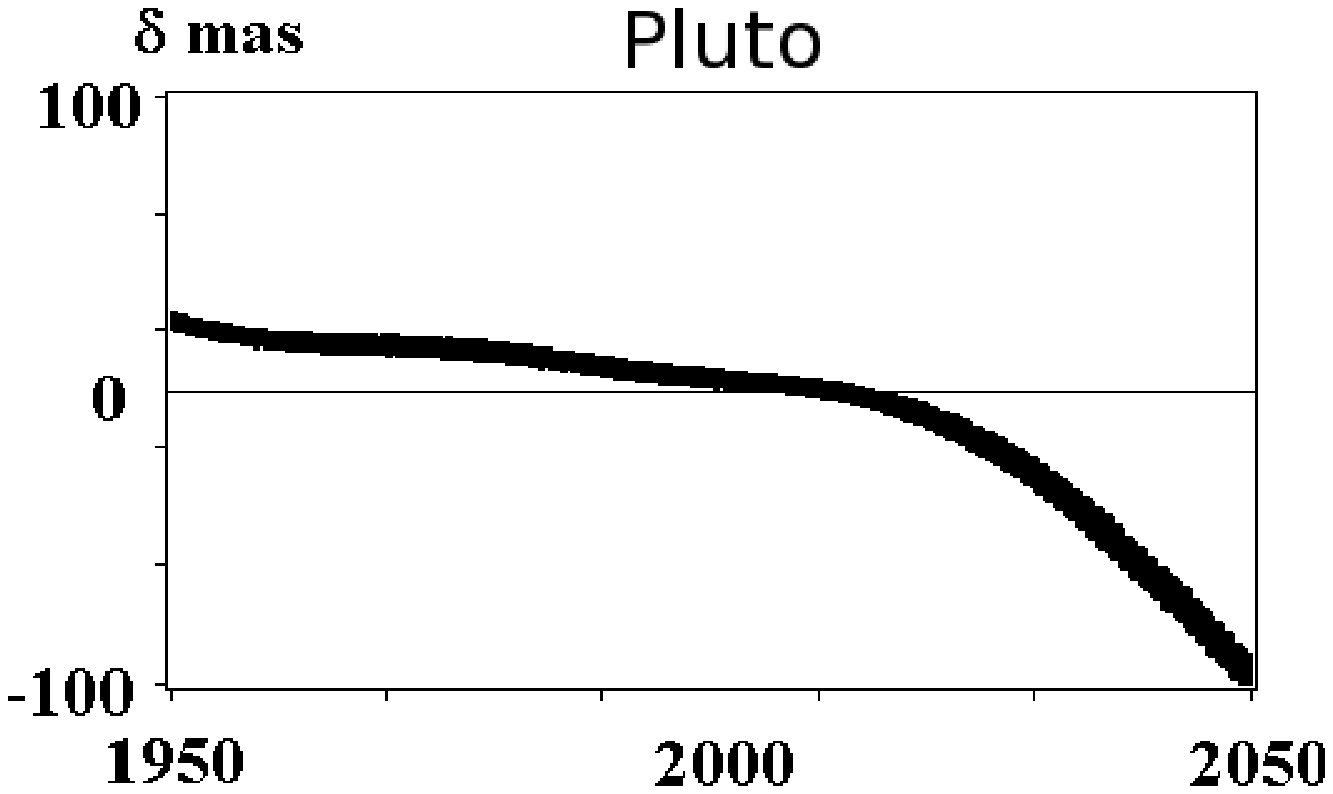}
}
\end{center}
\centerline{ {\bf Fig. 7.} Differences EPM2011--DE424 in geocentric distances 
($D$), right ascensions ($\alpha$),}
\centerline{and declinations ($\delta$) of the outer planets over a 100-year 
interval (from 1950 to 2050).}
\end{figure}

\vskip -0.5cm

In a paper by Fienga et al. (2011) the differences in geocentric distances, right 
ascensions, and declinations of planets over a 120-year interval (from 1900 to 
2020) are also shown for the INPOP10a and DE421 ephemerides. The comparison of 
results on the common interval (1950--2020) shows that all the differences in 
$D, \ \alpha, \ \delta$ for the EPM2011 and DE424 ephemerides are lower than the 
corresponding differences for the INPOP10a and DE421 ephemerides. This may be 
attributed to the use of the new version of the EPM ephemerides. Specifically, 
the INPOP10a ephemerides included the observations of the outer planets that were 
carried out not later than 2008, whereas the EPM2011 ephemerides include data up 
to the year 2011. The sole exception is the distance for Jupiter in the EPM2011 
and DE424 ephemerides. The distances for Jupiter are determined for the most part
by a few radar observations carried out from 1974 to 2000. All 7 such 
observations, weighted according to their accuracy, are used in the EPM2011 
ephemerides, while the other ephemerides include only the five most accurate ones.

It is interesting to look at the comparison of the same values given in a paper 
by Standish (2004) for the DE200 and DE409 ephemerides over a 50-year interval 
(from 1970 to 2020). It can be seen that all the differences are at least an 
order of magnitude larger. This leads to the conclusion that modern ephemerides 
have made great progress in terms of accuracy compared to DE200 (1982).

The comparison with modern observations and the DE ephemerides verifies that the 
planetary part of the EPM ephemerides is sufficiently accurate.

\bigskip
\centerline{THE USE OF THE EPM EPHEMERIDES IN SCIENTIFIC RESEARCH}
\smallskip

The potential to construct and maintain fundamental ephemerides of the major 
planets, the Sun, and the Moon may be viewed as one of the characteristics of a 
technologically mature state. The reason for this lies in the fact that these 
ephemerides have various practical applications. Specifically, they serve as an
important element of terrestrial~-~, marine~-~, and space-based navigational systems. 
Nowadays the DE/LE series of ephemerides that are developed in the United States 
and serve, first and foremost, to support the American space research program are 
adopted as the international standard of fundamental ephemerides. The high 
accuracy of these ephemerides is preconditioned by the fact that enormous 
high-quality sets of observational data obtained using terrestrial observatories 
and spacecraft are utilized in creating the DE/LE ephemerides. However, the use 
of the American DE/LE ephemerides may present some difficulties. Among them are 
the problems with licensing (the IAU did not issue recommendations for the use of 
any DE ephemerides except DE200), openness (not all the algorithms are described 
in detail), possible delays (the access to new versions of the DE/LE ephemerides 
may remain restricted for a certain period of time), and reliability. Since 
domestic ephemerides are not subject to these problems, the IAA RAS, developed 
its own EPM ephemerides and uses them when preparing the {\it Astronomical 
Yearbook} (starting from 2006), the {\it Nautical Astronomical Yearbook}, and the 
{\it Nautical Astronomical Almanac}. Besides that, it is planned to use these 
ephemerides in GLONASS and LUNA-RESURS programs.

The EPM ephemerides lie at the basis of much scientific research. For six years 
(1976--1982) the {\it Viking 1} and {\it Viking 2} landers were observed from 
California,
Madrid, and Canberra, while the {\it Pathfinder} lander was observed for three months 
in 1997. These unique observations made it possible to define more precisely the 
rotation of Mars when constructing the EPM ephemerides. The determination of the 
more precise values of the parameters of rotation of Mars is important for 
understanding its geophysics. Firstly, the comparison of the observed and the 
calculated precessions of Mars coupled with the oblateness coefficient of Mars
makes it possible to calculate the normalized polar moment of inertia that allows 
researchers to evaluate the density variations within the planet. Secondly, the
comparison of the determined amplitudes of short-period nutation terms with the 
theoretical predictions enables exploration of the question of the distinctions
between Mars and a rigid body. The observations of Martian landers (on the basis 
of the EPM ephemerides) made it possible to determine the coordinates of all the
three landers and define all parameters of rotation of Mars (precession, nutation, 
and seasonal rotation terms governed by melting and condensation of carbonic acid 
at the polar caps) and the polar moment of inertia, corresponding to the speed of 
precession of Mars (Pitjeva, 1999), more precisely. The parameters of rotation of 
Mars and their accuracies were found to be close to the corresponding values 
taken from a paper by Yoder and Standish (1997).

Asteroids exert a significant influence on the motion of planets (especially 
Mars); therefore, the masses of the largest asteroids (in the present work we
examined the 21 largest ones) and the total mass of the main asteroid belt may be 
estimated from radio observations. Hundreds of trans-Neptunian objects, which
also exert influence on the motions of planets, were discovered recently; their 
total mass may also be estimated, as was already done by Pitjeva (2010a). The
knowledge of such characteristics is important not only for devising a more 
precise description of the forces acting in the Solar System, but also for 
understanding the general dynamics of the Solar System and the processes 
associated with its formation.

The passage of photons and motion of planets in the gravitational filed of the 
Sun allow us to view the Solar System as a sufficiently convenient laboratory for 
testing gravity theories. Modern radar observations of planets and spacecraft, 
that have meter-level accuracy, make it possible to explore relativistic effects,
estimate the value of the heliocentric gravitational constant $GM_{\odot}$ (the 
Sun's mass) and its possible variation, and estimate the solar oblateness. The 
comparison of the results of determination of additional motion of the perihelia 
of planets, which is not modeled by Newtonian interaction and GR, the PPN 
parameters ($\beta, \ \gamma$), the quadrupole moment of the Sun, and 
$GM_{\odot}$, that were cited in previous works (Krasinsky et al., 1986; Pitjeva, 
1993; 2005b; 2010b) and obtained in the present work:
$$ \beta-1 = -0.00002 \pm 0.00003, \ \  \gamma-1 = +0.00004 \pm 0.00006, \ \
J_2 = (2.0 \pm 0.2) \cdot 10^{-7}, $$
shows that, firstly, the uncertainties of these parameters did decrease 
significantly (at least by an order of magnitude). This substantial progress may 
be attributed to the increase in accuracy of the dynamical models of motion and 
the methods of reduction of observations and to the improvement of observational
data (i.e., boost in precision and widening of the observational time interval). 
Secondly, the reduction of the uncertainties of these parameters constrains the
possible values of relativistic parameters and imposes increasingly tight 
restrictions on the gravity theories that are competing with GR.

For the first time, the variation of the heliocentric gravitational constant 
$$\dot {GM_{\odot}}/GM_{\odot} = (-5.0 \pm 4.1) \cdot 10^{-14}$$
per year ($3\sigma$) has been deduced through the analysis of various types 
(mostly radio) of positional observations of planets and spacecraft. The value
obtained, coupled with the known upper limits on the possible variation of the 
Sun's mass, allow us to place tighter restrictions on the variation of the 
gravitational constant and infer that its annual value falls in the interval 
$$-4.2 \cdot10^{-14} < \dot G/G < +7.5 \cdot10^{-14}$$ 
with a probability of $95\%$. The $GM_{\odot}$ variation is seemingly associated 
not with the variation of $G$, but with the variation of the Sun's mass. 
Therefore, the variation of $M_{\odot}$, is reflective of the balance between the 
mass lost through radiation and solar wind and the material falling onto the Sun 
(Pitjeva and Pitjev, 2012).

Besides that, the search for and the estimation of a possible gravitational 
influence of dark matter in the Solar System on the motion of planets has been 
carried out on the basis of the EPM2011 planetary theory by studying the 
additional motion of the perihelia of planets and the estimates of the 
heliocentric gravitational constant obtained through the analysis of observations 
of certain planets. The estimates obtained of the density and mass of dark matter 
at different distances from the Sun are, as a rule, exceeded by their errors 
($\sigma$). This points to the fact that the density of dark matter 
$\rho_{\hbox{dm}}$ (if any) is very low and resides well below the errors of 
determination of such parameters achievable nowadays. It was found that 
$\rho_{\hbox{dm}}$ at the distance of the orbits of Saturn, Mars, and the Earth 
should be lower than $1.1\cdot10^{-20}$ g/cm3, $1.4 \cdot 10^{-20}$ g/cm3, and 
$1.4\cdot10^{-19}$ g/cm3, respectively. The possibility of dark matter 
concentrating at the center of the Solar System was also considered, and it was
found that the mass of dark matter located in the sphere inside the Saturn's 
orbit would still not exceed $1.7 \cdot 10^{-10}M_{\odot}$ (Pitjev and Pitjeva, 
2013).

\bigskip
\centerline{CONCLUSIONS}
\smallskip

The EPM series of high-precision ephemerides of planets and the Moon that is 
faithful to modern observations and comparable in terms of accuracy with the
latest versions of the well-known DE ephemerides (JPL) was created at the IAA RAS. 
The use of a more accurate dynamical model of planetary motion and a large number 
of additional high-precision observations allows us to assert that the latest 
versions (EPM2004--EPM2011) of the EPM ephemerides are more accurate than the 
DE405 ephemerides, which are adopted as an international standard. The EPM 
ephemerides have the following advantages over the DE ones while using EPM for 
Russian astronavigation:
\begin{itemize}
\itemsep -3mm
\item They are constructed using independent and constantly updated software.
\item They are promptly updated and improved according to incoming new data.
\item The clients (GLONASS programs) may request additional needed data in any 
format.
\end{itemize}

Convenient access procedures (Bratseva et al., 2010) for external users were 
recently devised at the IAA RAS. The users may access the EPM ephemerides of 
planets and the Moon together with the corresponding differences TT$-$TDB, as 
well as the ephemerides, computed simultaneously with the EPM ones, of seven 
additional objects (Ceres, Pallas, Vesta, Eris, Haumea, Makemake, and Sedna) that 
are provisionally called dwarf planets. The EPM ephemerides are available at 
ftp://quasar.ipa.nw.ru/incoming/EPM/.

The constructed EPM ephemerides used in practice form the basis of the 
{\it Astronomical Yearbook}, and are needed to fulfill the GLONASS Federal  
Program and to carry out space experiments in the Solar System. They also help us 
to solve some of the problems of fundamental astrometry, including the 
determination of the dynamical structure of the Solar System and a number of 
astronomical constants.

\bigskip
\centerline{ACKNOWLEDGMENTS}
\smallskip

This work was supported by a grant from the RAS Presidium Program 22 
``Fundamental Problems of Research and Exploration of the Solar System''.

\bigskip
\centerline{REFERENCES}
\smallskip
{
\parindent = 0.0truecm

\hangindent = 0.7truecm
\hangafter = 1
{\it Akim E.L., Brumberg V.A., Kislik M.D., et al.} A relativistic theory of 
motion of inner planets // in {\it Proc. 114th IAU Symp. Relativity in Celestial 
Mechanics and Astrometry} / Eds. Kovalevsky J. and Brumberg V.A., Dordrecht: 
Kluver Acad. Publ, 1986. pp.~63-68.

\hangindent = 0.7truecm
\hangafter = 1
{\it Aleshkina E.Yu., Krasinsky G.A., Vasilyev M.V.} Analysis of LLR data by the 
program system ERA //in {\it Proc. 165th IAU Coll. Dynamics and astrometry of 
natural and artificial celestial bodies.} / Eds Wytrzyszczak I.M.,
Lieske J.H., Feldman R.A. Dordrecht: Kluwer Acad. Publ, 1997. pp.~227-238.

\hangindent = 0.7truecm
\hangafter = 1
{\it Bratseva O.A., Novikov F.A., Pitjeva E.V., et al.} Software support of 
IAA planetary and the Moon ephemerides, in {\it Trudy Inst. Appl. Astron., Russ. 
Acad. Sci.}, 2010, vol. 21, pp.~201-204.

\hangindent = 0.7truecm
\hangafter = 1
{\it Fienga A., Manche H., Laskar J., Gastineau M.} INPOP06: A new numerical 
planetary ephemeris // {\it Astron. and Astrophys.} 2008, vol. 477, pp.~315-327.

\hangindent = 0.7truecm
\hangafter = 1
{\it Fienga A., Laskar J., Kuchynka P., et al.} The INPOP10a planetary ephemeris 
and its applications in fundamental physics // {\it Celest. Mech. and Dyn. 
Astron.} 2011, vol. 111, pp.~363-385.

\hangindent = 0.7truecm
\hangafter = 1
{\it Folkner W.M.} Planetary ephemeris DE423 fit to Messenger encounters
with Mercury // {\it JPL Interoffice Memorandum}, Pasadena 2010, vol. 343-10-001,
pp.~1-15.

\hangindent = 0.7truecm
\hangafter = 1
{\it Hilton J.L., Hohenkerk C.Y.} A comparison of the high accuracy planetary 
ephemerides DE421, EPM2008, and INPOP08 // in {\it Systems de reference 
spatio-temporels. Journees-2010} /Ed. Capitaine N., Observatoire de Paris, 2011, 
pp.~77-80.

\hangindent = 0.7truecm
\hangafter = 1
{\it Jones D.L., Fomalont E., Dhawan V., et al.} Very long baseline array
astrometric observations of the Cassini spacecraft at Saturn // {\it Astron. J.} 
2011, vol. 141, Issue 2, Article~29.

\hangindent = 0.7truecm
\hangafter = 1
{\it Klioner S.A., Gerlach E., Soffel M.H.} Relativistic aspects of rotational 
motion of celestial bodies // in {\it Proc. 165th IAU Symp. Giant Step: from 
Milli- to Micro-arcsecond Astrometry} / Eds. Klioner S.A., Seidelmann P.K., 
Soffel M., Cambridge Univ., 2010, pp.~112-123.

\hangindent = 0.7truecm
\hangafter = 1
{\it  Konopliv A.S.,  Asmar S.W.,  Folkner W.M., et al.} Mars high
resolution gravity field from MRO, Mars seasonal gravity, and other
dynamical parameters // {\it Icarus}, 2011, vol. 211, pp.~401-428.

\hangindent = 0.7truecm
\hangafter = 1
{\it Krasinsky G.A., Aleshkina E.Yu., Pitjeva E.V., Sveshnikov M.L.} Relativistic 
effects from planetary and lunar observations of the XVIII-XX centuries // in 
{\it Proc. 114th IAU Symp. Relativity in celestial mechanics and astrometry}, / Eds. 
Kovalevsky J., Brumberg V.A., Dordrecht: Kluver Acad. Publ, 1986, pp.~315-328.

\hangindent = 0.7truecm
\hangafter = 1
{\it Krasinsky G.A.,  Pitjeva E.V.,  Sveshnikov M.L., Chunajeva L.I.} The motion 
of major planets from observations 1769-1988 and some astronomical constants // 
{\it Celest. Mech. and Dyn. Astron.} 1993, vol. 55, pp.~1-23.

\hangindent = 0.7truecm
\hangafter = 1
{\it Krasinsky G.A., Vasilyev M.V.}  ERA: knowledge base for ephemeris and 
dynamic astronomy // in {\it Proc. 165th IAU Coll. Dynamics and astrometry of 
natural and artificial celestial bodies} / Eds. Wytrzyszczak I.M., Lieske J.H., 
Feldman R.A., Dordrecht: Kluwer Acad. Publ, 1997, pp.~239-244.

\hangindent = 0.7truecm
\hangafter = 1
{\it Krasinsky G.A.} Selenodynamical parameters from analysis of LLR observations 
of 1970-2001 // {\it Commun. IAA RAS}, 2002a, vol. 148, pp.~1-27.

\hangindent = 0.7truecm
\hangafter = 1
{\it Krasinsky G.A., Pitjeva E.V., Vasilyev M. V., Yagudina E. I.} Hidden mass in 
the asteroid belt // {\it Icarus}, 2002b, vol. 158, pp.~98-105.

\hangindent = 0.7truecm
\hangafter = 1
{\it Luzum B., Capitaine N., Fienga A., et al.} The IAU 2009 system of 
astronomical constants: The report of the IAU working group on numerical 
standards for fundamental astronomy // {\it Celest. Mech. and Dyn. Astron.}, 2011, 
vol. 110, pp.~293-304.

\hangindent = 0.7truecm
\hangafter = 1
{\it Pitjev N.P. and Pitjeva E.V.} Constraints on dark matter in the solar system, 
{\it Astron. Lett.}, 2013, vol. 39, no. 3, pp.~141-149.

\hangindent = 0.7truecm
\hangafter = 1
{\it Pitjeva E.V.} Experimental testing of relativistic effects, variability of 
the gravitational constant and topography of Mercury surface from radar 
observations 1964-1989 // {\it Celest. Mech.}, 1993. vol. 55, pp.~313-321.

\hangindent = 0.7truecm
\hangafter = 1
{\it Pitjeva E.V.} ERM98: the new numerical motion theory for planets and its 
comparison with DE403 ephemerede of Jet Propulsion Laboratory, {\it Trudy Inst. 
Appl. Astron., Russ. Acad. Sci.}, 1998, vol. 3, pp.~5-23.

\hangindent = 0.7truecm
\hangafter = 1
{\it Pitjeva E.V.} Study of Mars dynamics from analysis of Viking and Pathfinder 
lander radar data // {\it Trudy Inst. Appl. Astron., Russ. Acad. Sci.}, 1999, 
vol. 4, pp.~22-35.

\hangindent = 0.7truecm
\hangafter = 1
{\it Pitjeva E.V.} Modern numerical ephemerides of planets and the importance of 
ranging observations for their creation //{\it Celest. Mech. and Dyn. Astron.}, 
2001, vol. 80, pp.~249-271.

\hangindent = 0.7truecm
\hangafter = 1
{\it Pitjeva E.V.} High-precision ephemeredes of planets -- EPM and determination 
of some astronomical constants, {\it Solar Syst. Res.}, 2005a, vol. 39, no. 3, 
pp.~176-186.

\hangindent = 0.7truecm
\hangafter = 1
{\it Pitjeva E.V.} Relativistic effects and solar oblateness from radar 
observations of planets and spacecraft // {\it Astron. Lett.}, 2005b, vol. 31, 
No. 5, pp.~340-349.

\hangindent = 0.7truecm
\hangafter = 1
{\it Pitjeva E.V.} Highly accurate planetary ephemeredes of IAA RAS - EPM2008 and 
their orientation in ICRF system, {\it Trudy Inst. Appl. Astron., Russ. Acad. Sci.}, 
2009a, vol. 20, pp.~463-466. 

\hangindent = 0.7truecm
\hangafter = 1
{\it Pitjeva E.V., Standish E.M.} Proposal1s for the masses of the three largest 
asteroids, the Moon-Earth mass ratio and the Astronomical Unit // {\it Celest. 
Mech. and Dyn. Astron.}, 2009. vol. 103, pp.~365-372.

\hangindent = 0.7truecm
\hangafter = 1
{\it Pitjeva E.V.} EPM ephemerides and relativity // in {\it Proc. 165th IAU Symp. 
Giant Step: from Milli- to Micro-arcsecond Astrometry}, / Eds. Klioner S.A., 
Seidelmann P.K., Soffel M., Cambridge: Univ. Press, 2010a, pp.~170-178.

\hangindent = 0.7truecm
\hangafter = 1
{\it Pitjeva  E. V.} Influence of asteroids and trans-Neptunian objects on motion 
of major planets and masses of the asteroid main belt and TNO ring // {\it Proc. 
Int. Conf. Asteroid-comet hazard-2009}/ Eds. Finkelstein A., Huebner W., Shor V., 
Saint-Peterburg: Nauka, 2010b, pp.~237-241.

\hangindent = 0.7truecm
\hangafter = 1
{\it Pitjeva E.V.} Fundamental national ephemeredes of planets and the Moon (EPM)
by the Institute of Applied Astronomy of the Russian Academy of Sciences: 
dynamical model, parameters, accuracy, {\it Trudy Inst. Appl. Astron., Russ. Acad. 
Sci.}, 2012a, vol. 23, pp.~149-157.

\hangindent = 0.7truecm
\hangafter = 1
{\it Pitjeva E.V., Pitjev N.P.} Changes in the Sun's mass and gravitational 
constant estimated using modern observations of planets and spacecraft // {\it 
Solar Syst. Res.}, 2012b, vol. 46, No. 1, pp.~78-87.

\hangindent = 0.7truecm
\hangafter = 1
{\it Poroshina A., Kosmodamianskiy G., Zamarashkina M.} Construction of the 
Numerical motion theories for the main satellites of Mars, Jupiter, Saturn, and 
Uranus in IAA RAS // {\it Proc. JENAM-2011}, St. Petersburg, {\it Trudy Inst. Appl. 
Astron., Russ. Acad. Sci.}, 2012, vol. 26, pp.~75-87.

\hangindent = 0.7truecm
\hangafter = 1
{\it Russell C.T., Raymond C.A., A. Coradini C.A., et al.} Dawn at Vesta: Testing 
the protoplanetary paradigm // {\it Science}, 2012, vol. 336, pp.~684-686.

\hangindent = 0.7truecm
\hangafter = 1
{\it Standish E.M., Newhall XX, Williams J.G., Folkner W.M.} JPL Planetary and 
Lunar Ephemerides, DE403/LE403 // {\it JPL Interoffice Memorandum}, Pasadena 1995, 
vol.~314.10-127, pp.~1-22.

\hangindent = 0.7truecm
\hangafter = 1
{\it Standish E.M.} JPL Planetary and Lunar Ephemerides, DE405/LE405 // {\it JPL 
Interoffice Memorandum}, Pasadena 1998, vol.~312.F-98-048, pp.~1-18.

\hangindent = 0.7truecm
\hangafter = 1
{\it Standish E.M.,  Fienga A.} Accuracy limit of modern ephemerides imposed by 
the uncertainties in asteroid masses // {\it Astron. and Astrophys.}, 2002,
vol.~384, pp.~322-328.

\hangindent = 0.7truecm
\hangafter = 1
{\it Standish E.M.} An approximation to the errors in the planetary ephemerides 
of the Astronomical Almanac // {\it Astron. and Astrophys.}, 2004, vol.~417. 
pp.~1165-1171.

\hangindent = 0.7truecm
\hangafter = 1
{\it Williams J.G.} Determining asteroid masses from perturbations on Mars // 
{\it Icarus}, 1984, vol.~57, pp.~1-13.

\hangindent = 0.7truecm
\hangafter = 1
{\it Williams J.G.} Harmonic analysis // {\it Bull. Am. Astron. Soc.}, 1989, 
vol.~21, pp.~1009-1010.

\hangindent = 0.7truecm
\hangafter = 1
{\it Yagudina E.I., Krasinskii G.A., Prokhorenko S.O.} EPM-ERA ephemeride of the 
Moon and selenodynamical parameters from LLR observations processing, {\it Trudy 
Inst. Appl. Astron., Russ. Acad. Sci.}, 2012, vol.~23, pp.~165-171. 

\hangindent = 0.7truecm
\hangafter = 1
{\it Yoder C.F., Standish E.M.} Martian precession and rotation from
 Viking lander range data // {\it J. Geophys. Res.} 1997, vol. 102, pp.~4065-4080.

\end{document}